\documentclass[letterpaper,11pt]{article}
\pdfoutput=1

\usepackage{jheppub}

\usepackage{multirow,subfig,xspace,xcolor}
\usepackage[countmax]{subfloat}

\definecolor{darkblue}{rgb}{0,0,0.5}

\definecolor{darkgreen}{rgb}{0,0.5,0}

\DeclareRobustCommand{\Sec}[1]{Sec.~\ref{#1}}
\DeclareRobustCommand{\Secs}[2]{Secs.~\ref{#1} and \ref{#2}}
\DeclareRobustCommand{\Secss}[3]{Secs.~\ref{#1}, ~\ref{#2}, and \ref{#3}}

\DeclareRobustCommand{\Fig}[1]{Fig.~\ref{#1}}
\DeclareRobustCommand{\Figs}[2]{Figs.~\ref{#1} and \ref{#2}}
\DeclareRobustCommand{\Eq}[1]{Eq.~(\ref{#1})}
\DeclareRobustCommand{\Eqs}[2]{Eqs.~(\ref{#1}) and (\ref{#2})}
\DeclareRobustCommand{\Ref}[1]{Ref.~\cite{#1}}
\DeclareRobustCommand{\Refs}[1]{Refs.~\cite{#1}}

\preprint{MIT--CTP 5049}

\title{A Theory of Quark vs.~Gluon Discrimination}

\author[1]{Andrew J. Larkoski}
\affiliation[1]{Physics Department, Reed College, Portland, OR 97202, USA}

\author[2,3]{and Eric M.~Metodiev}
\affiliation[2]{Center for Theoretical Physics, Massachusetts Institute of Technology, Cambridge, MA 02139, USA}
\affiliation[3]{Department of Physics, Harvard University, Cambridge, MA 02138, USA}

\emailAdd{larkoski@reed.edu}
\emailAdd{metodiev@mit.edu}

\abstract{
Understanding jets initiated by quarks and gluons is of fundamental importance in collider physics.
Efficient and robust techniques for quark versus gluon jet discrimination have consequences for new physics searches, precision $\alpha_s$ studies, parton distribution function extractions, and many other applications.
Numerous machine learning analyses have attacked the problem, demonstrating that good performance can be obtained but generally not providing an understanding for what properties of the jets are responsible for that separation power.
In this paper, we provide an extensive and detailed analysis of quark versus gluon discrimination from first-principles theoretical calculations.
Working in the strongly-ordered soft and collinear limits, we calculate probability distributions for fixed $N$-body kinematics within jets with up through three resolved emissions (${\cal O}(\alpha_s^3)$).
This enables explicit calculation of quantities central to machine learning such as the likelihood ratio, the area under the receiver operating characteristic curve, and reducibility factors within a well-defined approximation scheme.
Further, we relate the existence of a consistent power counting procedure for discrimination to ideas for operational flavor definitions, and we use this relationship to construct a power counting for quark versus gluon discrimination as an expansion in $e^{C_F-C_A}\ll1$, the exponential of the fundamental and adjoint Casimirs.
Our calculations provide insight into the discrimination performance of particle multiplicity and show how observables sensitive to all emissions in a jet are optimal.
We compare our predictions to the performance of individual observables and neural networks with parton shower event generators, validating that our predictions describe the features identified by machine learning.
}

\begin{document}
\maketitle

\section{Introduction}

High energy quarks and gluons fragment and hadronize into jets of particles through quantum chromodynamics (QCD).
Identifying light jets as arising from quarks or gluons is a fundamental challenge for collider physics at the Large Hadron Collider (LHC).
Many efforts have proposed new observables~\cite{Nilles:1980ys,Jones:1988ay,Fodor:1989ir,Jones:1990rz,Pumplin:1991kc,Gallicchio:2011xc,Gallicchio:2011xq,Gallicchio:2012ez,FerreiradeLima:2016gcz,Frye:2017yrw,Davighi:2017hok,Komiske:2018cqr} or jet flavor definitions \cite{Banfi:2006hf,Frye:2016aiz,Gras:2017jty,Metodiev:2018ftz,Komiske:2018vkc}, completed theoretical calculations \cite{Larkoski:2014pca,Bhattacherjee:2015psa,Mo:2017gzp,Sakaki:2018opq}, and used machine learning methods \cite{Lonnblad:1990qp,Komiske:2016rsd,Cheng:2017rdo,Luo:2017ncs,Kasieczka:2018lwf} to push the boundaries of the discrimination power between quark and gluon jets.
While these studies have led to steady improvements over time, they have been done with no clear organizing principle or agreed-upon ``best'' discrimination strategy.
Further, while machine learning methods have demonstrated the greatest discrimination power, no clear physical reason for that performance has been presented.
It is therefore desirable to construct a general theory of quark versus gluon discrimination which both explains and provides robust understanding of the discrimination power.

Previous studies have made progress in this direction.
For example, \Ref{Gallicchio:2012ez} was the first broad study of the quark versus gluon discrimination power of a large number of jet observables in simulation, including identifying those pairs of observables that improved discrimination power the most. 
\Ref{Larkoski:2014pca} introduced mutual information as a metric for useful discrimination information in distributions, applying it to pairs of generalized angularities \cite{Berger:2003iw,Almeida:2008yp,Ellis:2010rwa} measured on jets.
Resummed predictions of mutual information were performed and compared to simulation which concretely enabled identification of features that are both under theoretic control and well-described by simulation.
Nevertheless, this study was limited to observables that are first non-zero for jets with two particles in them.
In \Ref{Frye:2017yrw}, an infrared and collinear (IRC) safe definition of multiplicity was introduced, based on a generalization of the soft drop grooming algorithm \cite{Larkoski:2014wba}.
This observable, called soft drop multiplicity $n_\text{SD}$, counts the number of relatively hard, angular-ordered emissions off of the hard jet core. 
At leading-logarithmic accuracy, it can be proven that $n_\text{SD}$ is the optimal quark versus gluon discriminant, on the phase space of particles directly emitted off of the hard initiating particle of the jet.
However, this is not a proof that $n_\text{SD}$ is the optimal observable for quark versus gluon discrimination in general, because there are regions of phase space in which emissions live that may improve discrimination power, but to which $n_\text{SD}$ does not have access.

In this paper, we present a first systematic theoretical analysis of quark versus gluon discrimination.
Working in the strongly-ordered soft and collinear limits, we explicitly calculate the resummed probability distribution of multiple infrared and collinear safe observables on a jet.
These multiple observables enable a characterization of the emission phase space and evaluation of the optimal observable for discrimination.
We calculate the energy distributions for quark and gluon jets with up to three resolved emissions, though nothing prohibits continuing to arbitrary numbers of emissions.
Our approximations enable simple, recursive evaluation of the probability distribution as a product of conditional probability distributions.
Though simple, these calculations are sufficient to validate predictions and make several concrete conjectures regarding quark versus gluon discrimination to all-orders.

Our first step in developing a theory of quark versus gluon discrimination is to establish a robust power counting scheme that can be used to construct individual observables, strictly from general statements about the singular limits of QCD.
Power counting rules for observables useful for discriminating multi-prong substructure in jets has been extensively developed~\cite{Larkoski:2014gra,Larkoski:2014zma,Moult:2016cvt}.
An observable parametrically separates jet categories if power counting identifies arbitrarily pure samples at the phase space boundaries, enabling an unambiguous definition in a singular limit.
By a pure sample we mean that a formal region of phase space, however small, exclusively consists of one type or category of jet. 
For binary classification, the existence of such pure phase space regions provides a robust definition of the jet categories, which is referred to as ``mutual irreducibility''~\cite{Metodiev:2018ftz} of the samples being discriminated.
The complementary ideas of power counting and mutual irreducibility are powerful tools we exploit to identify pure phase space regions and quantify potential discrimination power.

The precise notion of mutual irreducibility is relatively new in particle physics, but the requirement that pure phase space regions are necessary to unambiguously define jet categories is well-understood.
Throughout this paper, we refer to ``signal'' and ``background'' jets in an idealized sense, assuming that we have perfect knowledge of the jet categories.
Then, on a restricted space of measurements on those jets, we study the possible discrimination power accessible by those measurements.
Thus, even if two jet samples are not mutually irreducible on some restricted observable phase space, we are still able to use our perfect knowledge to study their separation.
This notion is widely used in discrimination studies in jet physics, though often not explicitly stated.
For jet samples that are not able to be purified on phase space, a so-called reducibility factor $\kappa$ is defined as the accessible purity of signal or background phase space regions.
Further, reducibility factors are just the limiting values of the likelihood ratio and, as we will show, they quantify parametric discrimination power.  

As a first familiar example, we demonstrate mutual irreducibility for a problem in which power counting is well-understood: in the context of QCD jet versus hadronically-decaying, boosted $Z$ boson discrimination.
Power counting for quark and gluon jets is intrinsically more difficult because, as we demonstrate on any phase space with finitely-many resolved emissions, quark and gluon jets are not strictly mutually irreducible.
Thus a power counting scheme does not currently exist to identify robust phase space boundaries between both quark-pure and gluon-pure regions.
Nevertheless, because the rates of particle emission from quarks and gluons are controlled by the color Casimirs of the fundamental and adjoint representations with $C_F = 4/3 < C_A = 3$, gluon jets exhibit greater Sudakov suppression near the singular phase space boundaries, and so one can define a quark-pure region of phase space.
This motivates using the power counting parameter $e^{C_F-C_A}\simeq 0.189$ to identify such a phase space region, which we formally take to be parametrically smaller than 1.
A gluon-rich phase space region is then one for which Sudakov factors are irrelevant and approximately unity.

With explicit, analytic expressions for multi-differential cross sections measured on quark and gluon jets, we are able to calculate any of the quantities familiar from statistics and machine learning, but within the context of a well-defined approximation scheme, with no black boxes.
By the Neyman-Pearson lemma~\cite{Neyman289}, the optimal binary discrimination observable formed from the measurement of some collection of observables is the likelihood ratio.
This is simply the ratio of the corresponding probability distributions for quark and gluon jets, and will provide a benchmark when comparing to other observables.
The likelihood ratio is in general some complicated function of the phase space variables that does not enable a simple determination of the receiver operating characteristic (ROC) or signal versus background efficiency curve.
Nevertheless, the discrimination power of the likelihood ratio, or any observable, can be quantified by the area under the ROC curve (AUC).
We use a ROC convention where AUC $=0$ is perfect performance and a random classifier has AUC $=\frac12$.
The AUC can be calculated directly from an ordered integral of the product of quark and gluon probability distributions.
This also enables a variational approach to construct discrimination observables, whose parameters are chosen to minimize the AUC.

Our results enable a number of statements that we prove at this accuracy including:
\begin{itemize}
\item Due to Sudakov suppression and since $C_F < C_A$, the reducibility factor for quark jets is $\kappa_q=0$ for the measurement of any number of resolved emissions in the jets.  Pure quark jet phase space regions can essentially always be defined.

\item For jets on which measured observables resolve $n$ emissions, the reducibility factor for gluon jets $\kappa_g$ is
\begin{equation}
\kappa_g = \left(\frac{C_F}{C_A}\right)^n\,.
\end{equation}
A fully pure gluon jet phase space region can therefore only be exactly defined if all emissions are resolved.
The gluon-rich region of phase space is where Sudakov factors are irrelevant, and so is well-described at fixed-order.
This particular scaling comes from diagrams in which all particles in the jet are emitted off of the initiating hard particle, ensuring maximum sensitivity to the color Casimirs $C_F$ and $C_A$.

\item There is an upper limit on the quark vs.~gluon discrimination performance with $n$ resolved emissions of
\begin{equation}
\text{AUC} \ge \frac{\kappa_q+\kappa_g-2\kappa_q\kappa_g}{2-2\kappa_q \kappa_g} = \frac12 \left(\frac{C_F}{C_A}\right)^n\,,
\end{equation}
at this accuracy, with even stronger bounds for specific observables.
This bound follows from monotonicity of the ROC and its first derivative, and so the reducibility factors define a quadrilateral whose area is necessarily no larger than the AUC.
Analogous bounds on other measures of classification performance can also be derived.

\end{itemize}

We also are able to make a number of well-motivated conjectures that follow from our explicit calculations including:
\begin{itemize}
\item The reducibility factor of gluon jets does not improve by resolving the full $3n-4$ dimensional phase space for a jet with $n$ constituents.  One only needs to measure $n-1$ observables to resolve the existence of each emission off of the initiating gluon.

\item Multiplicity is a powerful discrimination observable because it is sensitive to every emission in the jet.  Because $C_F/C_A \simeq 0.444$, the gluon reducibility factor of multiplicity quickly converges to 0 as the number of particles in the jet increases.

\item The discrimination power of a single observable $\tau_n$ that is sensitive to $n$ emissions in a jet, such that its value is 0 if the jet has fewer than $n$ emissions, is bounded by multiplicity.  The performance of $\tau_n$ increases with $n$ for small $n$, and degrades when $n$ is comparable to the total number of particles in the jets.  An optimal value of $n$ occurs when $n$ is comparable to the minimal number of constituents of gluon jets.

\item Unlike the case for discrimination of jets with different multi-prong substructure, the likelihood for quark vs.~gluon discrimination is an IRC-safe observable.  By the established power counting, the most singular region of phase space is necessarily pure quark jet, and so contours of constant likelihood should be parallel to this boundary.  Therefore, the singular region of phase space is mapped to a unique value of the likelihood.  This means that the distribution of the likelihood ratio can be calculated in fixed-order perturbation theory.

\end{itemize}
We perform an analysis of quark versus gluon discrimination in a Monte Carlo parton shower to validate that these results describe the physics in simulation.

This paper is organized as follows.
In \Sec{sec:obsapp}, we establish the observables that we measure on jets and clearly lay out our approximations.
While this is not a precision QCD study, our approximations become increasingly accurate as the jet energy increases.
\Sec{app:pcmr} reviews and relates concepts from power counting and mutual irreducibility, outlining our general conceptual and mathematical approach.
In \Sec{sec:pc}, we construct the rules for power counting on the observable phase space for quark versus gluon jet discrimination. 
Several results then immediately follow from these rules, which we validate in later sections.
\Secss{sec:one}{sec:two}{sec:three} contain our explicit calculations for jets on which one, two, or three emissions are resolved, respectively.
For concreteness, we focus our calculations on $N$-subjettiness~\cite{Stewart:2010tn,Thaler:2010tr,Thaler:2011gf} observables, though to our accuracy identical results follow for other observables, such as (generalized) energy correlation functions \cite{Tkachov:1994as,Tkachov:1995kk,Larkoski:2013eya,Moult:2016cvt,Komiske:2017aww}.
We are able to construct an IRC safe definition of multiplicity that depends on a resolution parameter $\Lambda_0 > 0$. 
\Sec{sec:mult} is devoted to calculations of the distribution of this multiplicity observable and developing an understanding of the ``true'' multiplicity limit for $\Lambda_0 \to 0$.
Simulated events are analyzed in \Sec{sec:mc}, in which we both test our predictions and verify that simulation describes physics as expected.
For high dimensional phase space, we utilize machine learning techniques to approximate the likelihood and related discrimination observables in simulation.
We conclude in \Sec{sec:conc} and look forward to further advancements in probing and defining quark and gluon jets.  An appendix applies reducibility ideas to the problem of up vs.~down quark jet discrimination.

\section{Approximations and Observables}\label{sec:obsapp}

We work to leading-logarithmic accuracy in the strongly-ordered soft and collinear limits of QCD with fixed coupling.
This means that we will successfully resum all double logarithms, terms in the fixed-order cross section that scale as $\alpha_s^n \log^{2n} {\cal O}$, of the observables ${\cal O}$ that we measure on our quark and gluon jets.
While this approximation clearly has its limitations, it does enable explicit, analytic formulas for all of the cross sections that we present in this paper.
Further, Sudakov factors in the double logarithmic limit can be easily calculated from the areas of emission veto regions in the Lund plane~\cite{Andersson:1988gp}.
We briefly present results for calculations beyond this accuracy from elsewhere in the literature in \Sec{sec:one}.

At double logarithmic order, the hard, initiating parton defines the jet flavor and so there is no ambiguity in the definition of quark and gluon jets.  
The subtleties in defining a jet flavor beyond this accuracy have been addressed by the community in a review article~\cite{Gras:2017jty} and it remains an active research direction, with recent efforts to define quark and gluon jets based directly upon mutual irreducibility ideas~\cite{Metodiev:2018ftz,Komiske:2018vkc}.
We also do not include non-perturbative physics due to hadronization, for example, which would be needed for precision comparison to data.  For IRC safe observables, the effects of non-perturbative physics is suppressed by $\Lambda_\text{QCD}/Q$ where $Q$ is some characteristic high energy scale ($\sim 1$ TeV), so our calculations will have an increasingly large domain of applicability at higher energies.  Nevertheless, at any finite $Q$, there is always some region of phase space dominated by non-perturbative physics.

Given the double-logarithmic approximation, in this paper we choose to analyze sets of $N$-subjettiness observables measured on our jets.
$N$-subjettiness observables vanish for configurations of $n<N$ particles, and hence they probe the degree to which a jet can be described by $N$-subjets.
The definition of $N$-subjettiness $\tau_N^{(\beta)}$ that we use when measured on jets at a hadron collider is
\begin{equation}
\tau_N^{(\beta)} = \frac{1}{p_{TJ}R_0^\beta}\sum_{i\in J} p_{Ti}\min\left\{
R_{i1}^\beta,R_{i1}^\beta,\dotsc,R_{iN}^\beta
\right\}\,.
\end{equation}
Here, $p_{TJ}$ is the transverse momentum of the jet with respect to the colliding beam axis, $R_0$ is the jet radius, the sum runs over all particles $i$ in the jet $J$, $p_{Ti}$ is the transverse momentum of particle $i$, and $R_{iK}$ is the distance in the rapidity-azimuth plane from particle $i$ to subjet axis $K$ in the jet.
Specifically, $R_{iK}$ is
\begin{equation}
R_{iK} = \sqrt{(y_i-y_K)^2 + (\phi_i - \phi_K)^2}\,,
\end{equation}
in terms of the respective rapidity $y$ and azimuthal angle $\phi$ of the particle and axis about the colliding beam axis.
The $N$-subjettiness observables are IRC safe with the angular exponent $\beta > 0$ and to our approximation, any recoil-free axis definition suffices for our calculations of the discrimination power.
In fact, our calculations hold even with recoil for choices of $\beta$ where the same emission dominates both the axis position and the value of the observable.
However, we will have to make an explicit choice of axes in our simulation, which we will discuss in \Sec{sec:mc}.

$N$-subjettiness observables are nice for calculation both because they are IRC safe, and so are calculable at fixed-order in perturbation theory, and additive, and so can be resummed to double logarithmic accuracy simply.
Measuring a sufficient number of these observables can be used to completely specify the $3M-4$ dimensional phase space of a jet with $M$ particles~\cite{Datta:2017rhs}.
$N$-subjettiness is not unique in these points, but the linear computational complexity in the number of particles (after determining axes) means that calculating $\tau_N^{(\beta)}$ for large $N$ ($N\gtrsim 5$) is not computationally prohibitive within simulation.

Further, to the accuracy of our calculations, the angular exponent $\beta$ does not affect the discrimination power of the $N$-subjettiness observables that we measure on the jets.
Effectively, to double logarithmic accuracy, the angular exponent can be absorbed into a redefinition of the coupling $\alpha_s \to \alpha_s/\beta$, which is the same for quark and gluon jets.
Therefore, we typically will simply fix $\beta = 1$ in our calculations so that $N$-subjettiness measures the total momentum that is transverse to the $N$ subjet axes in the jet.
For compactness, we denote $\tau_N^{(\beta=1)} \equiv \tau_N$ throughout this article.
However, in \Sec{sec:one}, we will discuss the effects of measuring two $1$-subjettiness observables on jets and higher-order effects of the angular exponent, in which we maintain explicit $\beta$ dependence in the observable definition.

An observable which counts the number of resolvable, angular-ordered emissions off of a hard core was introduced in \Ref{Frye:2017yrw}, referred to as soft drop multiplicity $n_\text{SD}$.
There, it was argued that $n_\text{SD}$ is the optimal quark vs.~gluon discriminator at leading-logarithmic accuracy for observables on the phase space of those particles emitted directly off of the hard core of the jet.
For such emissions, the rate of emission is controlled by the appropriate color Casimir and the kinematic distribution of the emissions is identical between quarks and gluons. 
Thus, all discrimination information is contained in simply counting the emissions, with the kinematics adding no discrimination power.
In this paper, we will consider the more general case of jets with relevant emissions off of emitted particles.
In this more general case, the quark and gluon kinematic distributions on phase space are no longer equal, because there are different weights on the phase space regions in which such secondary emissions could live, depending on the quark or gluon color Casimirs.
We explicitly demonstrate that there is discrimination information in kinematic distributions, beyond just counting emissions.

\section{Power Counting and Mutual Irreducibility}
\label{app:pcmr}

Using power counting to identify optimal observables for classification~\cite{Larkoski:2014gra,Larkoski:2014zma,Moult:2016cvt} is a conceptual framework that has led to new jet substructure observables which have successfully been applied to analyses at the LHC~\cite{Aad:2015rpa,Aaboud:2016qgg,Aaboud:2018psm,Aaboud:2019aii}.
The key idea is to identify regions of phase space that parametrically separate signal and background.
An observable is then optimal in this framework if it separates the signal-dominated and background-dominated regions of phase space.
A robust power counting on a phase space of jet observables requires that the boundaries of that phase space define pure regions of the underlying categories.
That is, for power counting of discrimination observables as studied in earlier work, this implicitly requires that the two discriminated samples are ``mutually irreducible''.

Mutual (ir)reducibility was first introduced in a collider physics context to statistically disentangle or define different types of jets from mixed samples~\cite{Metodiev:2018ftz,Komiske:2018vkc}.
Signal and background categories are said to be mutually irreducible if there exist pure phase space regions, however small, for each of the categories.
Further, the degree to which two categories are mutually irreducible can be sharply quantified in terms of their reducibility factors:
\begin{equation}
\kappa_S \equiv \min_{\mathcal O} \frac{p_B(\mathcal O)}{p_S(\mathcal O)}\,,
\quad \quad \quad \quad
\kappa_B \equiv \min_{\mathcal O} \frac{p_S(\mathcal O)}{p_B(\mathcal O)}\,,
\end{equation}
where ${\cal O}$ is an observable or set of observables that define some phase space.
$p_S({\cal O})$ and $p_B({\cal O})$ are the probability distributions of the observable measured on signal and background, respectively.\footnote{In the notation of \Refs{Metodiev:2018ftz,Komiske:2018vkc}, our $\kappa_S$ and $\kappa_B$ are $\kappa_{BS}$ and $\kappa_{SB}$, respectively. While the two-index notation generalizes to more categories, we use our simplified notation for the two-class context of this paper.}
Evidently, if there is a region of phase space where signal dominates then its reducibility factor vanishes, $\kappa_S=0$.
Similarly, $\kappa_B = 0$ if and only if there is a region of phase space where background dominates.
Hence the categories are mutually irreducible only when $\kappa_S=\kappa_B=0$.

Here, we will use the language and mathematical machinery of mutual (ir)reducibility for a new purpose: as a technique to quantify the parametric separability of two calculated distributions.
The central importance of pure phase space regions is shared with power counting strategies.
In particular, these ideas will allow us to quantify the power counting ideas in a new way and apply them to quark versus gluon jet classification.
While previous studies have conjectured that quark vs.~gluon discrimination did not admit a power counting~\cite{Larkoski:2014gra}, later efforts have identified requirements on observables to go beyond the leading-order $C_A/C_F$ separation \cite{Moult:2016cvt}.
Our definition of power counting for quarks and gluons here will be much more general than previous considerations and enable analysis of arbitrary multi-differential probability distributions.

\subsection{Theoretically Bounding Classification Performance}

While we will have our quark vs.~gluon case in mind for the following discussion, we keep the signal vs.~background terminology general in order to highlight the broad applicability of this reasoning.
The signal and background reducibility factors are related to the derivatives of the ROC curve near its endpoints.
Note that the ROC curve is the background cumulative distribution evaluated at the inverse of the signal cumulative distribution:
\begin{equation}
\text{ROC}(x)=\Sigma_B\left(
\Sigma_S^{-1}(x)
\right)\,,
\end{equation}
for signal efficiency $x$. The derivative of the ROC curve is then
\begin{equation}
\frac{d}{dx}\text{ROC}(x) = \frac{d}{dx}\Sigma_B\left(
\Sigma_S^{-1}(x)
\right) =  \frac{p_B(\Sigma_S^{-1}(x))}{p_S(\Sigma_S^{-1}(x))} = \frac{p_B(\mathcal O(x))}{p_S(\mathcal O(x))}\,,
\end{equation}
which is precisely the signal-background likelihood ratio for the observable value $\mathcal O(x)$ giving rise to signal efficiency $x$.
The emergence of the likelihood ratio as centrally relevant highlights the close relationship between mutual (ir)reducibility, power counting ideas, and optimal classification.

\begin{figure}
\begin{center}
\includegraphics[width=7cm]{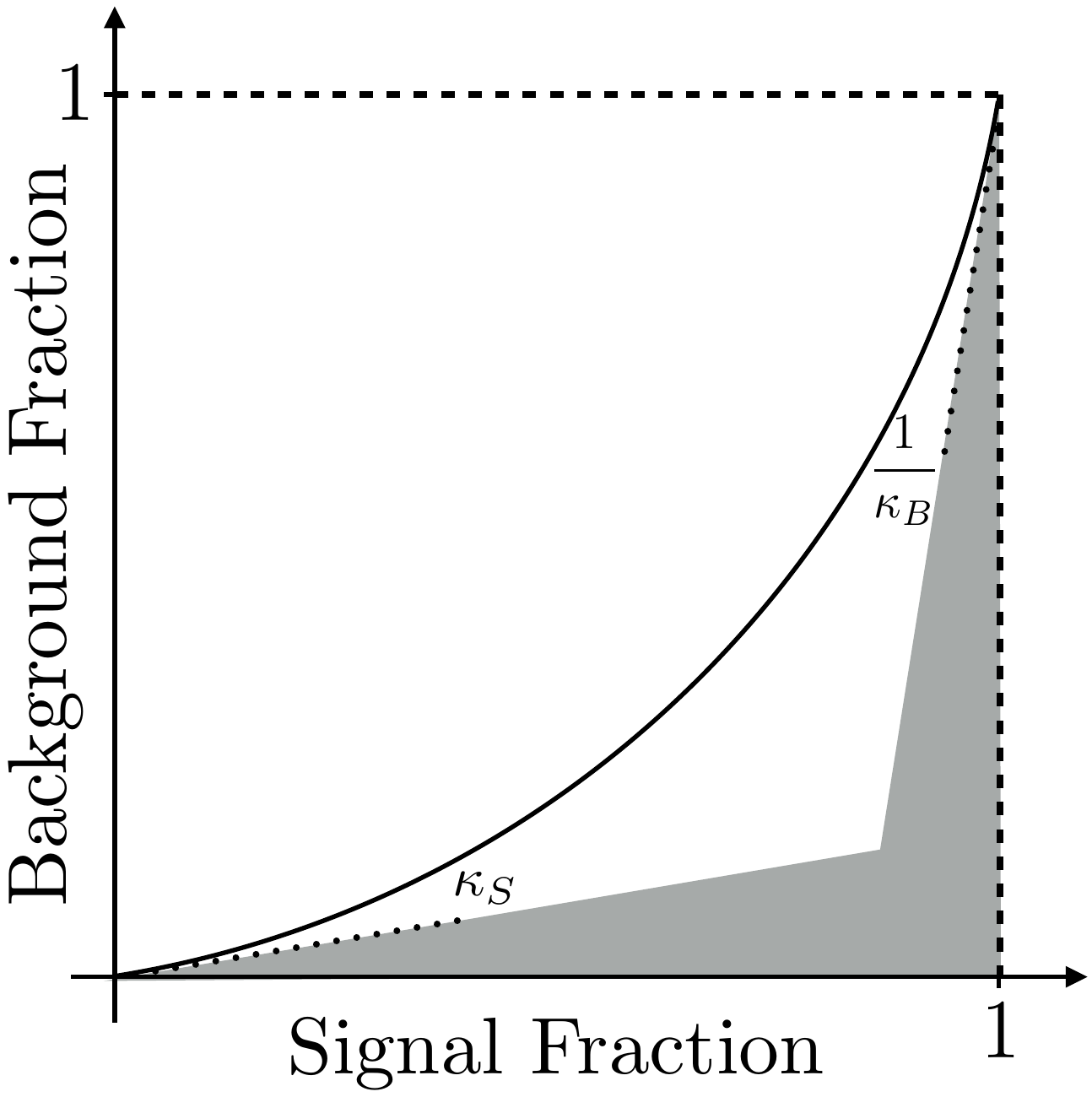}
\end{center}
\caption{An illustration of the bound on the ROC curve and its AUC from extrapolating the reducibility factor slopes $\kappa_S$ and $1/\kappa_B$ from the endpoints.
The ROC is monotonic and concave up and so the gray quadrilateral is always completely contained underneath the full ROC curve, yielding the bound.
}
\label{fig:aucbound}
\end{figure}

This relationship between reducibility factors and the ROC curve can be exploted further: we now prove a strict lower bound on the ROC curve and its AUC from the reducibility factors.
The ROC curve can be taken to be strictly monotonic with a positive first derivative because the value of the ROC curve between any two points can (at worst) be a random weighting of the values at those points.
The reducibility factors are then the slope (or its inverse) of the ROC curve at the appropriate endpoints.
Therefore, we can bound the area under the ROC curve by a quadrilateral, of which the angle of two of the vertices are set by the values of $\kappa_S$ and $\kappa_B$. 
An illustration of this quadrilateral for a general ROC curve is shown in \Fig{fig:aucbound}.
Its area is straightforward to compute, yielding the bound
\begin{equation}\label{eq:aucbound}
\text{AUC} \geq \frac{\kappa_S+\kappa_B-2\kappa_S\kappa_B}{2-2\kappa_S \kappa_B}\,.
\end{equation}

This bound only vanishes when $\kappa_S = \kappa_B = 0$, namely when the categories are mutually irreducible.
Thus when pure phase space regions do not exist, an intrinsic ceiling on classification performance at that accuracy can instead be obtained.
Further, as we shall show in later sections, the reducibility factors tend to isolate the dominant phase space regions and are thus typically significantly simpler to calculate than the full distributions of the the phase space observables.

The quadrilateral in \Fig{fig:aucbound} provides a bound to the overall signal vs.~background ROC curve.
Hence any measure of the classification performance can be bounded through the reducibility factors in this way, not solely the AUC.
To highlight this fact, we also derive a bound on another common measure of classification performance: the (inverse) background mistag rate $1/\varepsilon_B$ at a specified signal efficiency $\hat\varepsilon_S$.
Computing this bound, we find
\begin{equation}\label{eq:bgatsig}
 \left.\frac{1}{\varepsilon_B}\right|_{\varepsilon_S = \hat\varepsilon_S}  \le \left\{
     \begin{array}{lr}
       \frac{1}{\hat\varepsilon_S \kappa_S} & \text{ if }\, \hat\varepsilon_S \le \frac{1-\kappa_B}{1-\kappa_S\kappa_B}\\
       \frac{\kappa_B}{\hat\varepsilon_S+ \kappa_B - 1} & \text{otherwise},
     \end{array}
   \right.
\end{equation}
demonstrating again the relationship between parametric discrimination power and phase space purity, quantified through the reducibility factors.

\subsection{$Z$ Boson vs.~QCD Jets}
\label{sec:Zboson}

In this section, we calculate the reducibility factors for a discrimination problem in which a robust power-counting scheme has been defined and used~\cite{Larkoski:2014gra}.
Specifically, we study the discrimination of two-prong quark jets from hadronically-decaying boosted $Z$ bosons.
This will provide us with a concrete case study to explore the relationship between power counting optimality and mutual irreducibility in a known context before moving on to discuss quark vs.~gluon discrimination.
The calculations that follow were also presented in \Ref{Dasgupta:2015lxh}.

Unlike quark or gluon jets, $Z$ bosons are massive, which fixes a relationship between the energies of the $Z$ decay products and their opening angle.
Because the $Z$ boson has a fixed mass, we consider measuring $N$-subjettiness observables with angular exponent $\beta = 2$, which (approximately) corresponds to the ratio of mass to jet energy squared.
In particular, there is no soft singularity for the decay products of the $Z$ boson, so in the large boost limit, 1-subjettiness measured on the $Z$ boson is simply
\begin{equation}
\tau_1^{(2)} =z(1-z)\theta^2 = \frac{m_Z^2}{p_{T\,J}^2}\,, 
\end{equation}
where $z$ is the energy fraction of one of the quark decay products of the $Z$ boson and $\theta$ is the angle between decay products.
For unpolarized $Z$ bosons because there is no soft singularity, to leading power, the distribution of the energy fraction $z$ is uniform on $z\in[0,1]$.
To calculate the cross section of $\tau_2^{(2)}$ given this value of $\tau_1^{(2)}$, we consider the emission of a soft and collinear gluon off of either decay product of the $Z$ boson and find:
\begin{align}
\frac{d\sigma_Z(\tau_1^{(2)})}{d\tau_2^{(2)}} &= 4\frac{\alpha_s}{\pi}C_F \int_0^1 dz \int_0^1 \frac{dz_1}{z_1}\int_0^1 \frac{d\theta_1}{\theta_1}\, \delta\left(\tau_2^{(2)} - z z_1 \theta_1^2\right)\,\Theta\left(\frac{\tau_1^{(2)}}{z(1-z)}-\theta_1^2\right)\\
&=2\frac{\alpha_s}{\pi}\frac{C_F}{\tau_2^{(2)}} \log\frac{\tau_1^{(2)}}{\tau_2^{(2)}}\,,\nonumber
\end{align}
where we have neglected subleading terms in $\tau_2^{(2)}/\tau_1^{(2)}\ll 1$.

The corresponding conditional cross section for quark jets will be calculated in \Secs{sec:one}{sec:two}, and we will state them here to complete our argument.
We find
\begin{align}
\frac{d\sigma_q(\tau_1^{(2)})}{d\tau_2^{(2)}} 
&=-\frac{\alpha_s}{\pi}\frac{1}{ \tau_2^{(2)}}\left[C_F\log \tau_2^{(2)}+C_A \log\frac{\tau_2^{(2)}}{\tau_1^{(2)}}\right]\,.
\end{align}

To calculate the quark reducibility factor, we would in principle need the complete, normalized probability distributions for both quark and $Z$ boson jets.
However, these fixed order cross sections are sufficient, without the inclusion of exponential Sudakov factors, because the reducibility factor vanishes.
In the limit that $\tau_2^{(2)}\to \tau_1^{(2)}$, the quark reducibility factor can be found from taking the ratio of these two cross sections:
\begin{align}
\kappa_q &= \min_{\tau_2^{(2)}}\frac{\frac{d\sigma_Z(\tau_1^{(2)})}{d\tau_2^{(2)}}}{\frac{d\sigma_q(\tau_1^{(2)})}{d\tau_2^{(2)}}}=\left. -\frac{2C_F \log\frac{\tau_1^{(2)}}{\tau_2^{(2)}}}{C_F\log \tau_2^{(2)}+C_A \log\frac{\tau_2^{(2)}}{\tau_1^{(2)}}}\right|_{\tau_2^{(2)}\to \tau_1^{(2)}}=0\,.
\end{align}
The identified purifying phase space region of $\tau_2^{(2)} \to \tau_1^{(2)}$ suggests using an observable such as $\tau_2^{(2)}/\tau_1^{(1)}$ as a parametrically optimal classifier.  This has long been studied and identified from power counting arguments as the combination of $N$-subjettiness observables most sensitive to two-prong substructure, so it is pleasing to observe that reducibility arguments readily produce the same result.

To determine the $Z$ boson reducibility factor directly, we would need to include the appropriate Sudakov form factors for both quark and $Z$ boson jets.
The prediction of the resummed conditional probability for quark jets can be extracted from our later results in \Secs{sec:one}{sec:two} and $Z$ bosons require a new calculation.
We will not perform that calculation explicitly here, though it is relatively simple because the distribution of energy fractions of decay products from the $Z$ boson is simply uniform.
The $Z$ boson reducibility factor $\kappa_Z$ is also 0, because the quark jet Sudakov factor provides more exponential suppression in the limit that $\tau_2^{(2)}\to 0$ than for $Z$ bosons.
This is due to the fact that the two prongs of the quark jet are a quark and a gluon, while the two prongs of the $Z$ boson are both quarks.
Further, this reasoning also applies to gluon jets versus $Z$ bosons, through replacing the color factors in the quark jet distributions $C_F\to C_A$.
Because $C_A>C_F$, gluon jets and $Z$ boson jets are also mutually irreducible.

It is worth noting that higher order effects, such as $g\to q\bar q$, may spoil this mutual irreducibility and hence the parametric separation of the categories.
Calculating these effects requires working beyond leading logarithmic accuracy, at least including non-singular pieces of the splitting functions as well as the running of the strong coupling constant.
While we will not pursue this further here, we highlight that the reducibility factors allow for the investigation of optimal parametric separation at higher orders.
Developing collider observables which are optimal at next-to-leading and higher logarithmic accuracy is an interesting avenue for further exploration.

\section{Quark and Gluon Power Counting Rules}\label{sec:pc}

We now present power counting rules that can be applied to simply determine powerful observables for quark versus gluon discrimination.
As we justify in the following sections, resolving any finite number of emissions in a jet strictly prohibits the isolation of a gluon-pure phase space region.
Nevertheless, due to Sudakov suppression and the fact that the fundamental Casimir $C_F$ is smaller than the adjoint Casimir $C_A$, only quark jets survive deep in the infrared regions of phase space.
Thus, a quark-pure region of phase space can be defined, which motivates a power counting parameter and a definition of the gluon-rich region of phase space simply as that region for which the Sudakov factors are unity.  
Further, the power counting for quark and gluon jets is a bit different than that established for prong discrimination, for example.  In the quark versus gluon case, we construct a power counting scheme for the distribution of an observable (or multiple observables), and not for the observables themselves.  This enables us to identify the necessary properties of the distribution such that quarks and gluons are optimally separated.

With this motivation and within the stated caveats, we present the power counting rules for quark versus gluon discrimination:
\begin{enumerate}
\item Given a measured set of observables on the jets, such as $N$-subjettiness $\{\tau_N^{(\beta)}\}$, identify the corresponding phase space boundaries defined by these observables.

\item Formally take the power counting 
\begin{equation}
e^{C_F-C_A}\ll1\,.
\end{equation}
With this power counting, the boundaries of phase space where any ratio of a pair of measured observables $\{\tau_N^{(\beta)}\}$ becomes large (or small) are dominantly populated by quarks.
This is because Sudakov form factors exponentially suppress the gluon jet cross section beyond that of quarks.  The boundaries on which all  observable ratios are order 1 are dominantly populated by gluons.

\item Construct a function of the observables $\{\tau_N^{(\beta)}\}$ whose constant values define hypersurfaces for which, for example, when the function is 1 only the gluon region is selected, and when the function is 0, only the quark region is selected.  The resulting function is guaranteed to be a powerful quark/gluon discriminant.
\end{enumerate}

Our explicit calculations in the following sections will justify these rules in a concrete context.
Additionally, there are numerous immediate consequences.
For a given set of observables, the observable that is directly sensitive to the most emissions in the jet satisfies the power counting requirements.
In the context of $N$-subjettiness observables, $\tau_N^{(\beta)}$ is necessarily smaller than $\tau_{N-1}^{(\beta)}$.
This means that $\tau_N^{(\beta)}$ is a better quark/gluon discriminant than $\tau_{N-1}^{(\beta)}$ in the limit that parametrically approaches the phase space boundaries.
Further, the multiplicity observable is obviously sensitive to all particles in a jet, hence it will also be a very good quark/gluon discriminant.

Perhaps the most surprising consequence of these power counting rules is that good quark versus gluon discrimination observables are IRC safe.  By ``IRC safe'' we mean that the region of phase space in which cross sections calculated at fixed-order in perturbation theory diverge are mapped to a single value of the observable.  This is a bit more of an abstract definition of IRC safety than is typically stated (see for example \Ref{Ellis:1991qj}), but is equivalent to the heuristic that the observable is insensitive to exactly collinear or zero energy emissions.

The argument for the IRC safety of good quark/gluon discriminants using the power counting rules is as follows.  
The regions of phase space on which any $N$-subjettiness observable ratio becomes large is the singular limit of perturbative QCD, in which the corresponding fixed-order cross section would diverge.
For powerful discrimination, we need constant hypersurfaces of the constructed observable to be approximately parallel to these boundaries; otherwise quark-pure and gluon-rich regions of phase space would be mixed by the observable.
Thus, all singular phase space regions must be mapped onto the same value of the discrimination observable.
As such, all real and virtual divergences can be correspondingly cancelled order-by-order.
Because the $N$-subjettiness observables can form a complete basis of $M$-dimensional phase space for any $M$, the optimal quark versus gluon discrimination observable is some IRC safe combination of (many) $N$-subjettiness observables.
We emphasize that this is purely a perturbative argument, as IRC safety is only relevant within perturbation theory.  Nevertheless, this suggests a guiding principle for constructing quark/gluon discriminants and attempting to understand the output of high-dimensional machine learning studies.

We also note that this observation is not vacuous, as it is not true that IRC safe observables are optimal for all jet discrimination problems.
For example, in the case of discrimination of jets with different numbers of prongs, such as QCD jets versus boosted top quarks, it has been argued that optimal observables are not IRC safe.
Power counting in the two- versus one-prong or three- versus one-prong jet cases motivates ratio observables such as $\tau_2^{(\beta)}/\tau_1^{(\beta)}$, $D_2^{(\beta)}$, or $\tau_3^{(\beta)}/\tau_2^{(\beta)}$~\cite{Larkoski:2014gra,Larkoski:2014zma}, which are not IRC safe \cite{Soyez:2012hv}.
Of course, these ratios can become IRC safe if combined with a constraint on other observables, such as the jet mass.
Nonetheless, it is interesting that the optimal discriminants for multi-prong tagging are indeed not IRC safe without such a restriction.
This is in contrast to what we have established for quark vs.~gluon discrimination, where the likelihood ratio is always IRC safe.

\section{Resolving One Emission}\label{sec:one}

We now present explicit calculations of collections of $N$-subjettiness observables on jets, resolving one, two, or three emissions within the jet.
In this section, we showcase results for jets on which one emission is resolved and discuss their consequences, which will frame the calculations in the next two sections.
All results in this section have been calculated elsewhere in the literature \cite{Larkoski:2013paa,Larkoski:2014tva,Procura:2014cba,Procura:2018zpn}, so we will not present the details of the calculation.
We compile them to construct a complete picture of quark versus gluon discrimination on such jets.
The results in the following sections will be novel, in which complete calculations will be presented.

To double logarithmic accuracy, the normalized distribution of one-subjettiness $\tau_1^{(\beta)}$ for quark and gluons jets is
\begin{align}\label{eq:tau1qg}
p_q(\tau_1^{(\beta)}) &= -2 \frac{\alpha_s}{\pi}\frac{C_F}{\beta}\frac{\log \tau_1^{(\beta)}}{\tau_1^{(\beta)}} \exp\left[
-\frac{\alpha_s}{\pi}\frac{C_F}{\beta}\log^2\tau_1^{(\beta)}
\right]\,,\\
p_g(\tau_1^{(\beta)}) &= -2 \frac{\alpha_s}{\pi}\frac{C_A}{\beta}\frac{\log \tau_1^{(\beta)}}{\tau_1^{(\beta)}} \exp\left[
-\frac{\alpha_s}{\pi}\frac{C_A}{\beta}\log^2\tau_1^{(\beta)}
\right]\,.\nonumber
\end{align}
The corresponding cumulative distributions are
\begin{align}
\Sigma_q(\tau_1^{(\beta)}) &=  \exp\left[
-\frac{\alpha_s}{\pi}\frac{C_F}{\beta}\log^2\tau_1^{(\beta)}
\right]\,,\\
\Sigma_g(\tau_1^{(\beta)}) &=  \exp\left[
-\frac{\alpha_s}{\pi}\frac{C_A}{\beta}\log^2\tau_1^{(\beta)}
\right]=\left(
\Sigma_q(\tau_1^{(\beta)})
\right)^{C_A/C_F}\,,\nonumber
\end{align}
which are related by so-called Casimir scaling.
The quark/gluon ROC curve is thus:
\begin{equation}\label{eq:casroc}
\text{ROC}(x) = x^{C_A/C_F},
\end{equation}
 and its integral is the AUC, namely:
\begin{equation}\label{eq:casauc}
\text{AUC} = \int_0^1 dx\, x^{C_A/C_F} = \frac{1}{1+\frac{C_A}{C_F}} = \frac{4}{13}\simeq 0.308 \,.
\end{equation}
These results will provide a benchmark for discrimination performance that we will compare to in the following sections.

We now proceed to calculate the quark and gluon reducibility factors for the phase space of one resolved emission.
For Casimir-scaling observables, this was calculated in \Ref{Metodiev:2018ftz}, but we present the result here for completeness.  For the one-subjettiness distributions, the likelihood ratio is
\begin{equation}
\frac{p_g(\tau_1^{(\beta)})}{p_q(\tau_1^{(\beta)})} = \frac{C_A}{C_F}\exp\left[
-\frac{\alpha_s}{\pi}\frac{C_A-C_F}{\beta}\log^2\tau_1^{(\beta)}
\right]\,.
\end{equation}
Note the appearance of the power counting factor $\exp[C_F - C_A] \ll 1$ in this distribution.  Approaching the boundary where $\tau_1^{(\beta)}\to 0$, this small number is raised to a large positive power, demonstrating that the quark reducibility factor is 0.  The gluon reducibility factor $\kappa_g\left(\tau_1^{(\beta)}\right)$ is the inverse of the value of the likelihood for $\tau_1^{(\beta)}=1$ at which
\begin{equation}
\kappa_g\left(\tau_1^{(\beta)}\right)=\frac{p_q(\tau_1^{(\beta)}=1)}{p_g(\tau_1^{(\beta)}=1)} = \frac{C_F}{C_A}\,.
\end{equation}
That is, by just measuring $\tau_1^{(\beta)}$, any phase space region of gluon jets is always contaminated by quark jets, by a relative proportion of $C_F/C_A$ or greater.

\subsection{Resolving the One-Emission Phase Space}
\label{sec:2bodyps}

Measuring $\tau_1^{(\beta)}$ resolves one emission off of the hard jet core, and so effectively defines a jet with two particles.
Two-body phase space is two-dimensional, and this phase space can be defined by the relative energy fraction and angle of the emission.
Correspondingly, one can measure two one-subjettiness observables, $\tau_1^{(\alpha)}$ and $\tau_1^{(\beta)}$ with $\alpha > \beta$, to completely resolve two-body phase space.
To double logarithmic accuracy, this joint probability distribution was first calculated in \Ref{Larkoski:2013paa} and extended in \Refs{Larkoski:2014tva,Procura:2014cba,Procura:2018zpn} which found
\begin{align}
p_q\left(\tau_1^{(\alpha)},\tau_1^{(\beta)}\right) &= \frac{2\alpha_s}{\pi}\frac{C_F}{\alpha-\beta}\frac{1}{\tau_1^{(\alpha)}\tau_1^{(\beta)}}\left(1+\frac{2\alpha_s}{\pi}\frac{C_F}{\beta(\alpha-\beta)}\log\frac{\tau_1^{(\beta)}}{\tau_1^{(\alpha)}}\log\frac{{\tau_1^{(\alpha)}}^\beta}{{\tau_1^{(\beta)}}^\alpha}\right)\Delta_q\left(\tau_1^{(\alpha)},\tau_1^{(\beta)}\right)\,,\nonumber\\
p_g\left(\tau_1^{(\alpha)},\tau_1^{(\beta)}\right) &= \frac{2\alpha_s}{\pi}\frac{C_A}{\alpha-\beta}\frac{1}{\tau_1^{(\alpha)}\tau_1^{(\beta)}}\left(1+\frac{2\alpha_s}{\pi}\frac{C_A}{\beta(\alpha-\beta)}\log\frac{\tau_1^{(\beta)}}{\tau_1^{(\alpha)}}\log\frac{{\tau_1^{(\alpha)}}^\beta}{{\tau_1^{(\beta)}}^\alpha}\right)\Delta_g\left(\tau_1^{(\alpha)},\tau_1^{(\beta)}\right)\,.\nonumber
\end{align}
The Sudakov factor is
\begin{equation}
\Delta_i\left(\tau_1^{(\alpha)},\tau_1^{(\beta)}\right) = \exp\left[
-\frac{\alpha_s}{\pi}C_i\left(
\frac{1}{\beta}\log^2 \tau_1^{(\beta)} + \frac{1}{\alpha-\beta}\log^2 \frac{\tau_1^{(\alpha)}}{\tau_1^{(\beta)}}
\right)
\right]\,,
\end{equation}
where $C_i$ is the appropriate color factor. 
The physical phase space lies within the boundaries of $\tau_1^{(\alpha)} < \tau_1^{(\beta)}$ and ${\tau_1^{(\beta)}}^\alpha < {\tau_1^{(\alpha)}}^\beta$.

The likelihood ratio for the two one-subjettiness observables is then
\begin{equation}\label{eq:two1subLR}
\frac{p_g\left(\tau_1^{(\alpha)},\tau_1^{(\beta)}\right) }{p_q\left(\tau_1^{(\alpha)},\tau_1^{(\beta)}\right) } = \frac{C_A}{C_F}\frac{1+\frac{2\alpha_s}{\pi}\frac{C_A}{\beta(\alpha-\beta)}\log\frac{\tau_1^{(\beta)}}{\tau_1^{(\alpha)}}\log\frac{{\tau_1^{(\alpha)}}^\beta}{{\tau_1^{(\beta)}}^\alpha}}{1+\frac{2\alpha_s}{\pi}\frac{C_F}{\beta(\alpha-\beta)}\log\frac{\tau_1^{(\beta)}}{\tau_1^{(\alpha)}}\log\frac{{\tau_1^{(\alpha)}}^\beta}{{\tau_1^{(\beta)}}^\alpha}}\frac{\Delta_g\left(\tau_1^{(\alpha)},\tau_1^{(\beta)}\right)}{\Delta_q\left(\tau_1^{(\alpha)},\tau_1^{(\beta)}\right)}\,.
\end{equation}
The quark reducibility factor is still 0, due to the exponential suppression of the Sudakov factors.
Further, the gluon reducibility factor is still $C_F/C_A$; completely resolving the one-emission phase space does not improve gluon jet purity.
From power counting arguments, this then implies that completely resolving the phase space does not parametrically improve discrimination power.
It is most important to measure observables to demonstrate that a particular number of emissions exist in the jet.

The likelihood ratio is the optimal observable for discrimination, and it is straightforward to demonstrate that it is in this case indeed IRC safe, as claimed from our power counting arguments.
Due to the phase space boundaries, there is only one point on phase space that corresponds to the singular limit: when $\tau_1^{(\alpha)}=\tau_1^{(\beta)}=0$.
The only way that the likelihood can vanish is if the ratio of Sudakov factors vanish; the prefactor formed from a ratio of logarithms is always positive on the physical phase space.
However, the Sudakov factor can only vanish if its exponent diverges, corresponding to at least one of the one-subjettiness observables going to 0.
By the phase space constraints, if one goes to 0 the other must as well, and so the only point on phase space that makes the likelihood vanish is the singular point $\tau_1^{(\alpha)}=\tau_1^{(\beta)}=0$.
Therefore, all divergences on phase space are isolated to a single point in the likelihood, and thus it is IRC safe.

\subsection{Higher Order Effects}

For this case of one resolved emission, we also briefly discuss higher-order corrections.
In \Ref{Larkoski:2013eya}, a calculation of recoil-insensitive one-emission observables was presented at next-to-leading logarithmic accuracy.
For the one-subjettiness observables considered here, this would correspond to defining the jet axis with a recoil-free scheme, such as the broadening \cite{Larkoski:2014uqa} or winner-take-all \cite{Bertolini:2013iqa,salamunp} axis.
For the ROC curve, they found the following relationship between the gluon and quark cumulative distributions (with arguments suppressed):
\begin{align}
\label{eq:summaryequation}
\log \Sigma_g \simeq& \ \frac{C_A}{C_F}\left( 1 + \frac{n_f-C_A}{3C_A}\sqrt{\frac{\alpha_s
            C_F}{\pi \beta \log 1/\Sigma_q}}  + \frac{n_f-C_A}{C_A} \frac{\alpha_s}{36\pi} \frac{b_0}{\beta}(2-\beta) \right.\nonumber \\
            & \qquad\quad \left.+ \  \frac{\alpha_s\pi }{3}\frac{C_A-C_F}{\beta} - \frac{17}{36}\frac{\alpha_s}{\pi}\frac{C_F}{C_A}\frac{n_f - C_A}{\beta\log 1/\Sigma_q}  + \ldots \right)\log \Sigma_q \,,
\end{align}
where $b_0 = \frac{11}{3}C_A -\frac{2}{3}n_f$ is the one-loop $\beta$-function coefficient with $n_f$ active fermions.
The lowest-order relationship is simply the overall $C_A/C_F$ Casimir scaling, but effects like running coupling, hard collinear radiation, and multiple emissions all affect the discrimination power at higher orders.
In general, discrimination power improves as the angular exponent $\beta$ decreases, due to these higher order effects.
This is directly observed in simulations, suggesting that one should use as small an angular exponent as possible, while maintaining theoretical control.\footnote{However, this is not observed in experiment; see for example \Ref{Aad:2014gea}.}
These higher order effects could be explored for more resolved emissions, but we leave that to future work.
In the following sections, we will focus on the calculations at double logarithmic accuracy.

\section{Resolving Two Emissions}\label{sec:two}

We now turn to calculations for jets on which two emissions are resolved.
While some of the calculations that we present are included in parts of various other calculations in the literature \cite{Dasgupta:2015lxh,Salam:2016yht,Napoletano:2018ohv}, to our knowledge, these complete expressions have never appeared for quark versus gluon discrimination.
Therefore, we present a detailed discussion of the calculations that follow.
Further, as discussed in \Sec{sec:obsapp}, we simplify our analysis and strictly consider measuring $N$-subjettiness observables with an angular exponent $\beta = 1$.
As higher-order corrections in the one emission case demonstrate, there is likely discrimination power to be gained by changing the angular exponent.
However, we do not consider that here as even this simple analysis will enable significant understanding.

\subsection{Fixed-Order Analysis}

We begin with a calculation of the cross section for quarks jets on which both $\tau_1$ and $\tau_2$ have been measured.
The phase space restrictions demand that $\tau_2 \leq \tau_1\leq 1$ and at leading order, there are two possibilities for the orientation of emissions in the jet.
Either the gluons that set $\tau_1$ and $\tau_2$ could be sequentially emitted from the initiating quark, or the gluon that sets $\tau_2$ is emitted off of the gluon that sets $\tau_1$.
In the first case, the color factor is $C_F^2$ and the contribution to the cross section to double logarithmic accuracy is
\begin{align}
\frac{1}{\sigma_0}\frac{d^2\sigma_q^{C_F^2}}{d\tau_1 \, d\tau_2} &= \left(2\frac{\alpha_s}{\pi}\right)^2 C_F^2 \int_0^1\frac{dz_1}{z_1}\int_0^1\frac{d\theta_1}{\theta_1}\int_0^1\frac{dz_2}{z_2}\int_0^{1}\frac{d\theta_2}{\theta_2} \, \delta(\tau_1-z_1\theta_1)\delta(\tau_2-z_2\theta_2)\\
&=\left(2\frac{\alpha_s}{\pi}\right)^2 C_F^2\frac{\log \tau_1\, \log\tau_2}{\tau_1 \tau_2}
\,.\nonumber
\end{align}
In the second case, the color factor is $C_F C_A$ and we must account for the fact that the gluon that sets $\tau_2$ can neither have more energy than the gluon that sets $\tau_1$ nor be at larger angle.
In this color channel, the cross section is then
\begin{align}
\frac{1}{\sigma_0}\frac{d^2\sigma_q^{C_F C_A}}{d\tau_1 \, d\tau_2} &= \left(2\frac{\alpha_s}{\pi}\right)^2 C_FC_A \int_0^1\frac{dz_1}{z_1}\int_0^1\frac{d\theta_1}{\theta_1}\int_0^1\frac{dz_2}{z_2}\int_0^{\theta_1}\frac{d\theta_2}{\theta_2} \, \delta(\tau_1-z_1\theta_1)\delta(\tau_2-z_1z_2\theta_2)\nonumber \\
&=\left(2\frac{\alpha_s}{\pi}\right)^2 C_F\frac{\log \tau_1}{\tau_1 \tau_2}\left[C_A \log\frac{\tau_2}{\tau_1}\right]\,.
\end{align}
Combining these results, the leading-order double differential cross section in the double logarithmic limit for quark jets is
\begin{align}\label{eq:fot2q}
\frac{1}{\sigma_0}\frac{d^2\sigma_q}{d\tau_1 \, d\tau_2} =\left(2\frac{\alpha_s}{\pi}\right)^2 C_F\frac{\log \tau_1}{\tau_1 \tau_2}\left[C_F\log \tau_2+C_A \log\frac{\tau_2}{\tau_1}\right]\,.
\end{align}
This agrees with the results of \Ref{Dasgupta:2015lxh}, in which they calculate the distribution of $\tau_1$ when there is a cut on the ratio $\tau_2/\tau_1$.  The result for gluon jets can be found by simply replacing $C_F \to C_A$:
\begin{align}\label{eq:fot2g}
\frac{1}{\sigma_0}\frac{d^2\sigma_g}{d\tau_1 \, d\tau_2} =\left(2\frac{\alpha_s}{\pi}\right)^2 C_A^2\frac{\log \tau_1}{\tau_1 \tau_2}\log\frac{\tau_2^2}{\tau_1}\,.
\end{align}

While only evaluated at fixed-order, these results are not probability distributions, and so we cannot use them to determine likelihood ratios.
However, the gluon reducibility factor is the ratio of the cross sections in the region where the Sudakov factors are unity; that is, the gluon reducibility factor can be calculated strictly from fixed order results.
The ratio of the quark to gluon cross sections is
\begin{equation}
\frac{\frac{d^2\sigma_q}{d\tau_1 \, d\tau_2} }{\frac{d^2\sigma_g}{d\tau_1 \, d\tau_2} } = \frac{C_F}{C_A}\frac{C_F\log \tau_2+C_A \log\frac{\tau_2}{\tau_1}}{C_A \log\frac{\tau_2^2}{\tau_1}}\,.
\end{equation}
This ratio is minimized in the ordered limits in which first $\tau_2\to \tau_1$ and then $\tau_1\to 1$.
The second limit is required to remain in the fixed-order regime and neglect the Sudakov factor.
In these limits, the gluon reducibility factor $\kappa_g(\tau_1,\tau_2)$ is then
\begin{equation}\label{eq:t1t2redfac}
\kappa_g(\tau_1,\tau_2) = \frac{C_F^2}{C_A^2}\simeq 0.198\,.
\end{equation}
This is significantly smaller than the reducibility factor of $C_F/C_A\simeq 0.444$ with only one resolved emission, demonstrating that purer gluon phase space can be isolated through additional measurements.

With the fixed-order cross section in hand, we can additionally integrate over $\tau_1$ to determine the cross section for jets on which $\tau_2$ is measured alone.
From the power counting arguments, $\tau_2$ should be a good discriminant itself, because it vanishes in the singular phase space regions, where the Sudakov factors exponentially suppress the cross section, and when $\tau_2\to 1$, then necessarily $\tau_1\to 1$.
Integrating over $\tau_1$, we find the quark jet cross section singly-differential in $\tau_2$ to be
\begin{align}
\frac{1}{\sigma_0}\frac{d\sigma_q}{d \tau_2} = -\left(2\frac{\alpha_s}{\pi}\right)^2 \frac{3C_F^2+C_FC_A}{6}\frac{\log^3 \tau_2}{\tau_2}\,.
\end{align}
As before, the cross section for gluon jets can be found by replacing $C_F \to C_A$:
\begin{align}
\frac{1}{\sigma_0}\frac{d\sigma_g}{d \tau_2} = -\left(2\frac{\alpha_s}{\pi}\right)^2 \frac{2C_A^2}{3}\frac{\log^3 \tau_2}{\tau_2}\,.
\end{align}
The gluon reducibility factor for jets on which just $\tau_2$ is measured is then the ratio of these two cross sections:
\begin{equation}
\kappa_g(\tau_2) = \frac{3}{4}\frac{C_F^2}{C_A^2}+\frac{1}{4}\frac{C_F}{C_A}\simeq 0.259\,.
\end{equation}
While this reducibility factor is definitely larger than the case in which both $\tau_1$ and $\tau_2$ are measured, it is still significantly smaller than the Casimir-scaling result of $C_F/C_A \simeq 0.444$.  Therefore, as predicted by power counting, just measuring $\tau_2$ enables an increased purity of gluon jets and therefore improved discrimination power over just measuring $\tau_1$.

\subsection{Including Resummation}

For a thorough analysis, however, we need to calculate the joint probability distribution of $\tau_1$ and $\tau_2$ on jets.
To calculate this, we will employ the expression for the joint probability distributions expressed in terms of conditional probabilities.
For the joint probability distribution $p(\tau_1,\tau_2)$, we can express it as
\begin{equation}
p(\tau_1,\tau_2) = \int dz_1\, p(\tau_1)\, p(z_1|\tau_1)\, p(\tau_2| \tau_1,z_1)\,.
\end{equation}
Here, $z_1$ is the energy fraction of the gluon that sets the value of $\tau_1$.
It is necessary to include it in an intermediate step to correctly enforce angular ordering, as we will discuss shortly.
The probability distribution of $\tau_1$, $p(\tau_1)$, was presented for quark and gluon jets in \Eq{eq:tau1qg}.
The conditional distribution of the energy fraction $p(z_1|\tau_1)$ is found by noting that to double logarithmic accuracy, $\log 1/z_1$ is just distributed uniformly from $0$ to $\log 1/\tau_1$.
That is, the conditional distribution is
\begin{equation}
p(z_1|\tau_1) = -\frac{1}{z_1 \log\tau_1}\Theta(z_1-\tau_1)\,.
\end{equation}
This integrates to 1 on $z_1\in [\tau_1,1]$.

\begin{figure}
\begin{center}
\includegraphics[width=7cm]{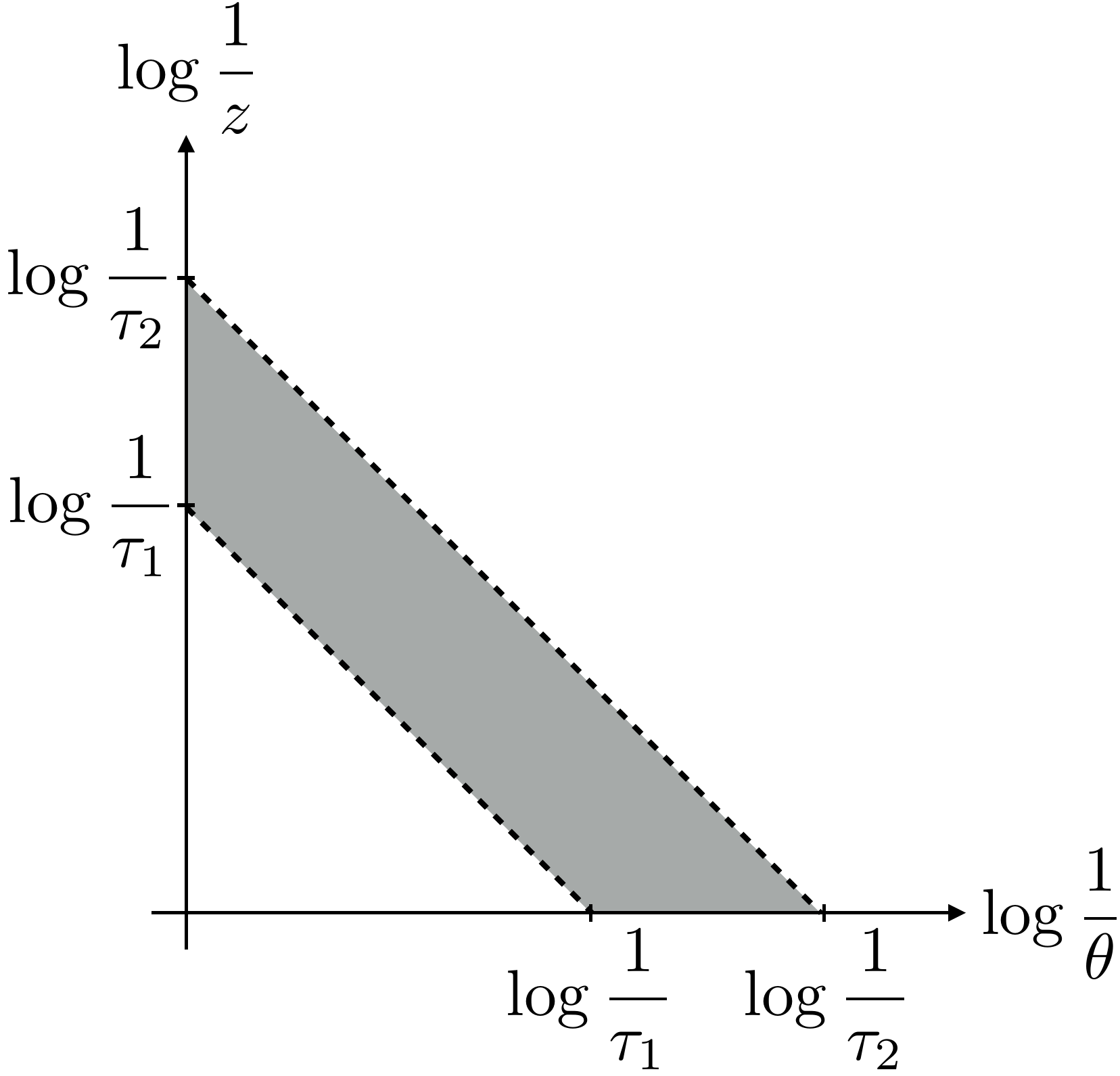}
\hfill
\includegraphics[width=7cm]{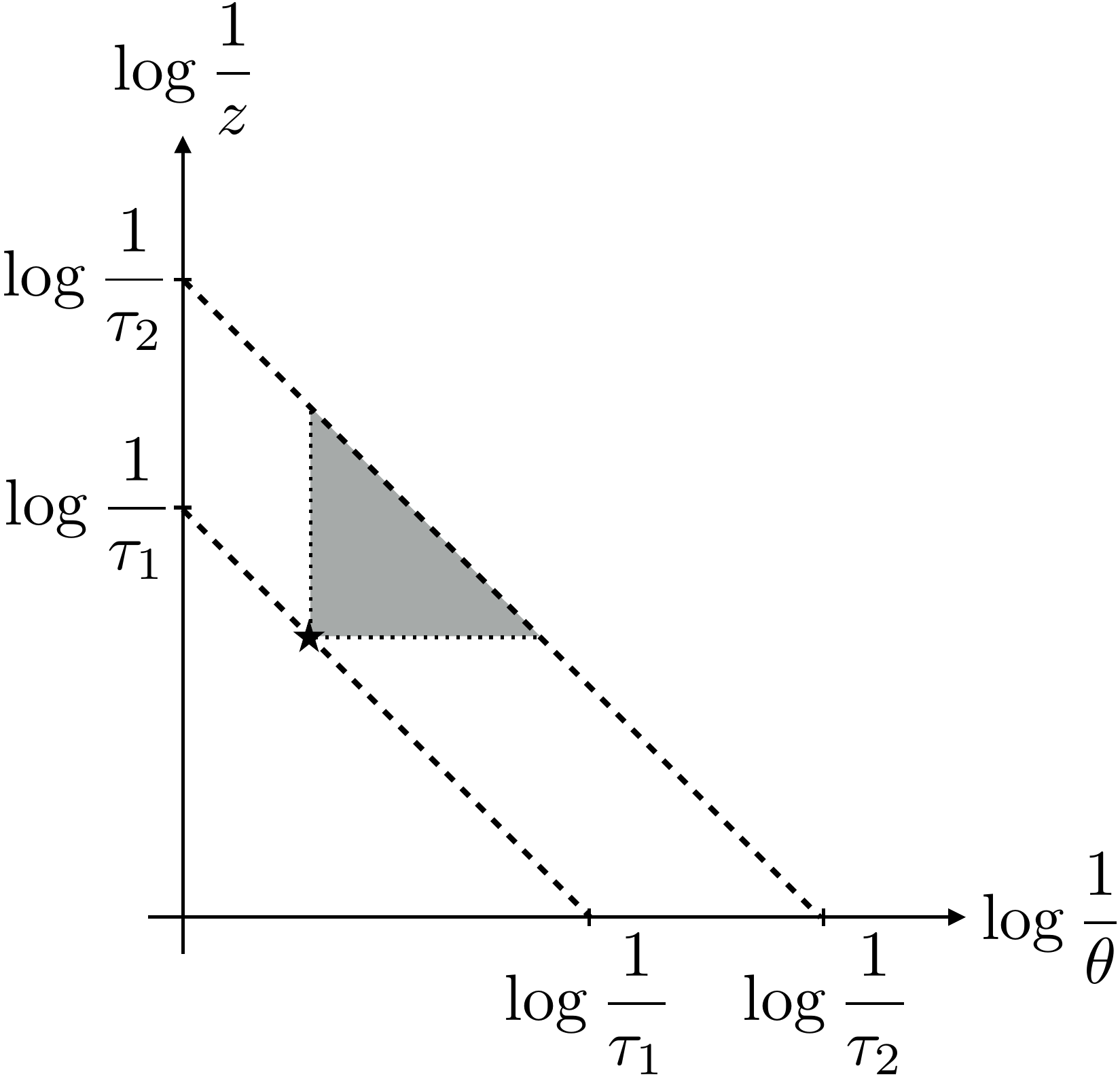}
\end{center}
\caption{Illustrations of the forbidden regions (grayed) for gluon emission that sets the value of $\tau_2$, given a value of $\tau_1$.  The location of the emission that sets the value of $\tau_1$ in the Lund plane is illustrated by the star.  On the left is the forbidden region if the gluon is emitted off of the initiating quark; the only requirement is that the gluon must enforce $\tau_2 < \tau_1$.  On the right is the forbidden region if the gluon is emitted off of the gluon that sets the value of $\tau_1$: it must be both at smaller angle and have smaller energy than the first emitted gluon.}
\label{fig:regt2}
\end{figure}

To calculate the conditional probability distribution for $\tau_2$, $p(\tau_2| \tau_1,z_1)$, we first calculate its cumulative distribution, $\Sigma(\tau_2| \tau_1,z_1)$.
To calculate this distribution requires identifying the regions in the Lund plane which are forbidden, given the measured value of $\tau_2$.
There are two possibilities for how the emission that sets $\tau_2$ was formed, and that produces two different no emission regions.
These regions are illustrated in \Fig{fig:regt2} in gray.  First, if the gluon that sets $\tau_2$ is emitted off of the quark, the only restriction on its phase space to this accuracy is that $\tau_2< \tau_1$.
This area, multiplied by the appropriate color and coupling factors, is
\begin{equation}
\text{Area}_{C_F} = \frac{\alpha_s}{\pi}C_F \left(
\log^2\tau_2 - \log^2\tau_1
\right)\,.
\end{equation}
The no emission region in the case in which the gluon that sets $\tau_2$ is emitted off of the gluon that sets $\tau_1$ is required to both be at smaller angle and smaller energy than the first emission.
This demonstrates why the energy fraction $z_1$ is measured, as this enables an identification of the angular-ordered phase space region.
The area of this region, including color and coupling factors, is
\begin{equation}
\text{Area}_{C_A} = \frac{\alpha_s}{\pi}C_A \log^2\frac{\tau_2}{\tau_1}\,.
\end{equation}

With these results, the cumulative conditional probability distribution is just the exponential of these areas, as follows from considering gluon emission as a Poisson process:
\begin{equation}
\Sigma_q(\tau_2|\tau_1,z_1) = \exp\left[
-\frac{\alpha_s}{\pi}\left(
C_F\log^2\tau_2-C_F\log^2\tau_1+C_A\log^2\frac{\tau_2}{\tau_1}
\right)
\right]\,.
\end{equation}
The conditional probability distribution is then just the derivative of this expression:
\begin{align}
p_q(\tau_2|\tau_1,z_1)&=\frac{\partial}{\partial \tau_2}\Sigma_q(\tau_2|\tau_1,z_1)\\
&=-2\frac{\alpha_s}{\pi}\frac{C_F \log \tau_2+C_A\log\frac{\tau_2}{\tau_1}}{\tau_2} \nonumber \\
&\quad\quad\times\exp\left[
-\frac{\alpha_s}{\pi}\left(
C_F\log^2\tau_2-C_F\log^2\tau_1+C_A\log^2\frac{\tau_2}{\tau_1}
\right)
\right]\,.\nonumber
\end{align}
For gluon jets, the whole analysis is identical, we just replace $C_F\to C_A$ and find
\begin{align}
p_g(\tau_2|\tau_1,z_1)&=\frac{\partial}{\partial \tau_2}\Sigma_g(\tau_2|\tau_1,z_1)\\
&=-2\frac{\alpha_s}{\pi}C_A\frac{ \log\frac{\tau_2^2}{\tau_1}}{\tau_2} \exp\left[
-\frac{\alpha_s}{\pi}C_A\left(
\log^2\tau_2-\log^2\tau_1+\log^2\frac{\tau_2}{\tau_1}
\right)
\right]\,.\nonumber
\end{align}

We can then multiply the distributions together and integrate over $z_1\in [\tau_1,1]$ to find the double differential probability distribution to resolve two emissions off of a quark.
We find
\begin{align}\label{eq:pqt12}
p_q(\tau_1,\tau_2) &=\left(
2\frac{\alpha_s}{\pi}
\right)^2C_F \frac{\log \tau_1}{\tau_1\tau_2}\left(
C_F \log\tau_2 + C_A \log\frac{\tau_2}{\tau_1}
\right)e^{-\frac{\alpha_s}{\pi}\left(
C_F \log^2\tau_2 + C_A \log^2\frac{\tau_2}{\tau_1}
\right)}\,.
\end{align}
The corresponding distribution for gluons is found by making the replacement $C_F \to C_A$:
\begin{align}\label{eq:pgt12}
p_g(\tau_1,\tau_2) &=\left(
2\frac{\alpha_s}{\pi}
\right)^2C_A^2 \frac{\log \tau_1}{\tau_1\tau_2}
\log\frac{\tau_2^2}{\tau_1}
e^{-\frac{\alpha_s}{\pi}C_A\left(
 \log^2\tau_2 +\log^2\frac{\tau_2}{\tau_1}
\right)}\,.
\end{align}
It is straightforward to see that these expressions reduce at lowest order in $\alpha_s$ to \Eqs{eq:fot2q}{eq:fot2g}.

\subsection{IRC Safety of the Likelihood}

These expressions for the quark and gluon probability distributions can be used to construct the likelihood ratio and demonstrate that it is IRC safe, as claimed. 
The likelihood ratio ${\cal L}(\tau_1,\tau_2)$ is
\begin{equation}
{\cal L}(\tau_1,\tau_2) = \frac{p_g(\tau_1,\tau_2)}{p_q(\tau_1,\tau_2)} = \frac{C_A^2}{C_F^2}\frac{\log\frac{\tau_2^2}{\tau_1}}{\log\tau_2+\frac{C_A}{C_F}\log\frac{\tau_2}{\tau_1}}\,e^{-\frac{\alpha_s}{\pi}(C_A-C_F)\log^2\tau_2}\,.
\end{equation}
The non-exponential prefactor never vanishes on the physical phase space where $\tau_2 < \tau_1$.
Because $C_A > C_F$, the exponential factor vanishes as $\tau_2 \to 0$, which is also the entire region of phase space on which fixed-order cross sections diverge.
Therefore, because the entire singular region of phase space is mapped to a single point, the likelihood ratio ${\cal L}(\tau_1,\tau_2)$ is indeed IRC safe.

%

\subsection{AUC Evaluation}
\label{sec:auc2}

To quantify the absolute discrimination power of the likelihood ${\cal L}(\tau_1,\tau_2)$, we could attempt to construct its complete ROC curve.
However, the likelihood is a complicated function of the observables $\tau_1$ and $\tau_2$, which doesn't enable a convenient inversion.
Therefore, we take a different route: instead of calculating the full functional form of the ROC curve, we just calculate its integral, the AUC.
Improved discrimination power corresponds to decreasing the value of the AUC, so we are able to compare directly between the AUC calculated with different numbers of resolved emissions.

What makes the AUC so convenient as a discrimination metric, even without an explicit form of the ROC curve, is that it can be expressed as an ordered integral over the probability distributions.  For signal and background distributions $p_s(x)$ and $p_b(x)$ of a random variable $x$, the AUC that corresponds to measurement of the variable $x$ is
\begin{equation}
\text{AUC} = \int_{-\infty}^\infty dx_s\int_{-\infty}^\infty dx_b \, p_s(x_s) \, p_b(x_b)\, \Theta(x_b-x_s)\,.
\end{equation}
Translated to the evaluation of the AUC of the likelihood for quark and gluon jets on which $\tau_1$ and $\tau_2$ are measured, we have
\begin{align}
&\text{AUC} = \int_0^1 \hspace{-1mm}d\tau_{1q} \int_0^{\tau_{1q}}\hspace{-1mm} d\tau_{2q}\int_0^1 \hspace{-1mm}d\tau_{1g} \int_0^{\tau_{1g}}\hspace{-1mm} d\tau_{2g}\, p_q(\tau_{1q},\tau_{2q})p_g(\tau_{1g},\tau_{2g})\,\Theta\left({\cal L}(\tau_{1q},\tau_{2q})-{\cal L}(\tau_{1g},\tau_{2g})\right)\,.\nonumber
\end{align}
To perform the integral to calculate the AUC, we use the implementation of {\tt Vegas} within Cuba 4.2 \cite{Hahn:2004fe}.  Using $C_F = 4/3$ and $C_A = 3$, we find that the AUC of the likelihood is
\begin{equation}
\text{AUC} \simeq 0.256 < \frac{1}{1+\frac{C_A}{C_F}}\simeq 0.308\,.
\end{equation}
On the right, we compare to the AUC for resolving one emission, just measuring $\tau_1$, \Eq{eq:casauc}.
Because the coupling $\alpha_s$ dependence enters in the exact same way for quarks and gluons, the AUC is independent of the particular value of the coupling, which we verified.

An additional benefit of the AUC as a measure of discrimination power is that it enables a simple, concrete variational algorithm to determine other observables.
Consider an observable ${\cal O}(\alpha_1,\alpha_2,\dotsc,\alpha_n)$ that is some function of the $N$-subjettiness observables, that depends on some set of parameters $\{\alpha_i\}$.
We can calculate the AUC for this observable and then fix the parameters to minimize the AUC.
Of course, the value of the AUC for such an observable is bounded from below by the likelihood.
However, this procedure provides an approximation to the likelihood that may have a significantly simpler functional form.

We can construct such an observable with this technique.
For illustration, we just consider the observable formed from a product of powers of $\tau_1$ and $\tau_2$:
\begin{equation}
{\cal O} = \tau_1^{\alpha_1} \tau_2^{\alpha_2}\,.
\end{equation}
In general, $\alpha_1$ and $\alpha_2$ are real numbers, but the observation that the likelihood is IRC safe helps to dramatically constrain the observables.
First, because the likelihood vanishes as $\tau_2\to 0$, we want our constructed observable to map the entire $\tau_2 = 0$ line to the point ${\cal O} = 0$.  This ensures that quark-pure and gluon-rich regions of phase space are still not mixed by ${\cal O}$.  We enforce this on ${\cal O}$ by requiring the power $\alpha_2>0$.
Any monotonic function of an observable has the same discrimination power, so the IRC safety of this observable enables us, with impunity, to set $\alpha_2 = 1$.
Further, the likelihood vanishes in the ordered limit $\tau_2\to\tau_1$ and $\tau_1\to 0$, and this requires $\alpha_1 > -1$.  That is, the observable that we consider is just
\begin{equation}
{\cal O} = \tau_1^\alpha \tau_2\,,
\end{equation}
with $\alpha>-1$.
While this ratio seems potentially ambiguous when $\alpha < 0$ for a jet with a single particle, it is nevertheless still IRC safe. The potential $0/0$ ambiguity can be eliminated and a well-defined result obtained by first taking $\tau_2 \to \tau_1$ and then $\tau_1 \to 0$.  The exponent $\alpha$ can then be determined by the value that minimizes the AUC.

\begin{figure}
\begin{center}
\includegraphics[width=9cm]{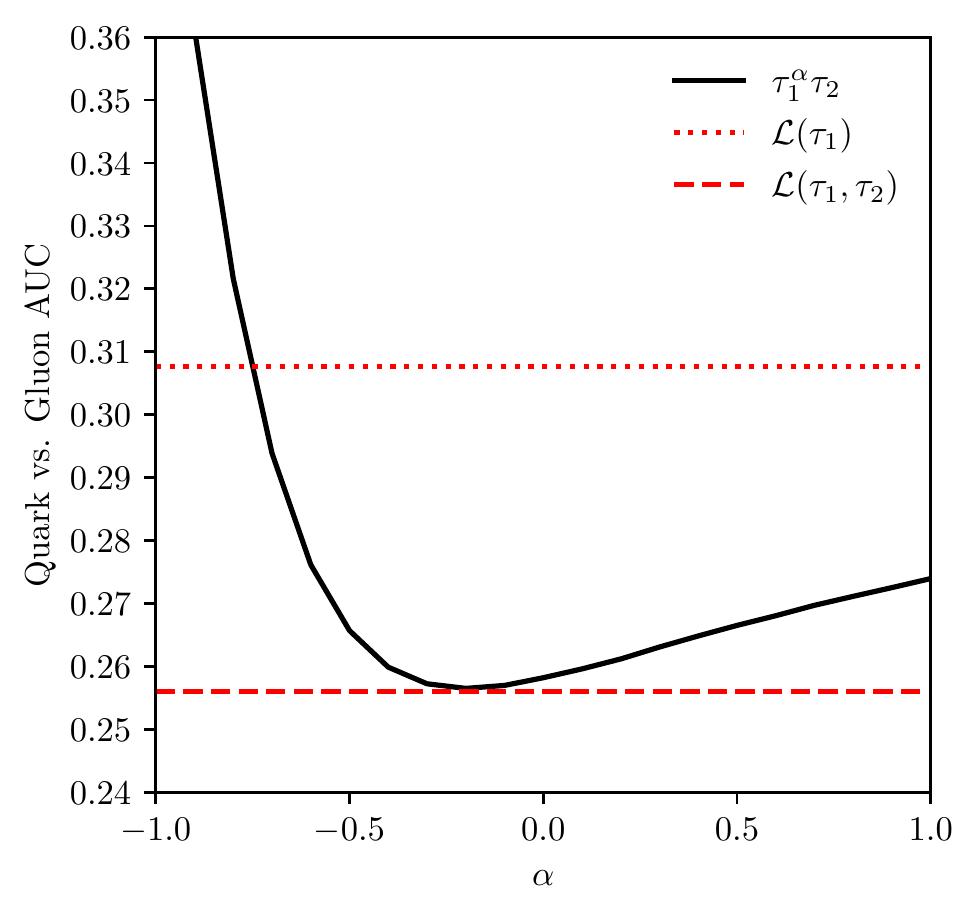}
\end{center}
\caption{Results of a scan over $\alpha$ of the AUC for the observable ${\cal O} = \tau_1^\alpha \tau_2$.  The AUC of the likelihood for jets on which only $\tau_1$ is measured is the dotted line and for jets on which both $\tau_1$ and $\tau_2$ are measured is the dashed line, for comparison.  The AUC for the observable is minimized at $\alpha = -0.2$ where it takes the value $0.256$.
}
\label{fig:auct2func}
\end{figure}

To do the minimization, we simply scan through $\alpha \in[-1,1]$, and plot the AUC as a function of $\alpha$.
The result of this scan is plotted in \Fig{fig:auct2func}.
Also shown on this plot are the AUC values of the likelihood for jets on which $\tau_1$ is measured and $\tau_1$ and $\tau_2$ are measured.
The AUC for the variational observable is minimized when $\alpha = -0.2$, corresponding to an observable that is ${\cal O}=\tau_1^{-0.2}\tau_2$.
To three significant figures, the value of the AUC at this point is $0.256$, which is significantly lower than that for just $\tau_1$, and well within 1\% of the AUC value of the two-resolved-emission likelihood. 
That the minimum AUC exists near $\alpha = 0$ can be understood in the following way.
As argued earlier, the likelihood is an IRC safe observable, and when $\alpha \leq -1$, the observable ${\cal O}$ is no longer IRC safe.
On the other hand, if $\alpha$ is very large, then the discrimination power of the observable ${\cal O}$ is essentially entirely controlled by $\tau_1$.
Because $\tau_2$ is directly sensitive to more emissions in the jet than $\tau_1$, it should have better discrimination power.
This suggests that the power $\alpha$ should be relatively close to 0 to maximize discrimination.
%

\section{Resolving Three Emissions}\label{sec:three}

We now present calculations for resolving three emissions off of a hard jet core, by e.g. measuring $\tau_1$, $\tau_2$, and $\tau_3$.
To our knowledge, these calculations are novel, even in the double logarithmic limit, and have application to top quark tagging.
In addition to the explicit fixed-order and resummed calculations, we also discuss properties that hold for an arbitrary number of emissions.
We prove that the gluon reducibility factor when $n$ emissions is resolved is $(C_F/C_A)^n$ and provide a robust lower bound on the AUC exclusively in terms of reducibility factors.

\subsection{Fixed-Order Analysis}

Starting with the fixed-order calculation of the triple-differential cross section of $\tau_1$, $\tau_2$, and $\tau_3$ in the double logarithmic limit, there are three separate color channels that contribute.
As in earlier sections, we start with the calculation for a quark jet, and then simply make the replacement $C_F \to C_A$ for gluon jets.
The $C_F^3$ color channel means that all three emissions that set these observables are sequentially emitted off of the quark and we find
\begin{align}
\frac{1}{\sigma_0}\frac{d^3\sigma_q^{C_F^3}}{d\tau_1\,d\tau_2\,d\tau_3} &=\left(2\frac{\alpha_s}{\pi}\right)^3 C_F^3 \int_0^1\frac{dz_1}{z_1}\int_0^1\frac{d\theta_1}{\theta_1}\int_0^1\frac{dz_2}{z_2}\int_0^1\frac{d\theta_2}{\theta_2}\int_0^1\frac{dz_3}{z_3}\int_0^1\frac{d\theta_3}{\theta_3}\\
&
\hspace{4cm}
\times \delta(\tau_1-z_1\theta_1)\delta(\tau_2-z_2\theta_2)\delta(\tau_3-z_3\theta_3)\nonumber\\
&=-\left(2\frac{\alpha_s}{\pi}\right)^3  C_F^3\,\frac{\log \tau_1 \, \log\tau_2 \,\log \tau_3}{\tau_1\tau_2\tau_3}\nonumber\,.
\end{align}
The $C_F^2 C_A$ channel has two emissions off of the quark and the third off of one of the secondary gluons.  There are three ways this can occur yielding
\begin{align}
\frac{1}{\sigma_0}\frac{d^3\sigma_q^{C_F^2 C_A}}{d\tau_1\,d\tau_2\,d\tau_3} &=\left(2\frac{\alpha_s}{\pi}\right)^3 C_F^2 C_A \int_0^1\frac{dz_1}{z_1}\int_0^1\frac{d\theta_1}{\theta_1}\int_0^1\frac{dz_2}{z_2}\int_0^1\frac{d\theta_2}{\theta_2}\int_0^1\frac{dz_3}{z_3}\int_0^1\frac{d\theta_3}{\theta_3}\\
&
\hspace{2cm}
\times \left[
\Theta(\theta_1-\theta_2)\delta(\tau_1-z_1\theta_1)\delta(\tau_2-z_1z_2\theta_2)\delta(\tau_3-z_3\theta_3)\right.\nonumber\\
&
\hspace{3cm}
+ \Theta(\theta_2-\theta_3)\delta(\tau_1-z_1\theta_1)\delta(\tau_2-z_2\theta_2)\delta(\tau_3-z_2z_3\theta_3)\nonumber\\
&\hspace{3cm}\left.
+\Theta(\theta_1-\theta_3)\delta(\tau_1-z_1\theta_1)\delta(\tau_2-z_2\theta_2)\delta(\tau_3-z_1z_3\theta_3)
\right]\nonumber\\
&=-\left(2\frac{\alpha_s}{\pi}\right)^3 C_F^2 C_A\frac{\log \tau_1}{\tau_1\tau_2\tau_3}\left(
\log\frac{\tau_2}{\tau_1}\log\tau_3 + \log\tau_2 \log\frac{\tau_3}{\tau_2}+\log\tau_2 \log\frac{\tau_3}{\tau_1}
\right)
\nonumber\,.
%
\end{align}
Finally, the $C_F C_A^2$ channel consists of the gluon that sets $\tau_1$ emitted off of the quark, and then the gluons that set $\tau_2$ and $\tau_3$ are subsequently emitted off of the the secondary gluon.  There are two possible ordering of emissions, which results in
\begin{align}
\frac{1}{\sigma_0}\frac{d^3\sigma_q^{C_F C_A^2}}{d\tau_1\,d\tau_2\,d\tau_3} &=\left(2\frac{\alpha_s}{\pi}\right)^3 C_F C_A^2 \int_0^1\frac{dz_1}{z_1}\int_0^1\frac{d\theta_1}{\theta_1}\int_0^1\frac{dz_2}{z_2}\int_0^1\frac{d\theta_2}{\theta_2}\int_0^1\frac{dz_3}{z_3}\int_0^1\frac{d\theta_3}{\theta_3}\\
&
\hspace{1cm}
\times \left[
\Theta(\theta_1-\theta_2)\Theta(\theta_2-\theta_3)\delta(\tau_1-z_1\theta_1)\delta(\tau_2-z_1z_2\theta_2)\delta(\tau_3-z_1z_2z_3\theta_3)\right.\nonumber\\
&\left.
\hspace{2cm}
+\Theta(\theta_1-\theta_2)\Theta(\theta_1-\theta_3)\delta(\tau_1-z_1\theta_1)\delta(\tau_2-z_1z_2\theta_2)\delta(\tau_3-z_1z_3\theta_3)
\right]\nonumber\\
&=-\left(2\frac{\alpha_s}{\pi}\right)^3 C_F C_A^2\frac{\log\tau_1}{\tau_1\tau_2\tau_3}\log\frac{\tau_2}{\tau_1}
\log\frac{\tau_3^2}{\tau_1\tau_2}
\nonumber\,.
\end{align}
The total cross section is then the sum of these three color channels.  For brevity, we will not write the combined result.  Further, the result for gluon jets to this approximation is found by making the replacement $C_F\to C_A$, though we also will not write that out explicitly.

These results are sufficient to calculate the gluon reducibility factor, corresponding to the smallest value of the likelihood formed from the ratio of the quark to gluon cross sections.
Motivated by the location of the likelihood minima in the case of the cross section for $\tau_1$ and $\tau_2$, we consider the ordered limit $\tau_3\to \tau_2\to\tau_1$.  In this limit, the cross sections in the $C_F^2 C_A$ and $C_F C_A^2$ vanish; only the $C_F^3$ channel is non-zero.  We therefore find
\begin{equation}
\left. \frac{1}{\sigma_0}\frac{d^3 \sigma_q}{d\tau_1\, d\tau_2\,d\tau_3}\right|_{\tau_3\to \tau_2\to\tau_1} = -\left(2\frac{\alpha_s}{\pi}\right)^3 C_F^3 \frac{\log^3 \tau_3}{\tau_1^3}\,.
\end{equation}
The corresponding limit for gluon jets is similar:
\begin{equation}
\left. \frac{1}{\sigma_0}\frac{d^3 \sigma_g}{d\tau_1\, d\tau_2\,d\tau_3}\right|_{\tau_3\to \tau_2\to\tau_1} = -\left(2\frac{\alpha_s}{\pi}\right)^3 C_A^3 \frac{\log^3 \tau_3}{\tau_1^3}\,.
\end{equation}
The reducibility factor for gluons is then the ratio of these cross sections, with $\tau_1\to 1$:
\begin{equation}
\kappa_g(\tau_1,\tau_2,\tau_3) = \left(\frac{C_F}{C_A}\right)^3 \simeq 0.0878\,.
\end{equation}

Marginalizing the cross section over $\tau_1$ and $\tau_2$ enables us to determine the distribution of $\tau_3$.  For quarks, we find
\begin{equation}
\frac{1}{\sigma_0}\frac{d\sigma_q}{d \tau_3} = -\frac{\alpha_s^3}{\pi^3}\left(C_F^3+C_F^2 C_A+\frac{4}{15}C_FC_A^2 \right) \frac{\log^5 \tau_3}{\tau_3}\,,
\end{equation}
Correspondingly, for gluons, we find
\begin{equation}
\frac{1}{\sigma_0}\frac{d\sigma_g}{d \tau_3} = -\frac{\alpha_s^3}{\pi^3}\frac{34}{15}C_A^3 \frac{\log^5 \tau_3}{\tau_3}\,,
\end{equation}
It then follows that the gluon reducibility factor with $\tau_3$ can be found from the ratio of these distributions:
\begin{equation}\label{eq:gluensubredfact}
\kappa_g(\tau_3)  = \frac{C_F^3+C_F^2 C_A+\frac{4}{15}C_FC_A^2}{\frac{34}{15}C_A^3} \simeq  0.178 < 0.259 = \kappa_g(\tau_2)\,.
\end{equation}
Note that this reducibility factor for just measuring $\tau_3$ is smaller than even the reducibility factor for measuring $\tau_1$ and $\tau_2$ in \Eq{eq:t1t2redfac}.  This suggests that just measuring $\tau_n$ for sufficiently large $n$ a pure sample of quarks and gluons can be defined.

\subsubsection{Calculation of Gluon Reducibility for Any Number of Emissions}

These results are evidence for the scaling of the reducibility factor of gluons to be $(C_F/C_A)^n$, if $n$ emissions in the jet are resolved by measuring the set of $N$-subjettiness observables $\tau_1,\tau_2,\dotsc,\tau_n$.
For this to be true, it must be that the contribution to the quark cross section from the mixed color channels $C_F^{n-i}C_A^i$ for $0<i<n$ vanishes at the point that the likelihood assumes its minimum value.
We will prove this from a direct calculation of the cross section in an arbitrary color channel, in the strongly-ordered soft and collinear limits.

The $n$-differential cross section for $N$-subjettiness observables measured on quark jets in the $C_F^{n-i}C_A^i$ color channel can be expressed as
\begin{align}
\hspace{-0.2cm}\frac{d^n\sigma_q^{C_F^{n-i}C_A^i}}{d\tau_1\,d\tau_2\,\cdots\, d\tau_n} = \left(
2\frac{\alpha_s}{\pi}
\right)^n C_F^{n-i}C_A^i \sum_{\sigma_k}\prod_{j=1}^n\left[\int_0^1 \frac{dz_j}{z_j}\int_0^{\theta_{j,\max}}\frac{d\theta_j}{\theta_j}  \delta\left(\tau_j - z_j\theta_j \prod_{k = 1}^m z_{\sigma_k}\right)\right]\,.
\end{align}
Here, the product runs over all $n$ emissions that set each of the $\tau_j$ values.
The outer sum runs over all possible orderings of the emission tree.
The upper bound on the angular integral $\theta_{j,\max}$ represents the appropriate maximum angle for $\theta_j$.  If the $j$th emission is from the hard core of the jet, $\theta_{j,\max}$ is just 1.  If $j$ is a secondary (or later) emission off of other emissions in the jet, then this is the appropriate angle  to enforce angular ordering.
Note that the particular ordering fixes the maximum energy of any given emission; this is expressed with the product of energy fractions within the $\delta$-functions.
In the strongly-ordered energy limit, only if a gluon is directly emitted off of the initiating quark does its energy range up to the total jet energy.

We now first assume that $0< i<n$, so that there is at least one gluon that is a secondary emission off of another gluon.
Now, set all $N$-subjettiness values $\tau_j$ equal to $\tau_1$, corresponding to the ordered limit $\tau_n \to \tau_{n-1} \to \cdots \to \tau_1$.
Then, as long as $0<i<n$, there will be at least one pair of $\delta$-functions in the differential cross section for $\tau_{j_1}$ and $\tau_{j_2}$, with $j_1>j_2$, whose arguments are of the form
\begin{equation}
\delta(\tau_{j_1} - z_{j_1} \theta_{j_1})\delta(\tau_{j_2} - z_{j_1} z_{j_2}\theta_{j_2})\to \delta(\tau_1 - z_{j_1} \theta_{j_1})\delta(\tau_1 - z_{j_1} z_{j_2}\theta_{j_2})\,.
\end{equation}
However, this then sets
\begin{equation}
\theta_{j_1} = z_{j_2}\theta_{j_2}\,.
\end{equation}
Choosing the appropriate pair $j_1$ and $j_2$ such that $\theta_{j_2,\max} = \theta_{j_1}$ means that $\theta_{j_1} > \theta_{j_2}$, but $z_{j_2}<1$, so these requirements are inconsistent.  Note that this choice of $j_1$ and $j_2$ can always be done: if $j_2$ is a secondary emission off of $j_1$, then both the energy and angle of $j_2$ are constrained by $j_1$.  Therefore, the $C_F^{n-i}C_A^{i}$ color channel of the quark cross section vanishes for $0<i<n$, in the ordered limit $\tau_n \to \tau_{n-1} \to \cdots \to \tau_1$.

By contrast, the cross section in the pure $C_F^n$ color channel does not vanish.  Every gluon that sets the value of the $N$-subjettiness observables in this color channel is emitted directly off of the initiating quark.  Therefore, the cross section in this channel is
\begin{align}
\frac{d^n\sigma_q^{C_F^n}}{d\tau_1\,d\tau_2\,\cdots\, d\tau_n} = \left(
2\frac{\alpha_s}{\pi}
\right)^n C_F^n \prod_{j=1}^n\int_0^1 \frac{dz_j}{z_j}\int_0^1\frac{d\theta_j}{\theta_j} \delta\left(\tau_j - z_j \theta_j \right)\,.
\end{align}
Setting all $\tau_j = \tau_1$, then this evaluates to
\begin{equation}
\left.\frac{d^n\sigma_q^{C_F^n}}{d\tau_1\,d\tau_2\,\cdots\, d\tau_n}\right|_{\tau_j = \tau_1} =(-1)^n \left(
2\frac{\alpha_s}{\pi}
\right)^n C_F^n\frac{\log^n \tau_1}{\tau_1^n}\,.
\end{equation}
The gluon cross section in this limit is found from $C_F\to C_A$:
\begin{equation}
\left.\frac{d^n\sigma_g^{C_A^n}}{d\tau_1\,d\tau_2\,\cdots\, d\tau_n}\right|_{\tau_j = \tau_1} =(-1)^n \left(
2\frac{\alpha_s}{\pi}
\right)^n C_A^n\frac{\log^n \tau_1}{\tau_1^n}\,.
\end{equation}
The gluon reducibility factor for measuring enough $N$-subjettiness observables to resolve $n$ emissions is then just the ratio of these cross sections:
\begin{equation}
\kappa_g(\tau_1,\tau_2,\dotsc,\tau_n) = \left(
\frac{C_F}{C_A}
\right)^n\,.
\end{equation}
Note that this is indeed the minimum value of the likelihood ratio; because $C_A > C_F$, a contribution to the quark cross section from any other $C_F^{n-i}C_A^i$ color channel would increase this ratio.
This completes the proof of the gluon reducibility factor for $n$ resolved emissions.

The arguments in this proof explicitly relied on the form of the cross section in the double logarithmic limit.  However, the region of phase space which is dominated by gluon jets, where $\tau_n \to \tau_{n-1}\to \cdots \to \tau_1\to 1$, is not accurately described by the double logarithmic approximation.  Higher-order resummation and fixed-order corrections are necessary to accurately describe this region, and those contributions do not necessarily have such a nice organization.  In the region of phase space dominated by fixed-order corrections, the matrix elements are smooth and exhibit no non-analytic structure.  Also, because $N_c = 3$ in QCD, the leading-color approximation is accurate, up to corrections of about 10\%.  These features of QCD and quark versus gluon discrimination suggest that the result for the reducibility factor for jets with $n$ resolved emissions derived in this section is a good approximation to what would be derived when all relevant effects are taken into account.

\subsection{Including Resummation}

To calculate the likelihood and related quantities, we further need to calculate the resummed probability distribution for $\tau_1$, $\tau_2$, and $\tau_3$ measured on jets. 
In a similar way to what was done in the case for just measuring $\tau_1$ and $\tau_2$, we can express the joint probability distribution as an integral over a product of conditional probabilities:
\begin{equation}\label{eq:tripdiffprob}
p(\tau_1,\tau_2,\tau_3)=\int_0^1 dz_1 \int_0^1 dz_2 \, p(\tau_1) p(z_1|\tau_1)p(\tau_2|z_1,\tau_1)p(z_2|\tau_2,z_1,\tau_1)p(\tau_3|z_2,\tau_2,z_1,\tau_1)\,.
\end{equation}
In the strongly-ordered limit, the first three of these probability distributions have already been calculated in the previous sections.
We only need to calculate $p(z_2|\tau_2,z_1,\tau_1)$ and $p(\tau_3|z_2,\tau_2,z_1,\tau_1)$.
The quark jet probability distribution for the energy fraction $z_2$ of the second gluon emission $p(z_2|\tau_2,z_1,\tau_1)$ can be extracted from the multi-differential fixed-order cross section, allowing the second gluon to be emitted either from the quark line or off of the primary gluon emission.  One then finds
\begin{equation}
p_q(z_2|\tau_2,z_1,\tau_1) = -\frac{1}{z_2}
\frac{C_F \Theta(1-z_2)\Theta(z_2-\tau_2)+C_A \Theta(z_1-z_2)\Theta\left(
z_2 - z_1 \frac{\tau_2}{\tau_1}
\right)}{C_F \log \tau_2 + C_A \log\frac{\tau_2}{\tau_1}}\,.
\end{equation}

\begin{figure}
\begin{center}
\includegraphics[width=7cm]{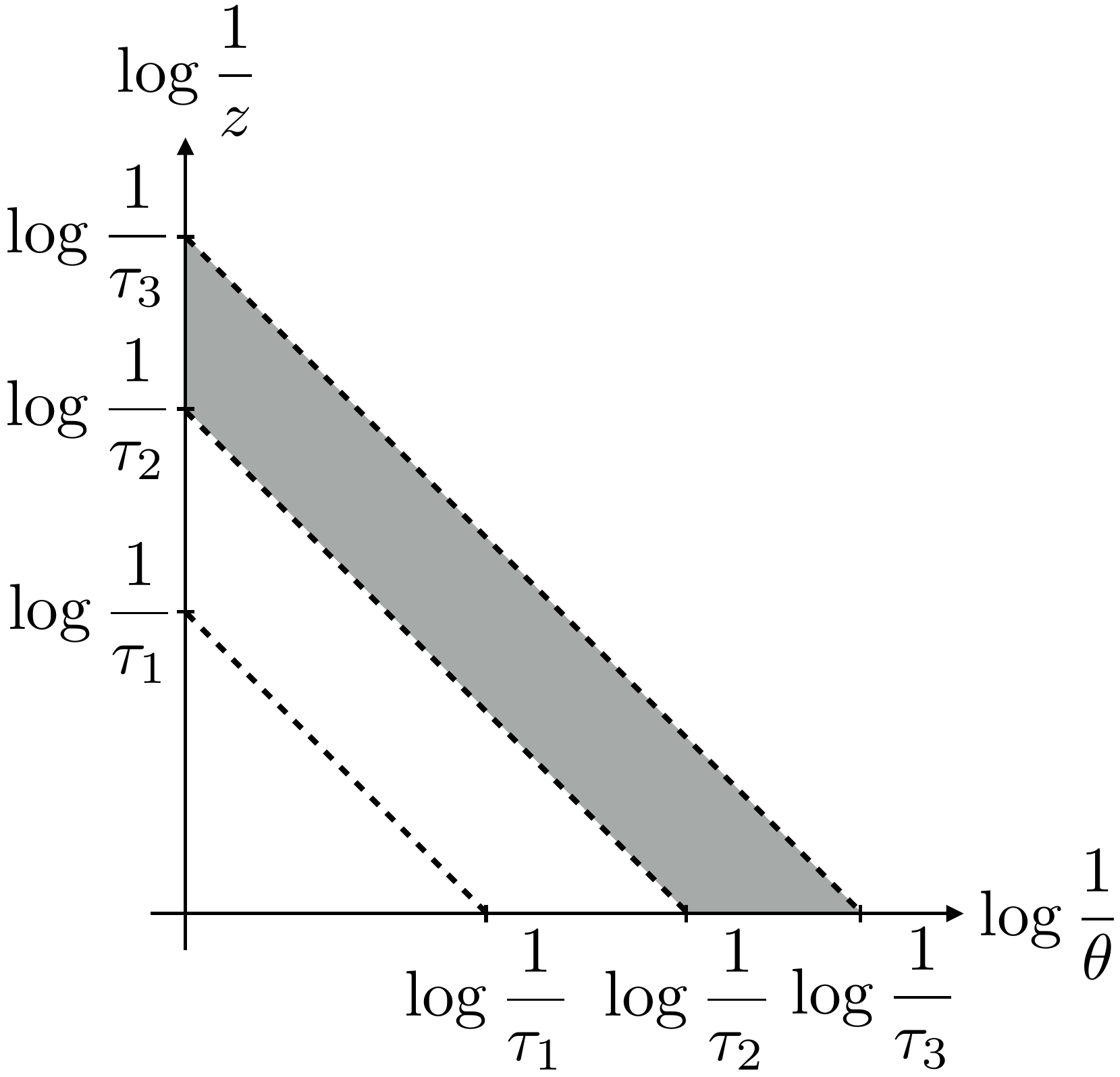}
\hfill
\includegraphics[width=7cm]{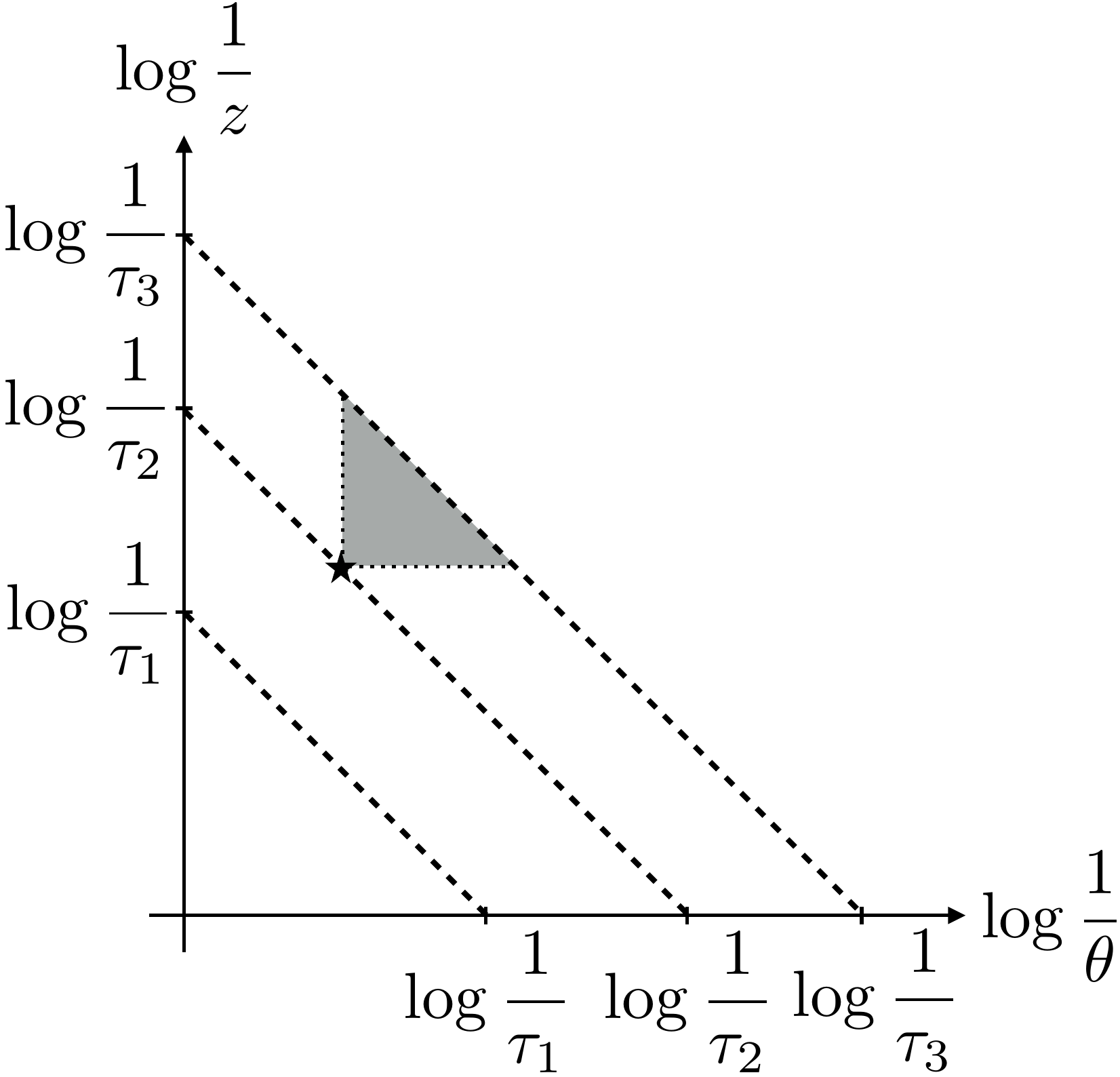}
\end{center}
\caption{Illustrations of the two of the forbidden regions (grayed) for gluon emission that sets the value of $\tau_3$, given a values of $\tau_1$ and $\tau_2$.  The location of the emission that sets the value of $\tau_2$ in the Lund plane is illustrated by the star.  On the left is the forbidden region if the gluon is emitted off of the initiating quark; the only restriction on the gluon is that it must set $\tau_3 < \tau_2$.  On the right is the forbidden region if the gluon is emitted off of the gluon that sets the value of $\tau_2$; it must be both at smaller angle and have smaller energy than the first emitted gluon.}
\label{fig:regt3}
\end{figure}

To calculate the quark jet resummed conditional distribution for $\tau_3$, we first consider its cumulative conditional distribution, $\Sigma_q(\tau_3|z_2,\tau_2,z_1,\tau_1)$.
This distribution is just the Sudakov form factor in the double logarithmic approximation, and so is just exponentiated areas on the Lund plane.
These areas are illustrated in \Figs{fig:regt3}{fig:regt3_new}.
First, on the left in \Fig{fig:regt3}, we can consider the forbidden emission area if the gluon that sets $\tau_3$ is emitted off of the quark.
With appropriate color and coupling factors, this area is:
\begin{equation}
\text{Area}_{C_FC_F} = \frac{\alpha_s}{\pi}C_F \left(
\log^2\tau_3-\log^2\tau_2
\right)\,.
\end{equation}
On the right of \Fig{fig:regt3} is the situation if the gluon is emitted off of the secondary gluon, the gluon that sets the value of $\tau_2$.
The forbidden emission area in this case is:
\begin{equation}
\text{Area}_{C_F C_A} = \frac{\alpha_s}{\pi}C_A \log^2\frac{\tau_3}{\tau_2}\,.
\end{equation}
Both of these areas are just the analogs of the corresponding situation in the two emission case of \Fig{fig:regt2}.

\begin{figure}
\begin{center}
\includegraphics[width=7cm]{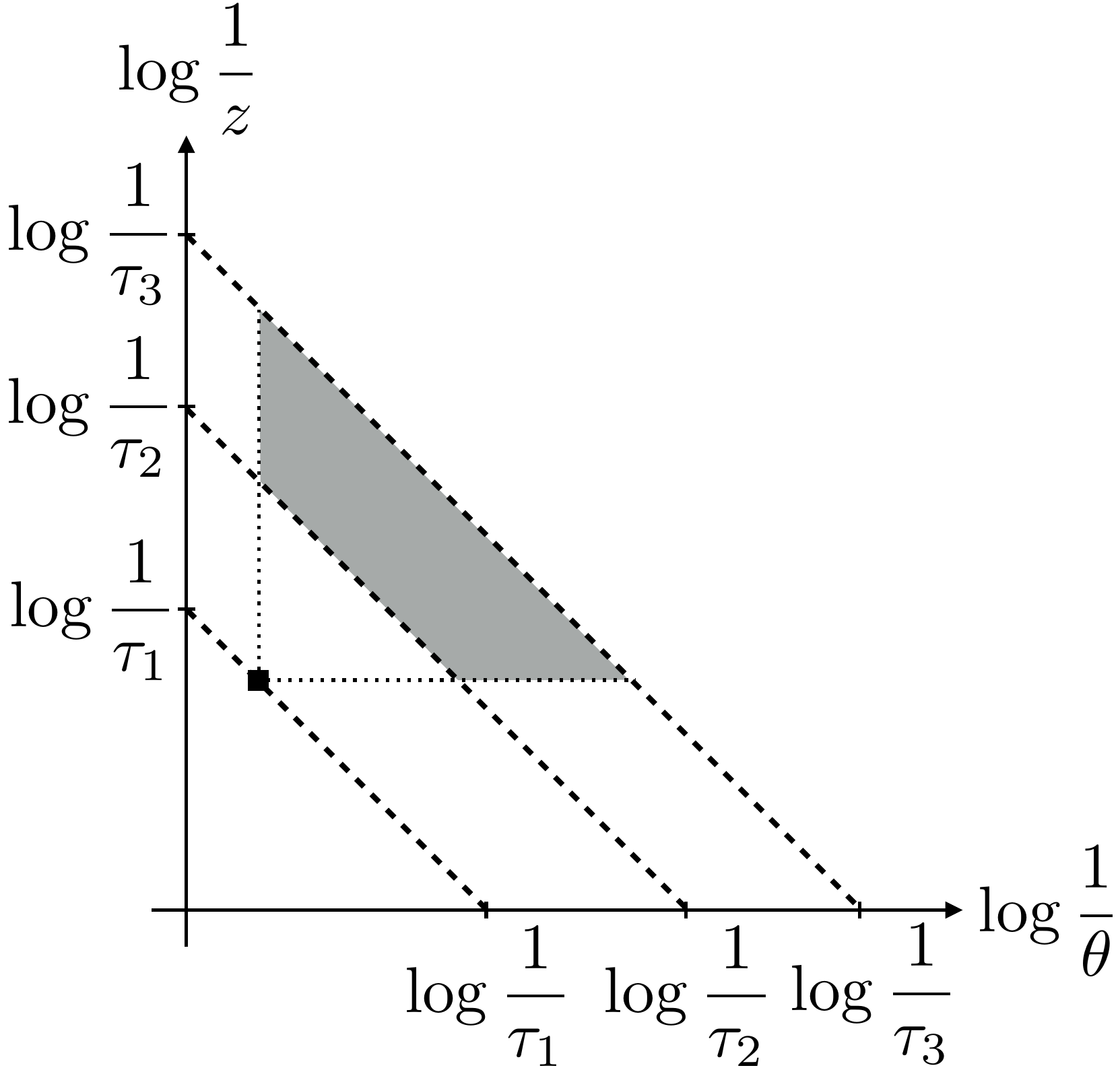}
\end{center}
\caption{Illustration of the third of the forbidden regions (grayed) for gluon emission that sets the value of $\tau_3$, given a values of $\tau_1$ and $\tau_2$.  The location of the emission that sets the value of $\tau_1$ in the Lund plane is illustrated by the square and the emission.  The forbidden region is constrained by the energy and angle of the primary gluon emission and by enforcing $\tau_2 < \tau_3$.}
\label{fig:regt3_new}
\end{figure}

If the emission that sets $\tau_3$ is off of the primary gluon, then the forbidden emission area is a bit more subtle.
This is illustrated in \Fig{fig:regt3_new}, and the area now depends on the energy fraction and angle of the primary gluon emission, as well as the value of $\tau_2$.  With color and coupling factors, this forbidden emission area is
\begin{equation}
\text{Area}_{C_AC_F} = \frac{\alpha_s}{\pi}C_A \left(\log^2\frac{\tau_3}{\tau_1}-\log^2\frac{\tau_2}{\tau_1}\right)\,.
\end{equation}
The Sudakov form factor is just the exponential of these areas.
For calculating the conditional probability, we differentiate the Sudakov form factor to find
\begin{align}
&
\hspace{-0.18cm}
p_q(\tau_3|z_2,\tau_2,z_1,\tau_1) = -2\frac{\alpha_s}{\pi}\frac{1}{\tau_3}\left(
C_F \log \tau_3 + C_A\log\frac{\tau_3^2}{\tau_1\tau_2}
\right)e^{-\text{Area}_{C_FC_F}-\text{Area}_{C_F C_A}-\text{Area}_{C_AC_F}}\,.
%
\end{align}
We leave the area factors in the exponential implicit for brevity and, as always, the result for gluon jets is found from replacing $C_F \to C_A$.

Unlike in previous sections, we will not explicitly write the triple differential distribution out, as it is now unwieldy.
At any rate, it can be calculated from the provided conditional probabilities and by integrating over the values of the primary and secondary emitted gluon energy fractions, $z_1$ and $z_2$, as in \Eq{eq:tripdiffprob}.
To lowest order, the resummed expression agrees with the fixed-order calculations from earlier in this section.
Additionally, we just note that the likelihood ratio
\begin{equation}
{\cal L}(\tau_1,\tau_2,\tau_3) = \frac{p_g(\tau_1,\tau_2,\tau_3)}{p_q(\tau_1,\tau_2,\tau_3)}
\end{equation}
is IRC safe, by a similar reasoning as we used in the previous section.  

\subsection{AUC Evaluation}

With the probability distributions and the likelihood calculated, we can then calculate the AUC for quark versus gluon discrimination when three emissions in jets are observed.
Extending the calculation for the AUC from \Sec{sec:auc2}, it can be expressed in this case as
\begin{align}
&\text{AUC} = \int_0^1 d\tau_{1q} \int_0^{\tau_{1q}} d\tau_{2q}\int_0^{\tau_{2q}} d\tau_{3q}\int_0^1 d\tau_{1g} \int_0^{\tau_{1g}} d\tau_{2g}\int_0^{\tau_{2g}} d\tau_{3g}\\
&
\hspace{3cm} 
\times\, p_q(\tau_{1q},\tau_{2q},\tau_{3q})p_g(\tau_{1g},\tau_{2g},\tau_{3g})\,\Theta\left({\cal L}(\tau_{1q},\tau_{2q},\tau_{3q})-{\cal L}(\tau_{1g},\tau_{2g},\tau_{3g})\right)\,.\nonumber
\end{align}
As before, to perform the integral to calculate the AUC, we use the implementation of {\tt Vegas} within Cuba 4.2.
Using $C_F = 4/3$ and $C_A = 3$, we find that the AUC of the triple differential likelihood is
\begin{equation}
\text{AUC} \simeq 0.231 <  0.256 < \frac{1}{1+\frac{C_A}{C_F}}\simeq 0.308\,.
\end{equation}
Going right we compare to the value of the AUC for resolving two emissions (0.256), and resolving just one emission (0.308).
So, the absolute discrimination power is definitely improved, but the size of the relative improvement in going from resolving two to three emissions has decreased from that of resolving one to two emissions.

\begin{figure}
\begin{center}
\includegraphics[width=9cm]{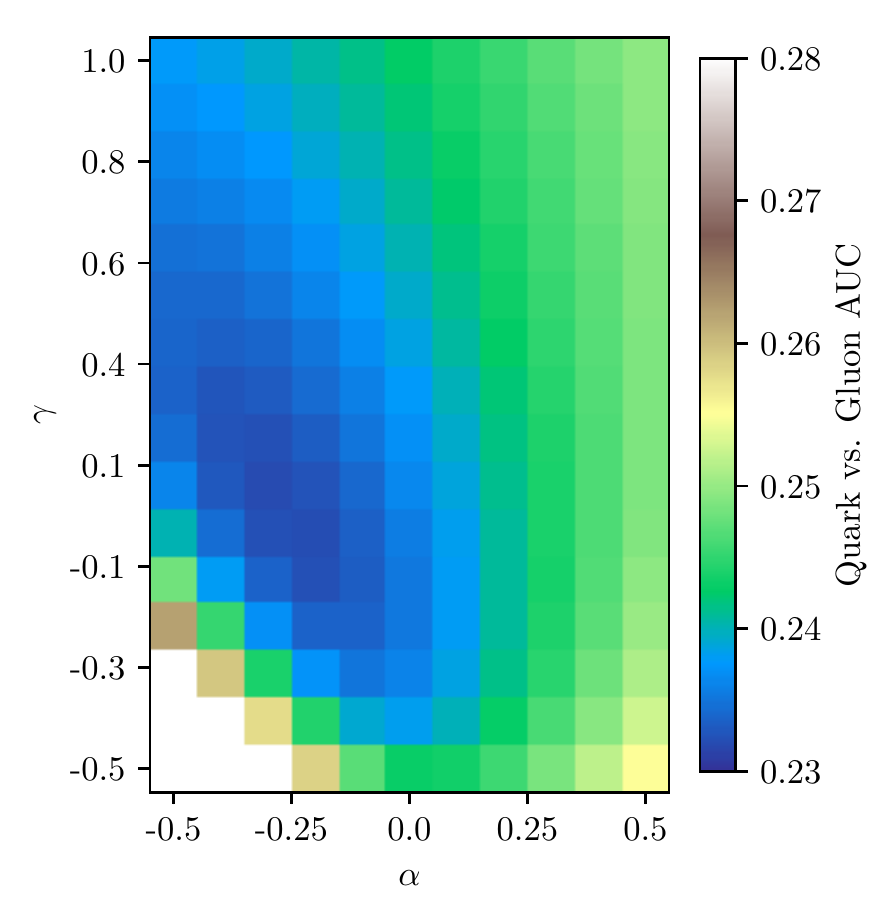}
\end{center}
\caption{Results of a scan over $\alpha$ of the AUC for the observable ${\cal O} = \tau_1^\alpha \tau_2^\gamma \tau_3$.  The AUC for the observable is minimized at $\alpha = -0.3$ and $\gamma = 0.1$ where it takes the value $0.232$.
}
\label{fig:auct3func}
\end{figure}

We can also extend the variational approach to construct a powerful discrimination observable whose functional form is much simpler than the full likelihood.
For illustration, we take a product form for an observable ${\cal O}$, where
\begin{equation}
{\cal O} = \tau_1^\alpha \tau_2^\gamma \tau_3^\delta\,.
\end{equation}
The likelihood vanishes in the $\tau_3\to 0$ limit, manifesting its IRC safety, and so we enforce $\delta > 0$.
Thus, without loss of generality, we can just set $\delta = 1$ and consider the observable
\begin{equation}
{\cal O} = \tau_1^\alpha \tau_2^\gamma \tau_3\,.
\end{equation}
In the ordered limits $\tau_3 \to \tau_2$ and $\tau_3 \to \tau_2 \to \tau_1$, IRC safety further enforces that $1+\gamma > 0$ and $1+\alpha+\gamma>0$.
To find the $\alpha$ and $\gamma$ values that yield the best discrimination power, we calculate the value of the AUC for an observable scan.
The results of this scan are shown in \Fig{fig:auct3func} for which the minimal AUC of 0.232 is achieved at $\alpha = -0.3$, $\gamma = 0.1$.
Perhaps a more complicated observable could be constructed that performed slightly closer to that of the likelihood, but we won't pursue that further here.

\subsubsection{Estimate of $n \to \infty$ AUC}

While we won't present further calculations of probability distributions to resolve four or more emissions in a jet in this paper, we can still make some robust statements about the discrimination power for any number of resolved emissions.

Our general analysis of the reducibility factors and their relationship to the ROC curve provided a bound on the AUC in \Eq{eq:aucbound}.
We can then apply this bound to our results for the quark and gluon reducibility factors.
We have shown that $\kappa_q = 0$ for any number of resolved emissions, while
\begin{equation}
\kappa_g = \left(\frac{C_F}{C_A}\right)^n\,,
\end{equation}
when $n$ emissions are resolved.  Plugging these values into the bounding formula we find
\begin{equation}
\text{AUC} \geq \frac{1}{2}\left(\frac{C_F}{C_A}\right)^n\,.
\end{equation}
That is, perfect discrimination power between quark and gluon jets is only possible if an infinite number of emissions are resolved.
As any physical jet contains only a finite number of particles in it, this suggests that there is an absolute lower bound on the AUC for the discrimination of physical quark and gluon jets.

This bound on the AUC is indeed satisfied by our results for one, two, and three resolved emissions.
Comparing to this bound, we had found
\begin{align}
&\text{AUC}_1 \simeq 0.308 > \frac{1}{2}\frac{C_F}{C_A} \simeq 0.222\,,\\
&\text{AUC}_2 \simeq 0.256 > \frac{1}{2}\left(\frac{C_F}{C_A}\right)^2 \simeq 0.0988\,,\nonumber\\
&\text{AUC}_3 \simeq 0.231 > \frac{1}{2}\left(\frac{C_F}{C_A}\right)^3 \simeq 0.0439\,.\nonumber
\end{align}
Here, the subscripts on the AUC represents the number of resolved emissions.
Because the rate of convergence to 0 observed in the complete calculations is so much slower than the bound would suggest, this might be evidence that any achievable AUC for a physical jet, even resolving all of its emissions, is relatively large.
So, our calculations suggest that there seems to be an inherent limitation to quark and gluon jet discrimination, beyond all of the subtleties regarding their fundamental, theoretical definition.

\section{IRC Safe Multiplicity}\label{sec:mult}

The way through which we defined resolved emissions in a jet, by measuring $N$-subjettiness, enables a simple, IRC safe, definition of resolved particle multiplicity in the jet.
Given an $n+1$ dimensional joint probability distribution, $p(\tau_1,\tau_2,\dotsc,\tau_{n+1})$, the probability that the jet has exactly $n$ resolved constituents is
\begin{align}\label{eq:ircsafemult}
p_n &= \int_0^1 d\tau_1 \int_0^{\tau_1}d\tau_2 \cdots \int_0^{\tau_{n-2}} d\tau_{n-1} \int_{\Lambda_0}^{\tau_{n-1}}d\tau_n \int_0^{\Lambda_0}d\tau_{n+1}\, p(\tau_1,\tau_2,\dotsc,\tau_{n+1})\\
&= \int_{\Lambda_0}^{1}d\tau_n \int_0^{\Lambda_0}d\tau_{n+1}\, p(\tau_n,\tau_{n+1})\nonumber\\
&= \int_0^1d\tau_n \int_0^{\Lambda_0}d\tau_{n+1}\, p(\tau_n,\tau_{n+1}) \,\Theta(\tau_n-\tau_{n+1})-\int_0^{\Lambda_0}d\tau_n \int_0^{\Lambda_0}d\tau_{n+1}\, p(\tau_n,\tau_{n+1}) \,\Theta(\tau_n-\tau_{n+1})\nonumber\\
&=\Sigma_{n+1}(\Lambda_0)-\Sigma_n(\Lambda_0)\,.\nonumber
\end{align}
Here, $\Lambda_0>0$ is some resolution cut that is responsible for the IRC safety of this multiplicity definition.  While we always assume the ordering of $N$-subjettiness observables $\tau_n > \tau_{n+1}$, we only first explicitly write it in the third line to connect to the expression in the final line.
In the final equation, $\Sigma_n(\Lambda_0)$ is shorthand for the cumulative distribution
\begin{equation}
\Sigma_n(\Lambda_0) \equiv \int_0^{\Lambda_0}d\tau_n \, p(\tau_n)=\int_0^{\Lambda_0}d\tau' \, p_n(\tau')\,.
\end{equation}
This multiplicity distribution is normalized when summed over all $n$:
\begin{align}
\sum_{n=0}^\infty p_n = \sum_{n=0}^\infty \left[
\Sigma_{n+1}(\Lambda_0)-\Sigma_n(\Lambda_0)
\right] = \lim_{n\to\infty} \Sigma_n(\Lambda_0) = 1\,.
\end{align}
Because the $N$-subjettiness variables are ordered $\tau_1 \geq \tau_2 \geq \cdots$, for sufficiently large $n$ and a fixed cutoff $\Lambda_0$, the value of $\tau_n$ will have probability 1 to be below $\Lambda_0$.

Note that the probability distribution of this multiplicity has strictly less information than the full joint probability distribution, because information in lost in doing the integral up to the scale of the resolution variable.
Therefore, the discrimination power of such a multiplicity is strictly less than that of the likelihood formed from the ratio of joint probability distributions $p_g(\tau_1,\tau_2,\dotsc,\tau_{n+1})/p_q(\tau_1,\tau_2,\dotsc,\tau_{n+1})$.
The AUC as a measure of the discrimination power of this IRC safe multiplicity can be calculated and one finds
\begin{align}
\text{AUC}&=\frac{1}{2}\sum_{i=0}^\infty \left[\Sigma_{q,i+1}(\Lambda_0)-\Sigma_{q,i}(\Lambda_0)\right]\left[\Sigma_{g,i}(\Lambda_0)+\Sigma_{g,i+1}(\Lambda_0)\right]\,.
\end{align}
This formula can be derived by summing over the area of the trapezoids that make up the ROC curve.  Additionally, we have assumed that the multiplicity is monotonic in the likelihood of multiplicity, which is expected.

More realistically, one only resolves up through $n$ emissions in the jet, inclusive over more emissions.
In our calculations, for example, we have only resolved up through three emissions in the jet, and so we can only say if the jet has 0, 1, 2, or three-or-more emissions.
In this case, to calculate the AUC, the sum must be truncated:
\begin{align}
\text{AUC} &= \frac{1}{2}\sum_{i=0}^{n-1} \left[\Sigma_{q,i+1}(\Lambda_0)-\Sigma_{q,i}(\Lambda_0)\right]\left[\Sigma_{g,i}(\Lambda_0)+\Sigma_{g,i+1}(\Lambda_0)\right]\,,
\end{align}
where $\Sigma_{q,n}(\Lambda_0)=1$.
Applying this formula to our multi-differential probability distribution $p(\tau_1,\tau_2,\tau_3)$ calculated in the previous section we found a minimum AUC value of about $0.274$ for $\Lambda_0 \simeq 0.005$.
Note that this is indeed larger than the AUC formed from the likelihood $p_g(\tau_1,\tau_2,\tau_3)/p_q(\tau_1,\tau_2,\tau_3)$.

The expression of the triple-joint probability distribution is complicated and does not provide an intuition for what physics controls the discrimination power of multiplicity.
In some cases, most notably through iterative soft drop \cite{Frye:2017yrw}, the multiplicity is approximately distributed as a Poisson random variable.
For such observables, the mean multiplicity of quark and gluon jets are related by their color factors:
\begin{align}
&\lambda_q = C_F \lambda\,,
&\lambda_g = C_A \lambda\,,
\end{align}
for some fiducial multiplicity $\lambda$.
The probability for $n$ resolved emissions distributed according to the Poisson distribution is then
\begin{equation}
p_n = \frac{\lambda_i^n}{n!}e^{-\lambda_i}\,,
\end{equation}
for a mean $\lambda_i$.
If all emissions in the jet are resolved, the quark and gluon reducibility factors of Poisson-multiplicity are
\begin{align}
&\kappa_q = e^{-\lambda_g+\lambda_q}= e^{-(C_A-C_F)\lambda}\,, &\kappa_g = 0\,.
\end{align}
From our expression on the lower bound on the AUC from reducibility factors, we find that
\begin{equation}
\text{AUC}\geq \frac{e^{-(C_A-C_F)\lambda}}{2}\,.
\end{equation}
Note that this lower bound only vanishes if the fiducial mean multiplicity $\lambda\to \infty$.

Iterated soft drop multiplicity was argued to be the optimal quark versus gluon discriminant at leading logarithmic accuracy \cite{Frye:2017yrw}.  This would seem to be at odds with our analysis here with collections of $N$-subjettiness observables.  However, there are a few differences.  First, at leading-logarithmic accuracy, iterated soft drop is only sensitive to emissions off of the hard core of the jet, while $N$-subjettiness (or related) observables can be sensitive to secondary emissions.  Thus, the leading-logarithmic phase space is different between these observables.  A sufficiently large collection of $N$-subjettiness observables completely resolves $M$-body phase space, but iterated soft drop at leading-logarithm could, in principle, remove an arbitrary number of emissions from the jet before identifying an emission that passes.  Thus, to directly compare, one would at least need to consider jets on which an arbitrary number of $N$-subjettiness observables are measured.  Further, $N$-subjettiness is a continuous variable while (any definition of) multiplicity is discrete, so comparing their discrimination power in practice is more challenging.

\subsection{Relationship of Multiplicity to Individual $N$-subjettiness Observables}\label{sec:multnsub}

This formulation of multiplicity suggests a new way of thinking about it that can provide insight into its discrimination performance in comparison to other observables.
In particular, in this section we compare the quark vs.~gluon discrimination performance of multiplicity to that of an individual $N$-subjettiness observable, $\tau_n$. 
Definitive statements about their relationship require more information about the multiplicity distribution, but we conjecture that $\tau_n$ has an AUC bounded from below by multiplicity.
Further, we conjecture that this inequality is saturated when $n$ is about the number of minimal constituents in a gluon jet.
The observation that $N$-subjettiness $\tau_n$ for large $n$ is a good quark vs.~gluon discriminant has been known for a long time \cite{Gallicchio:2012ez}, and we hope that the arguments presented here can be sharpened in the future.
In this section, we work beyond leading-logarithmic accuracy, and attempt to make general statements that hold even non-perturbatively regarding the relationship of multiplicity and $N$-subjettiness as quark vs.~gluon discriminants.

In practice, multiplicity is not defined with a cutoff; it is just a count of all those experimentally-resolved constituents of a jet.
We will not attempt at defining what ``experimentally-resolvable'' means nor attempt to include a finite cutoff representing the experimental limitations.  In this spirit, we will just write $\Lambda_0 = 0$ in the following with the caveat that ``0'' here may actually be a finite value.  At any rate, its value is not set within the applicability of perturbation theory so invalidates the conclusions made earlier.
True multiplicity is thus the $\Lambda_0\to 0$ limit of the IRC-safe multiplicity whose distribution we had defined in \Eq{eq:ircsafemult}:
\begin{align}
\text{AUC}_{\text{mult}}&=\frac{1}{2}\sum_{i=0}^{n_{q,\max}-1} \left[\Sigma_{q,i+1}(0)-\Sigma_{q,i}(0)\right]\left[\Sigma_{g,i}(0)+\Sigma_{g,i+1}(0)\right]\,.
\end{align}
Any realistic collection of jets will only have a finite number of constituents, and so the sum terminates once the maximum number of quark jet constituents $n_{q,\max}$ has been reached.
That is, once $i$ is at least $n_{q,\max}$ all jets have 0 value for $\tau_i$ or that
\begin{equation}
\Sigma_{q,i}(0)=1\,,
\end{equation}
for $i \geq n_{q,\max}$.  

Because we have defined multiplicity through properties of the $N$-subjettiness variables, this allows a convenient comparison to the discrimination power of an individual $N$-subjettiness $\tau_n$.
The AUC for $\tau_n$ can be approximated by:
\begin{align}
\text{AUC}_{\tau_n} &= \int_0^1 d\tau'\, p_{q,n}(\tau')\Sigma_{g,n}(\tau')\\
&\approx\frac{1}{2}\sum_{i=0}^{N-1} \left[\Sigma_{q,n}(x_{i+1})-\Sigma_{q,n}(x_i)\right]\left[\Sigma_{g,n}(x_i)+\Sigma_{g,n}(x_{i+1})\right]
\,.\nonumber
\end{align}
Here, the $\{x_i\}$ are a collection of points of $\tau_n\in[0,1]$ at which the ROC is evaluated.
To directly compare to multiplicity, it is convenient to set the number of bins in the ROC curve $N=n_{q,\max}$ and choose the locations of the bins to match that of multiplicity.
This means that we choose the points $x_i$ such that
\begin{equation}
\Sigma_{q,n}(x_i)=\Sigma_{q,i}(0)\,,
\end{equation}
or that 
\begin{equation}
x_i=\Sigma_{q,n}^{-1}(\Sigma_{q,i}(0))\,.
\end{equation}
Note also that due to $\Sigma_{q,i}(0) \geq \Sigma_{q,n}(0)$ for $i\geq n$ we have that $x_i = 0$ if $i\leq n$.  With this choice of points, the AUC of $\tau_n$ is then approximately
\begin{align}
&\text{AUC}_{\tau_n}\\
&
\hspace{0.2cm} \approx\frac{1}{2}\Sigma_{q,n}(0)\Sigma_{g,n}(0)+\frac{1}{2}\sum_{i=n}^{N-1} \left[\Sigma_{q,i+1}(0)-\Sigma_{q,i}(0)\right]\left[\Sigma_{g,n}(\Sigma_{q,n}^{-1}(\Sigma_{q,i}(0)))+\Sigma_{g,n}(\Sigma_{q,n}^{-1}(\Sigma_{q,i+1}(0)))\right]\,.\nonumber
\end{align}

It's then straightforward to evaluate the difference between the AUC for $\tau_n$ and multiplicity:
\begin{align}\label{eq:aucdiff}
&
\hspace{-0.25cm}\text{AUC}_{\tau_n}-\text{AUC}_{\text{mult}}\approx
\frac{1}{2}\Sigma_{q,n}(0)\Sigma_{g,n}(0)-\frac{1}{2}\sum_{i=0}^{n-1} \left[\Sigma_{q,i+1}(0)-\Sigma_{q,i}(0)\right]\left[\Sigma_{g,i}(0)+\Sigma_{g,i+1}(0)\right]\\
&
\hspace{-0.4cm}+
\frac{1}{2}\sum_{i=n}^{n_{q,\max}-1}\left[\Sigma_{q,i+1}(0)-\Sigma_{q,i}(0)\right]\left[\Sigma_{g,n}(\Sigma_{q,n}^{-1}(\Sigma_{q,i}(0)))-\Sigma_{g,i}(0)+\Sigma_{g,n}(\Sigma_{q,n}^{-1}(\Sigma_{q,i+1}(0)))-\Sigma_{g,i+1}(0)\right]
\,.\nonumber
\end{align}
The first term in this AUC difference is just the area of a right triangle with sides of length $\Sigma_{q,n}(0)$ and $\Sigma_{g,n}(0)$.  Because the ROC and its first derivative are both monotonically increasing, the difference of the first two terms is necessarily non-negative:
\begin{align}
\frac{1}{2}\Sigma_{q,n}(0)\Sigma_{g,n}(0)-\frac{1}{2}\sum_{i=0}^{n-1} \left[\Sigma_{q,i+1}(0)-\Sigma_{q,i}(0)\right]\left[\Sigma_{g,i}(0)+\Sigma_{g,i+1}(0)\right] \geq0\,.
\end{align}

\begin{figure}[t]
\begin{center}
\includegraphics[width=7cm]{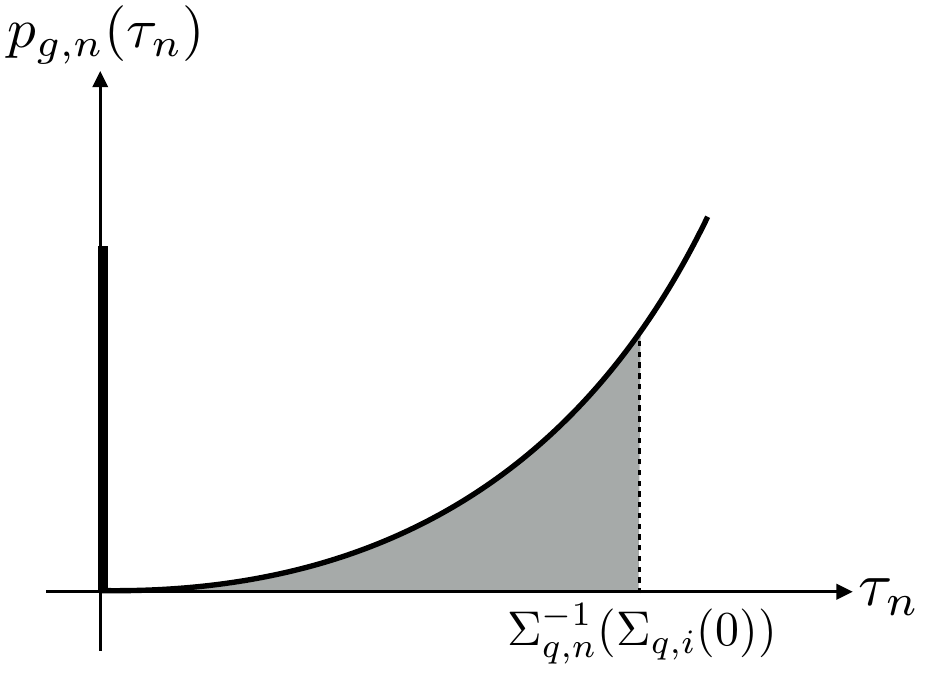}
\end{center}
\caption{\label{fig:aucdiff}
Illustration of the integration region over $N$-subjettiness $\tau_n$ of the gluon jet probability distribution that defines the quantity $\Sigma_{g,n}(\Sigma_{q,n}^{-1}(\Sigma_{q,i}(0)))$.  Note the $\delta$-function at $\tau_n=0$ for all those jets with $n$ or fewer constituents.
}
\end{figure}

Unfortunately, it is much more challenging to determine the sign of the sum on the second line of \Eq{eq:aucdiff}.  The sign of this term is set by the difference of gluon cumulative distributions:
\begin{equation}
\Sigma_{g,n}(\Sigma_{q,n}^{-1}(\Sigma_{q,i}(0)))-\Sigma_{g,i}(0)\,,
\end{equation}
for $i>n$.  For $i=n$, this difference is just 0.  The interpretation of the term $\Sigma_{g,n}(\Sigma_{q,n}^{-1}(\Sigma_{q,i}(0)))$ is the following.  First, $\Sigma_{q,i}(0)$ is the total integral of the quark jet events for which $\tau_i$ is zero.  Because we assume that $i>n$, note that $\Sigma_{q,n}(0)< \Sigma_{q,i}(0)$.  Then, there exists some $\epsilon>0$ such that $\Sigma_{q,n}(\epsilon)< \Sigma_{q,i}(0)$.  This $\epsilon$ then sets the region over which we integrate the distribution $p_{g,n}(\tau_n)$, which includes a $\delta$-function at $\tau_n=0$ for those jets with $n$ or fewer constituents.  This is illustrated in \Fig{fig:aucdiff}.  We then need to compare this term to $\Sigma_{g,i}(0)$.  We now make the following reasonable conjecture, but have not been able to prove it.  We assume that all gluon jets for which $\tau_i=0$ satisfy the inequality
\begin{equation}
\tau_n \leq \Sigma_{q,n}^{-1}(\Sigma_{q,i}(0))\,.
\end{equation}
With this assumption, it then follows that 
\begin{equation}
\Sigma_{g,n}(\Sigma_{q,n}^{-1}(\Sigma_{q,i}(0)))-\Sigma_{g,i}(0)\geq 0\,.
\end{equation}
While this is seems reasonable, we emphasize that we do not have a proof.

With this assumption, we then establish the approximate inequality
\begin{equation}
\text{AUC}_{\tau_n}\gtrsim\text{AUC}_{\text{mult}}\,.
\end{equation}
Note that, through our explicit calculation, we demonstrated that
\begin{equation}
\text{AUC}_{\tau_1} \geq \text{AUC}_{\tau_2} \geq \text{AUC}_{\tau_3}\,.
\end{equation}
Further, if $n$ is very large and approaching the maximal number of quark jet constituents $n_{q,\max}$, $\tau_n$ is just 0 for most quark and gluon jets.  So, at very large $n$, AUC$_{\tau_n}$ approaches $1/2$.  Therefore, there must be some $n$ at which $\text{AUC}_{\tau_n}$ is minimized, and is close to $\text{AUC}_{\text{mult}}$.  Gluon jets in our sample will have some minimal number of constituents, call it $n_{g,\min}$.  For all $n< n_{g,\min}$, $\tau_n = 0$ and then the first two terms of the difference in \Eq{eq:aucdiff} vanish:
\begin{equation}
\frac{1}{2}\Sigma_{q,n}(0)\Sigma_{g,n}(0)-\frac{1}{2}\sum_{i=0}^{n-1} \left[\Sigma_{q,i+1}(0)-\Sigma_{q,i}(0)\right]\left[\Sigma_{g,i}(0)+\Sigma_{g,i+1}(0)\right] = 0\,.
\end{equation}
This follows because $\Sigma_{g,i}(0) = 0$ for $i< n_{g,\min}$.  Therefore, the largest $n$ for which this term in the AUC difference is (approximately) 0 is when $n\gtrsim n_{g,\min}$.  This suggests that the difference between the $\tau_n$ and multiplicity AUCs is minimized when $ n_{g,\min} \lesssim n\ll n_{q,\max}$.  As we will see in our Monte Carlo studies, $n_{g,\min}$ is about 15, or so, suggesting that $\tau_{15}$ is about as good a discriminant as multiplicity.  However, in practice, the discrimination power of $\tau_n$ quickly saturates, even for $n$ as small as 5 or so.

\section{Comparison to Monte Carlo Simulation}\label{sec:mc}

In this section, we explore our calculations and conclusions in the context of simulated samples of quark and gluon jets.
While here we will use a manifestly unphysical flavor definition for quark and gluon jets, operational flavor definitions~\cite{Metodiev:2018ftz,Komiske:2018vkc} may be used in practice to study our conclusions directly in data.

Dijet events are generated with \textsc{Pythia 8.226}~\cite{Sjostrand:2006za,Sjostrand:2014zea} at $\sqrt{s}=14$~TeV with the default tunings and shower parameters, including hadronization and multiple parton interactions (i.e.~underlying event).
Final state non-neutrino particles are clustered into $R=0.4$ anti-$k_T$ jets~\cite{Cacciari:2008gp} with \textsc{FastJet 3.3.0}~\cite{Cacciari:2011ma}, keeping up to two jets with transverse momentum $p_T\in[1000,1100]$~GeV and rapidity $|y|<2.5$.
We compute $N$-subjettiness observables with $\beta\in\{0.5,1.0,2.0\}$ using \textsc{FastJet Contrib 1.029}~\cite{fjcontrib} with winner-take-all axes~\cite{Larkoski:2014uqa}.
Jets are labeled as ``quark'' or ``gluon'' based on the flavor of the closest parton in the hard process, required to be within $2R$ of the jet four-momentum.

\subsection{Quark vs.~Gluon Classification Performance}

\begin{figure}[t]
\centering
  \subfloat[]{
    \includegraphics[scale=0.775]{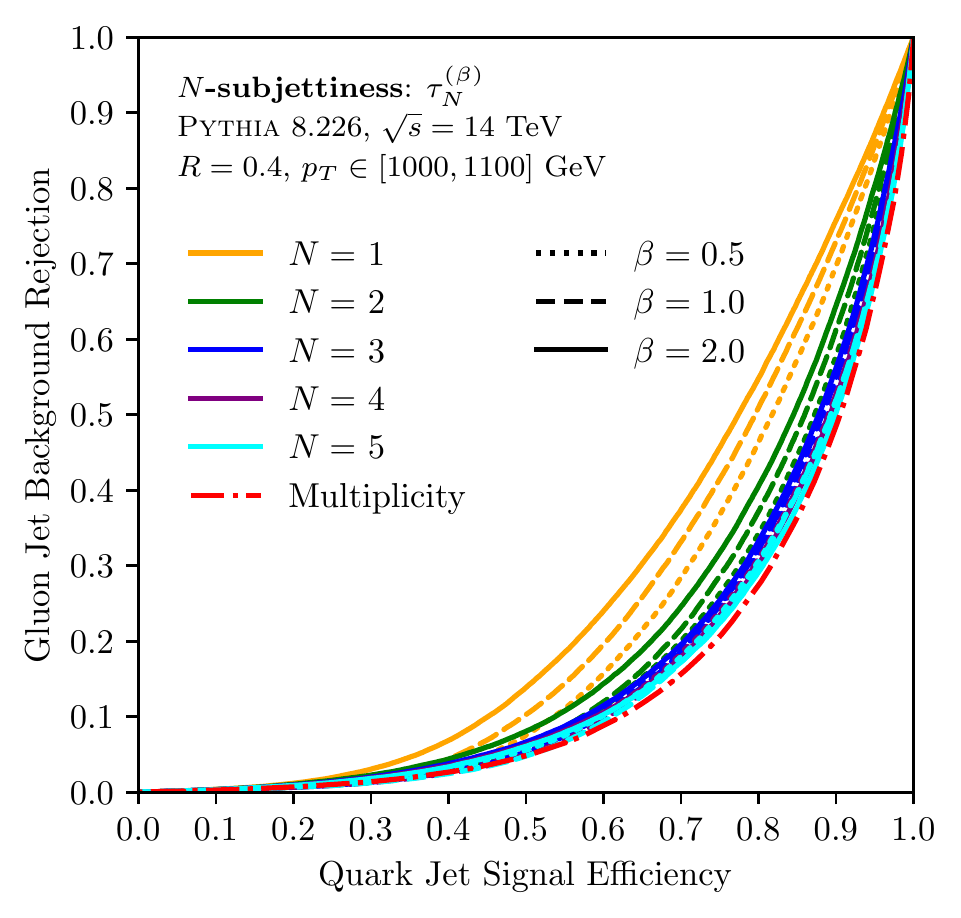}
	\label{fig:nsubauca}
	}
 \subfloat[]{
    \includegraphics[scale=0.775]{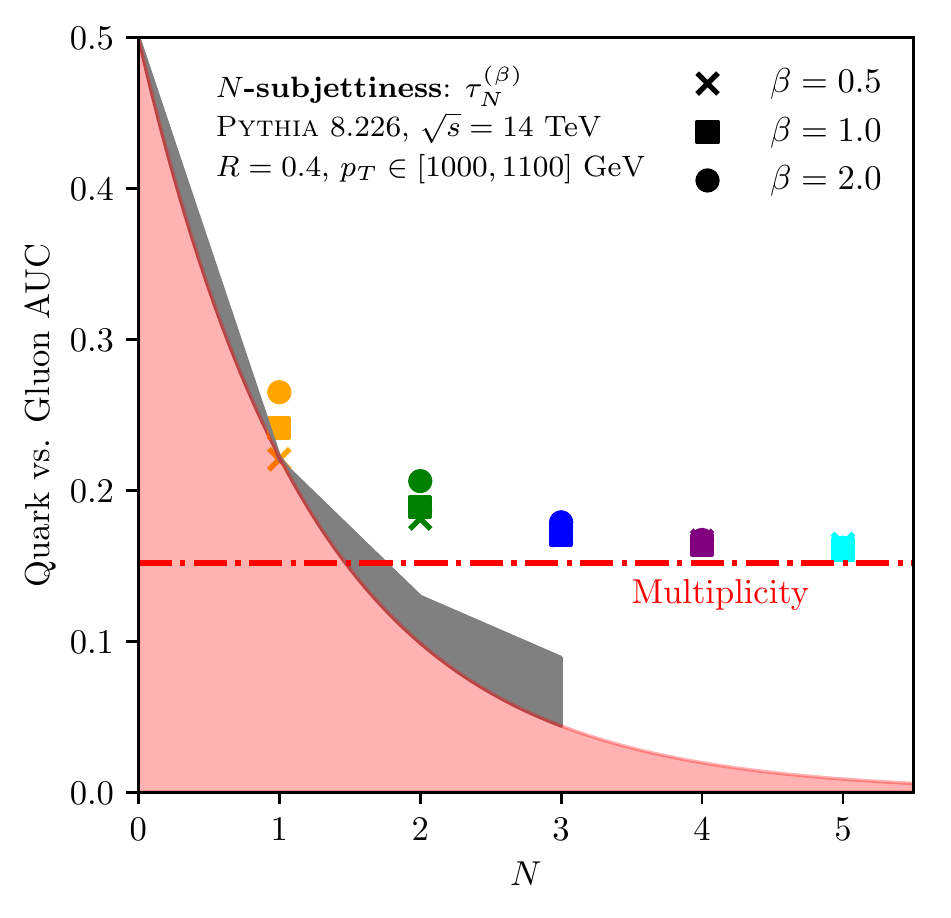}
	\label{fig:nsubaucb}
	}
\caption{\label{fig:ROCbounds} The (a) ROC curves and (b) AUCs for $N$-subjettiness observables for $N$ up to 5 and $\beta\in\{0.5,1.0,2.0\}$.
The AUCs indeed satisfy the predicted general $N$-emission bound $\frac12(C_F/C_A)^N$  (red) and specific calculated $N$-subjettiness bounds for $N\in\{1,2,3\}$ that follow from applying \Eq{eq:aucbound} to the results of \Eq{eq:gluensubredfact} (gray), with minimal $\beta$ dependence.
The performance of constitutent multiplicity is also shown, with $N$-subjettiness approaching the classification performance of multiplicity for large $N$.
}
\end{figure}

The quark vs.~gluon discrimination ROC curves of the $N$-subjettiness observables with $N$ up to 5 are shown in \Fig{fig:nsubauca}.
While the angular weighting parameter $\beta$ has no effect on our calculations at this accuracy, we show results for $\beta\in\{0.5,1.0,2.0\}$ to give a sense of the robustness of our predictions.
The AUCs of these observables are shown in \Fig{fig:nsubaucb}, along with the $N$-emission AUC bound of $\frac12(C_F/C_A)^N$ and the tighter $N$-subjettiness AUC bounds for $N\le3$.
The bounds are indeed borne out in practice, with only mild dependence on $\beta$, and the $N$-subjettiness AUC bound explains the majority of the performance ceiling for the computed $N$ values.
We find that smaller $\beta$ values tend to mildly improve the discrimination power of the individual observables, consistent with the overall conclusions of \Refs{Larkoski:2013eya,Komiske:2017aww}.
Further, \Fig{fig:ROCbounds} shows the ROC curve and AUC for the constituent multiplicity, which is an IRC unsafe observables that is known to be a good quark/gluon discriminant~\cite{Gallicchio:2012ez}.
We find that the $N$-subjettiness observables closely approach the performance of multiplicity for large values of $N$.

\begin{figure}[t]
\centering
  \subfloat[]{
    \includegraphics[height=0.46\columnwidth]{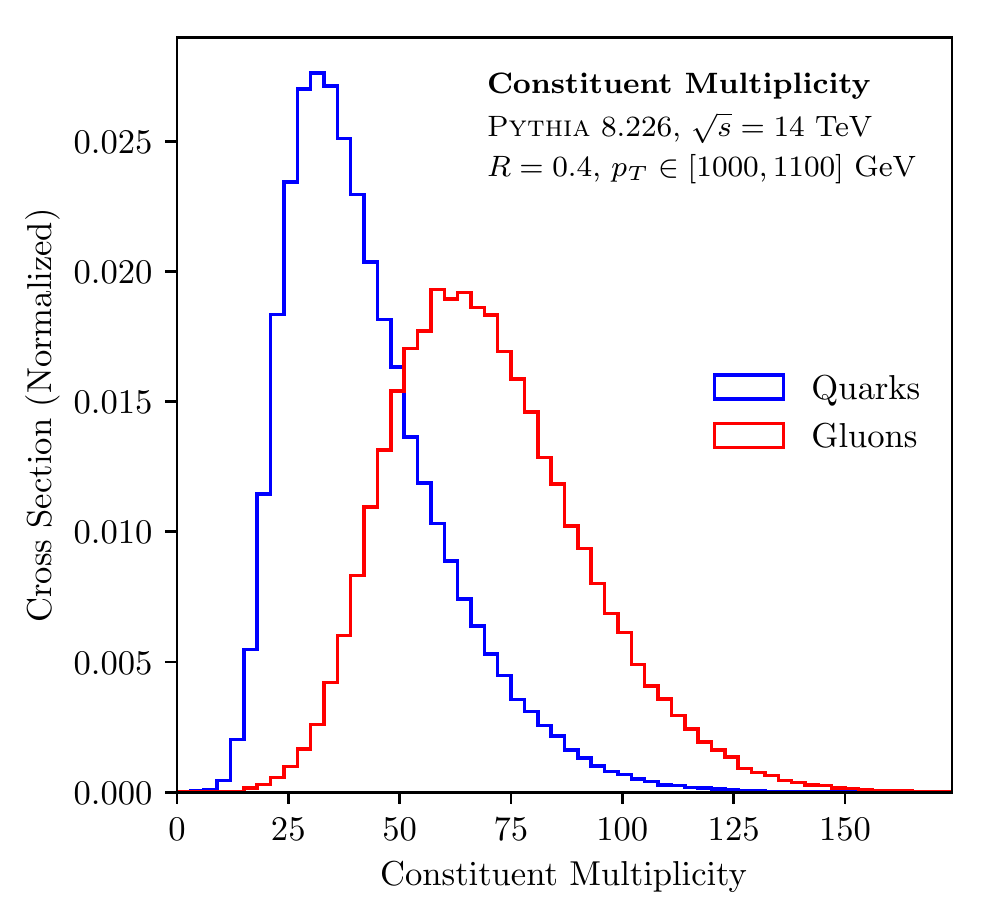}
	\label{fig:multdist}
	}
 \subfloat[]{
    \includegraphics[height=0.46\columnwidth]{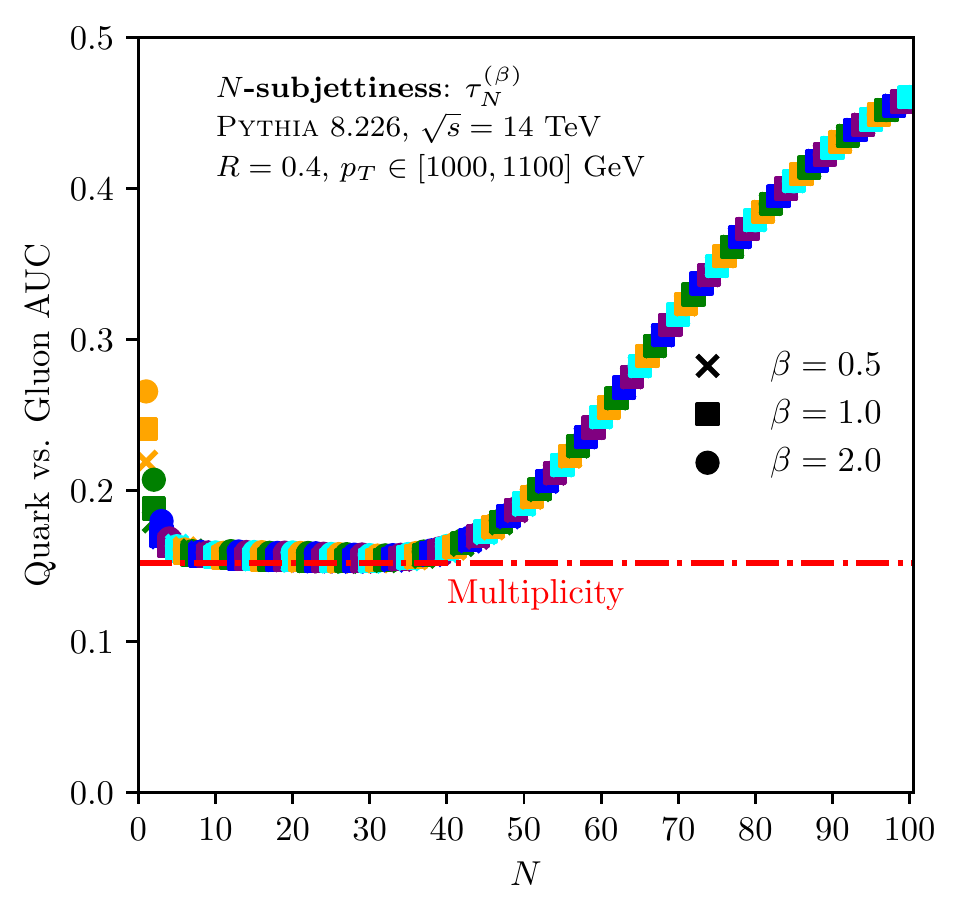}
	\label{fig:largensubauc}
	}
\caption{\label{fig:multnsub} The (a) distribution of constituent multiplicity in quark and gluon jets and (b) the AUCs for $N$-subjettiness observables for $N$ up to 100 and $\beta\in\{0.5,1.0,2.0\}$.
The AUCs of $N$-subjettiness observables quickly converge to the AUC of multiplicity for small $N$ and then begin to diverge from multiplicity once $N$ is comparable to and larger than the mean quark multiplicity of about $N\gtrsim 40$.  This is consistent with expectations from \Sec{sec:multnsub}.
}
\end{figure}

This can be studied in more detail following the discussion of multiplicity in \Sec{sec:multnsub}.
\Fig{fig:multdist} shows the distributions of constituent particle multiplicity for quark and gluons in our simulated jet samples.
On average, quark jets have fewer constituents than gluons, due to the smaller color factor, and the smallest nontrivially-populated bin (greater than about one part in $10^5$) for gluon jets is about 15 or so.
From our conjecture at the end of \Sec{sec:multnsub}, we then expect that $\tau_N$ from about $N\gtrsim 15$ or so to exhibit similar discrimination power to that of multiplicity.
This is demonstrated in \Fig{fig:largensubauc} in which we plot the AUC for $\tau_N^{(\beta)}$, for $\beta\in\{0.5,1.0,2.0\}$ and $N$ out to 100.
The AUC of individual $N$-subjettiness observables converges rapidly to the AUC of multiplicity, and remains comparable until $N\sim 40$ at which the AUC diverges, approaching $0.5$ as $N$ increases.
This $N$ of the divergence point is also approximately the mean multiplicity of quark jets, suggesting that once most of the quark jets have $\tau_N = 0$, the discrimination power of $\tau_N$ is no longer optimal.
Also, because multiplicity has (weak) jet $p_T$ dependence, these relationships will have some $p_T$ dependence. Nevertheless, we do expect, for any $p_T$, a wide range of $N$ for which $\tau_N$ and multiplicity have comparable discrimination performance.
As discussed earlier, these fascinating relationships between individual $N$-prong observables and multiplicity merit further study.

\begin{figure}[t]
\centering
\includegraphics[width=9cm]{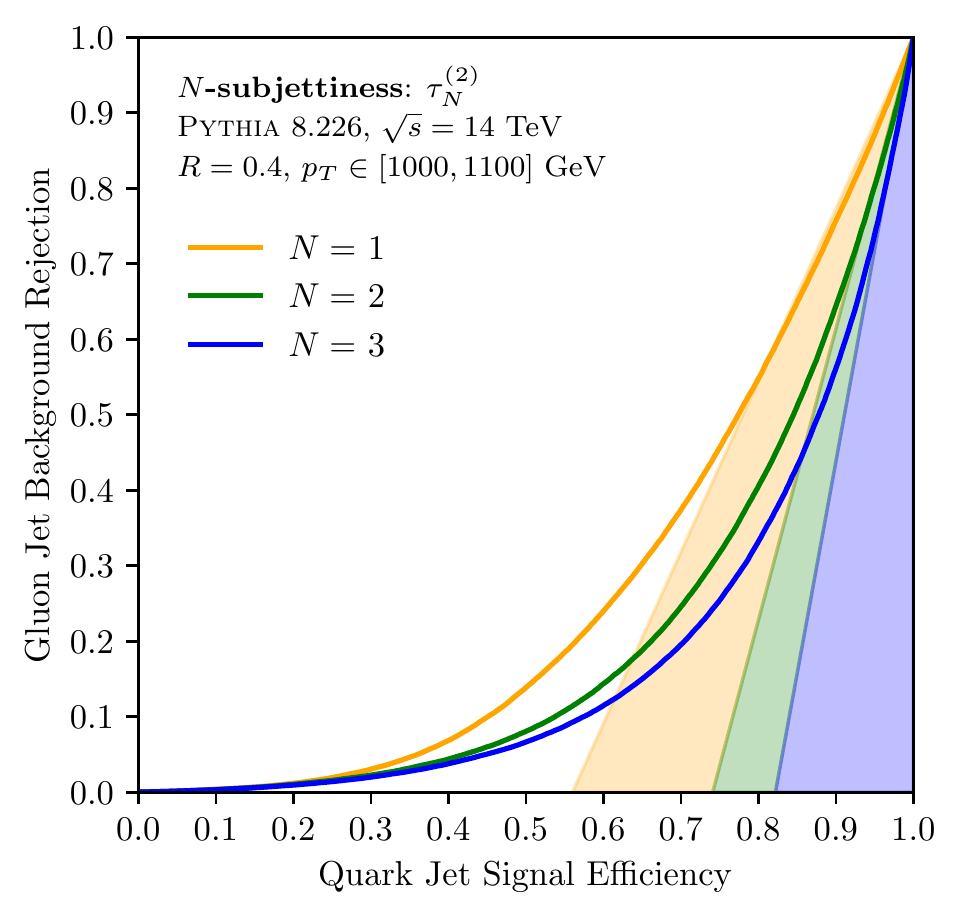}
\caption{\label{fig:ROCbounds2}
The ROC curves for $N$-subjettiness observables with $\beta=2$, together with the predictions for their asymptotic behavior.
The predictions are shown as shaded regions that are predicted to match the slope of the ROC curve in the high quark-efficiency region.
We also predict the slope of the ROC curve to approach zero in the low quark-efficiency region.
There is good agreement between the predictions and the observed classification performance.
}
\end{figure}

Beyond the AUC bounds, we also have predictions for the asymptotic ROC curve behaviors predicted by our power counting arguments.
\Fig{fig:ROCbounds2} shows the predicted asymptotic ROC curve behaviors in the high quark-efficiency region together with the ROC curve for $\beta=2$ $N$-subjettiness observables, where we have the best perturbative control.
We see good agreement with the analytical expectation, validating the applicability of the power counting reasoning to analyzing quark vs.~gluon discrimination.
We also predict that the ROC curve will have vanishing slope in the low quark-efficiency region, which is also borne out in these results.
These results indicate that $N$-subjettiness observables with large $N$ values may be a good candidates for data-driven quark/gluon definitions~\cite{Metodiev:2018ftz,Komiske:2018vkc}, due to their (near) mutual irreducibility while retaining analytic understanding and perturbative control.

To check the robustness of our analysis and conclusions to non-perturbative effects, we have also repeated the studies in this section at parton level, without hadronization.
Overall, we find a very similar story to the results presented in this section.
Differences include a decrease of quark/gluon discrimination power available at parton level and correspondingly smaller $N$ values for the performance saturation of $\tau_N$.

\subsection{Probing Machine Learning Strategies}

Our theoretical results allow us to explore and understand the behavior of machine learning strategies for jet or event classification in new ways, at least in a limited context.

We begin by considering classifiers formed via the product of observables.
This parameterization allows for the classification performance to be optimized while still producing a theoretically-understandable observable.
Such a strategy has been used successfully with products of $N$-subjettiness observables to optimize the performance of tasks such as $H\to b\bar b$ vs.~$g\to b\bar b$ using a brute force optimization of the product observable~\cite{Datta:2017lxt} as well as more sophisticated machine learning techniques~\cite{Datta:2019ndh}.
Here, we will consider this approach applied to quark versus gluon classification, with the product observable:
\begin{equation}
\mathcal O = \tau_1^{a_1} \tau_2^{a_2}\cdots \tau_N^{a_N},
\end{equation}
where the goal is to learn the parameters $a_1, \cdots, a_N$ to achieve optimal performance.
In general, we set $a_N=1$ by monotonically rescaling the observable without changing the classification performance.
Note that \Refs{Datta:2017lxt,Datta:2019ndh} used observables with three different $\beta$ values together in the product, whereas here we will consider observables with the same $\beta$ value for simplicity.

\begin{figure}[t]
\centering
\includegraphics[width=9cm]{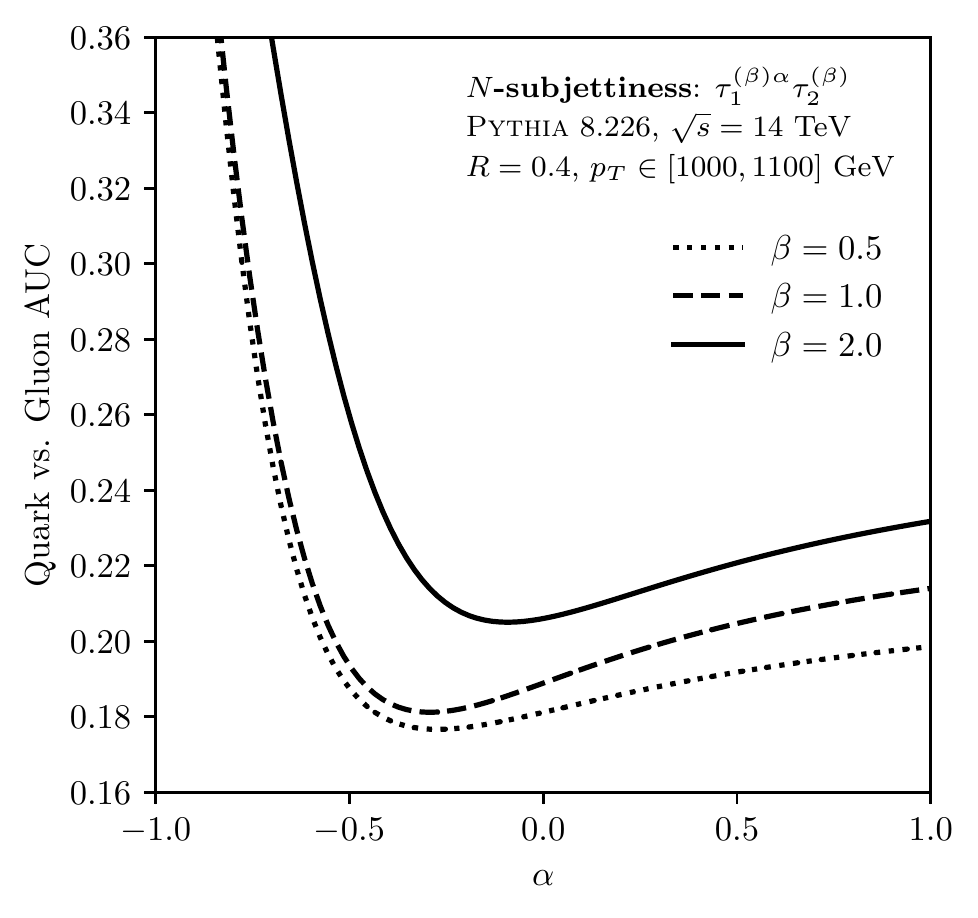}
\caption{\label{fig:productAUCs1}The AUC quark vs.~gluon discrimination performance of $N$-subjettiness product observable $\mathcal O = \tau_1^\alpha\tau_2$, sweeping over different parameter values.
We see good qualitative agreement with the predictions shown in \Fig{fig:auct2func}.
}
\end{figure}

\begin{figure}[t]
\centering
\includegraphics[width=9cm]{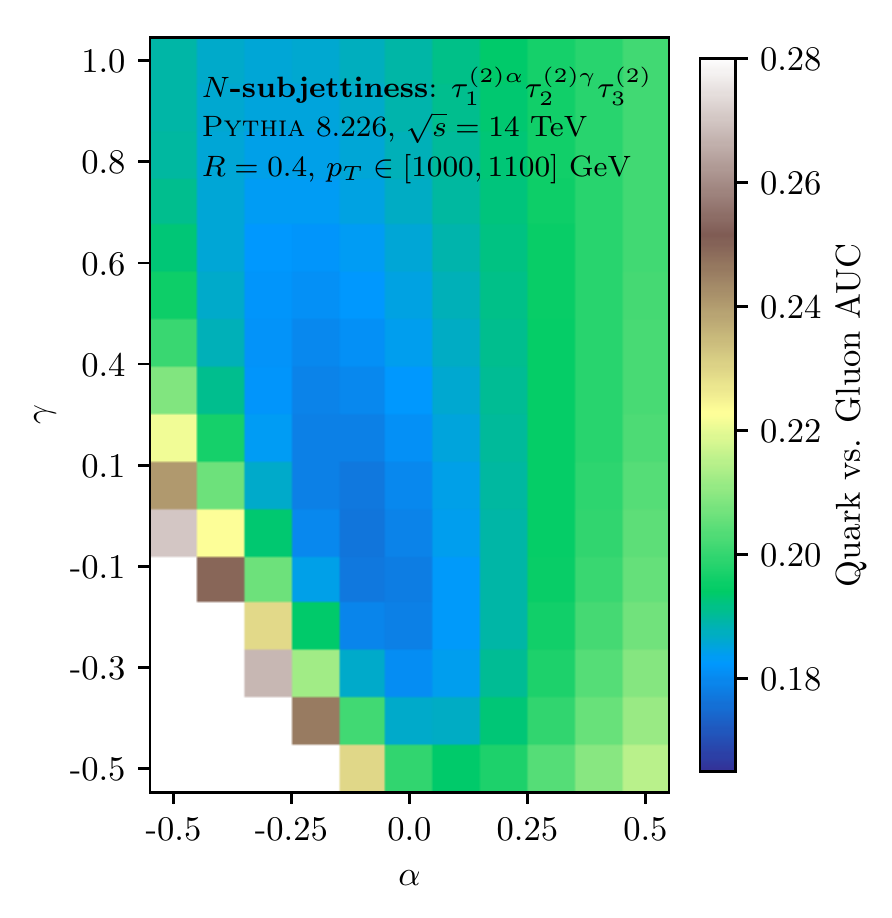}
\caption{\label{fig:productAUCs2}The AUC quark vs.~gluon discrimination performance of $N$-subjettiness product observable $\mathcal O = \tau_1^\alpha \tau_2^\gamma \tau_3$, sweeping over different parameter values.
We see good qualitative agreement with the predictions shown in \Fig{fig:auct3func}.
}
\end{figure}

The quark vs.~gluon AUC performance of the product observable $\mathcal O = \tau_1^\alpha \tau_2$ is shown in \Fig{fig:productAUCs1} over a sweep of $\alpha$ values with $\beta \in \{0.5,1.0,2.0\}$.
The qualitative features of this product agree rather well with the theoretical predictions of \Fig{fig:auct2func}, particularly for the case of $\beta=2$.
While the overall scale of the classification performance differs, the relative behavior of the different product observables is well-described by our calculation.
As with the prediction, the region near $\alpha=0$ is preferred to optimize the discrimination power.
Going further, the AUC performance of the product observable $\mathcal O = \tau_1^\alpha\tau_2^\gamma\tau_3$ is shown in \Fig{fig:productAUCs2} over a sweep of $\alpha$ and $\gamma$ parameter values with $\beta=2$.
Again, we see qualitative agreement with the predictions in \Fig{fig:auct3func}.
Both theoretically and in simulation, we see that a single $N$-subjettiness observable $\tau_N$ with the largest $N$ captures a great deal of the overall product classification performance.
These results suggest single $N$-subjettiness observables with large-$N$ as strong candidates for individual quark/gluon classification observables.
More broadly, these results are a significant step towards providing an analytic understanding of machine learning with product observables, such as those explored in \Refs{Datta:2017lxt,Datta:2019ndh}, from a first-principles multi-differential calculation.

A general strategy in machine learning for collider physics has been to combine the information from a collection of observables with dense neural networks (DNNs), boosted decision trees, or linear methods.
While these methods are intrinsically more opaque due to their black box nature, our theoretical understanding can nonetheless shed some light on the performance achieved by the model.
In particular, we will consider a relatively simple dense neural network consisting of two fully-connected layers of 100 nodes each.
All neural networks are implemented in Keras~\cite{keras} with the TensorFlow~\cite{tensorflow} backend on a sample of 200k jets with a 50k validation set and 50k test set.
A ReLU activation~\cite{relu} is used on each layer with He-uniform~\cite{heuniform} weight initializations, using a crossentropy loss function and the Adam optimization algorithm~\cite{adam}.
Models were trained with a batch size of 500 for 25 epochs.

\begin{figure}[t]
\centering
\includegraphics[width=9cm]{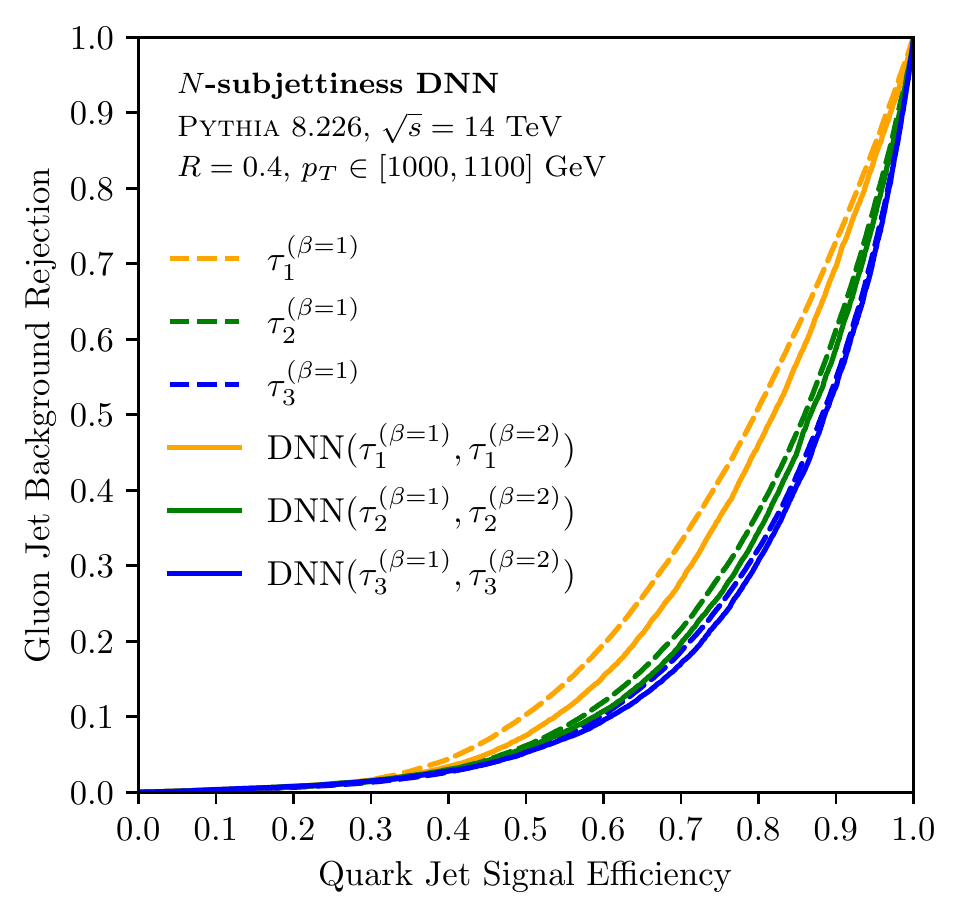}
\caption{\label{fig:nsubDNN} A comparison of the classification performance added by considering two $N$-subjettiness observables of the same $N$ and different $\beta$ values.
Combining $\beta=1$ and $\beta=2$ $N$-subjettiness observables with a DNN can increases the classification performance, particularly for $N=1$.
However, the parametric discrimination power, namely the ROC curve near the endpoints, is largely unchanged and approaches that for the $\beta=1$ $N$-subjettiness.
}
\end{figure}

To probe the information accessed by the model in combining observables, we begin by combining two $N$-subjettiness observables of the same $N$ and different $\beta$ values with a DNN.
The resulting ROC curves are shown in \Fig{fig:nsubDNN}.
Based on the analysis of \Sec{sec:2bodyps}, we do not anticipate two observables of the same $N$ to parametrically improve discrimination performance.
Indeed, the marginal improvements in performance are largely in the middle of the ROC curve with essentially unchanged parametric performance near the endpoints.
Thus on general grounds, in this simple case, we are able to understand the limits on the information probed by the network using the different observables without a requiring a detailed understanding of their multi-differential correlations.

The decisions made by the neural network can be understood in even more detail.
The output of a classifier trained with a crossentropy or mean squared error loss is optimally $S/(S+B)$, which is indeed monotonically related to the likelihood ratio.
In a feature space ${\bf x}$, the output of the trained classifier is optimally:
\begin{equation}
\text{NN}({\bf x}) = \frac{p_S({\bf x})}{p_S({\bf x}) + p_B({\bf x})},
\end{equation}
where in the case of a two softmaxed outputs, each component is optimally $S/(S+B)$ and $B/(S+B)$.
Assuming that the neural network is sufficiently trained to approach this limit, we can in principle predict its output and decision boundaries using our understanding of the signal and background distributions.

\begin{figure}[t]
\centering
\subfloat[]{
    \includegraphics[scale=0.675]{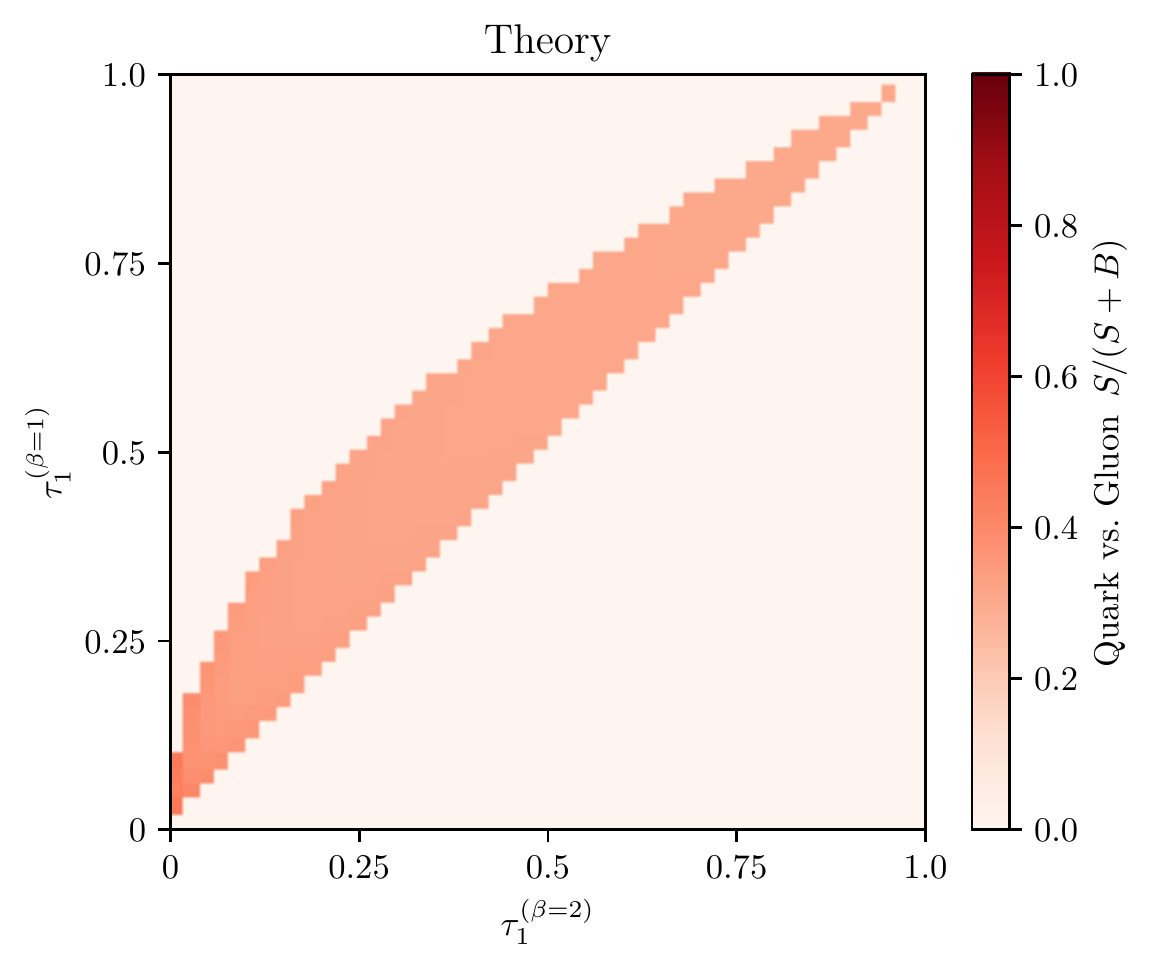}
	\label{fig:t1t2th}
	}
 \subfloat[]{
    \includegraphics[scale=0.675]{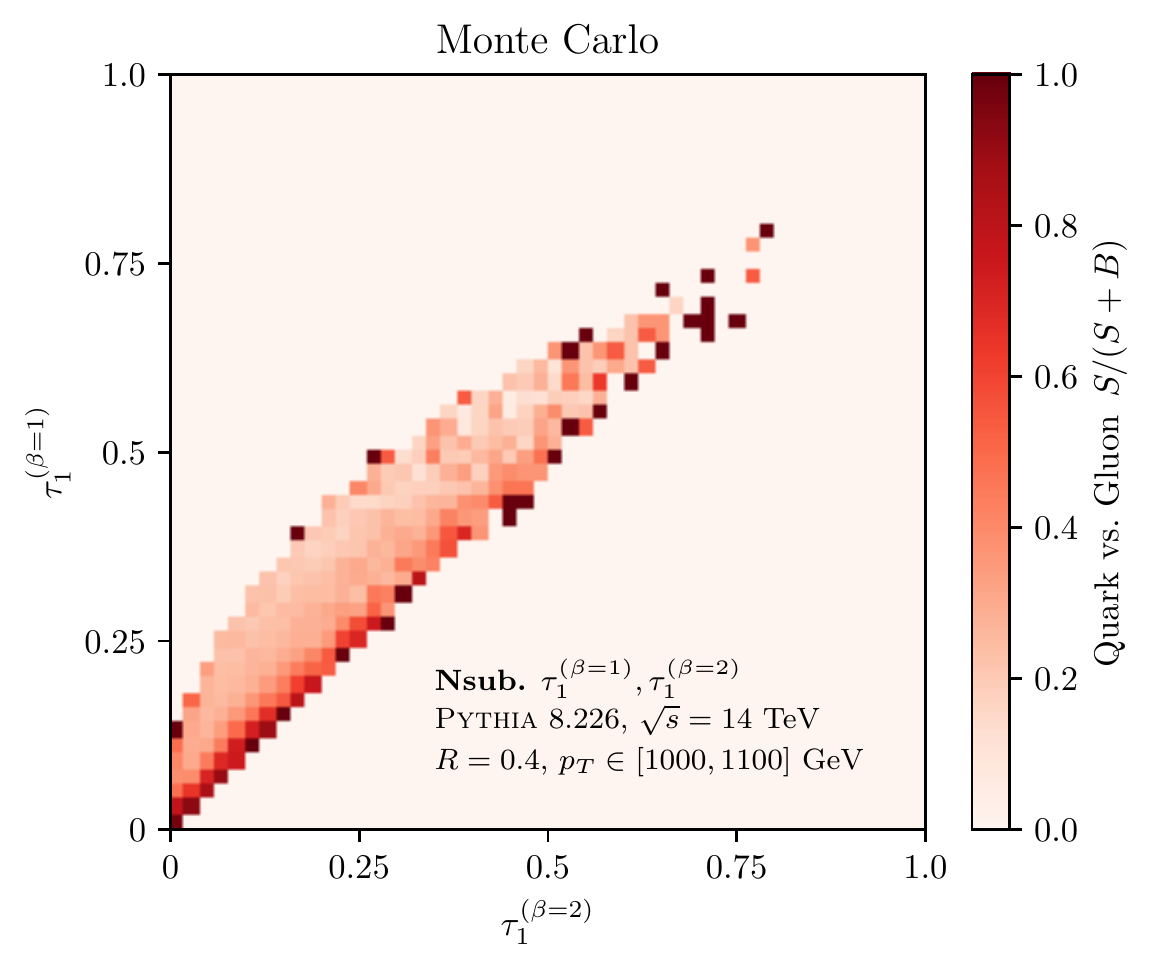}
	\label{fig:t1t2mc}
	}\\
 \subfloat[]{
    \includegraphics[scale=0.925]{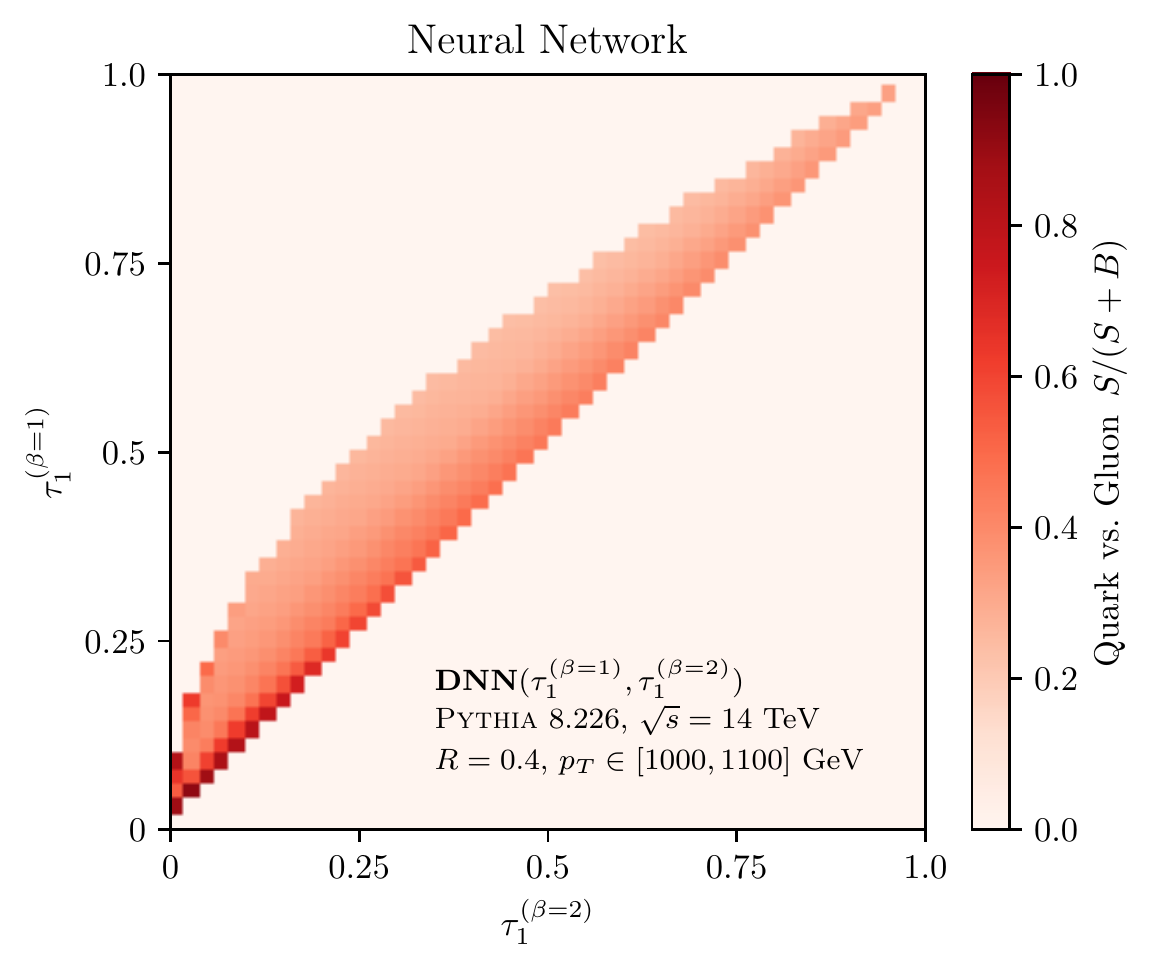}
	\label{fig:t1t2nn}
	}
\caption{\label{fig:tau1tau1ps} $S/(S+B)$ for quarks ($S$) vs.~gluons ($B$) in the $(\tau_1^{(\beta=1)},\tau_1^{(\beta=2)})$ phase space, determined by (a)  the prediction of \Eq{eq:two1subLR} using $\alpha_s=0.118$, (b) Monte Carlo histogram counts, and (c) the output of a neural network trained to classify quarks and gluons.
The prediction successfully captures the qualitative features of the neural network decision boundaries and correctly predicts its saturation around $C_F/(C_F+C_A)\simeq 0.308$.
}
\end{figure}

The output of a quark/gluon discrimination neural network that combines two $1$-subjettiness observables with different $\beta$ values is shown in \Fig{fig:tau1tau1ps}, compared with the theory prediction for $S/(S+B)$ and a Monte Carlo histogram estimate.
The values are shown only in the physical phase space of $\tau_1^{(\beta=2)} \le \tau_1^{(\beta=1)}$ and $(\tau_1^{(\beta=1)})^2 \le \tau_1^{(\beta=2)}$.
The neural network output is indeed well described by $S/(S+B)$, which can be verified based on its similarity to the binned histogram estimate of $S/(S+B)$.
Further, the network interpolates its output much more smoothly than the binned histogram estimate, which suffers from the curse of dimensionality.
The theory prediction captures the general scale of the neural network output and predicts is saturation around $C_F/(C_F+C_A)\simeq 0.308$.
Note that we use a fixed strong coupling constant for the prediction, where running coupling effects would provide an additional enhancement near the origin of phase space.

\begin{figure}[t]
\centering
\subfloat[]{
    \includegraphics[scale=0.65]{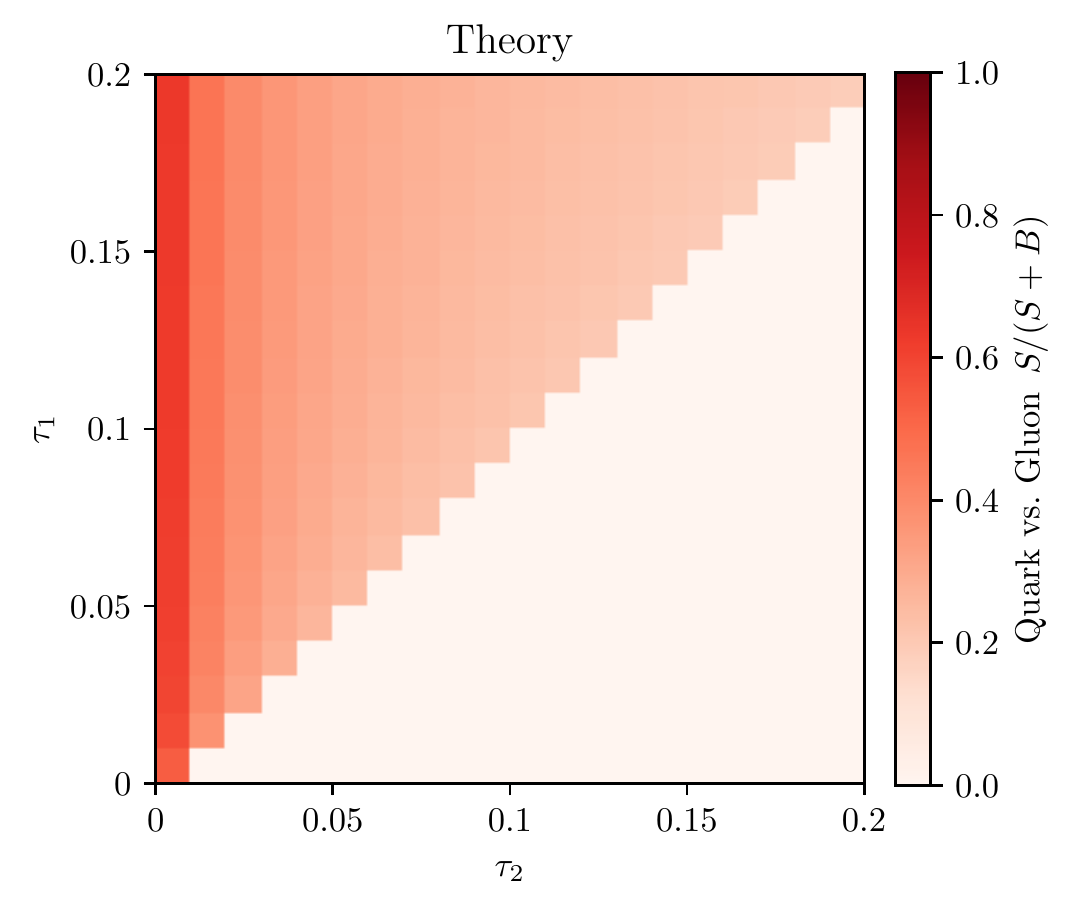}
	\label{fig:t1t2th}
	}
 \subfloat[]{
    \includegraphics[scale=0.65]{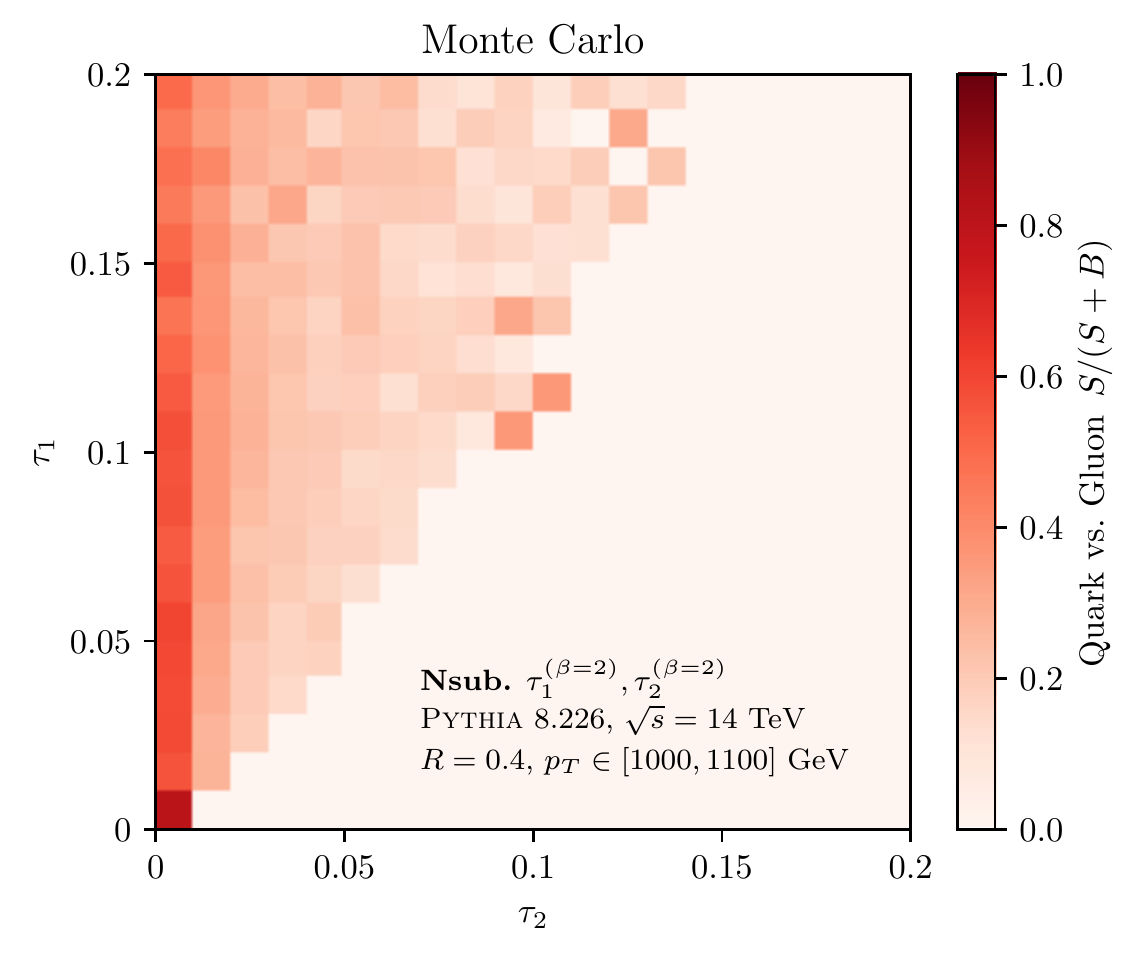}
	\label{fig:t1t2mc}
	}\\
 \subfloat[]{
    \includegraphics[scale=0.925]{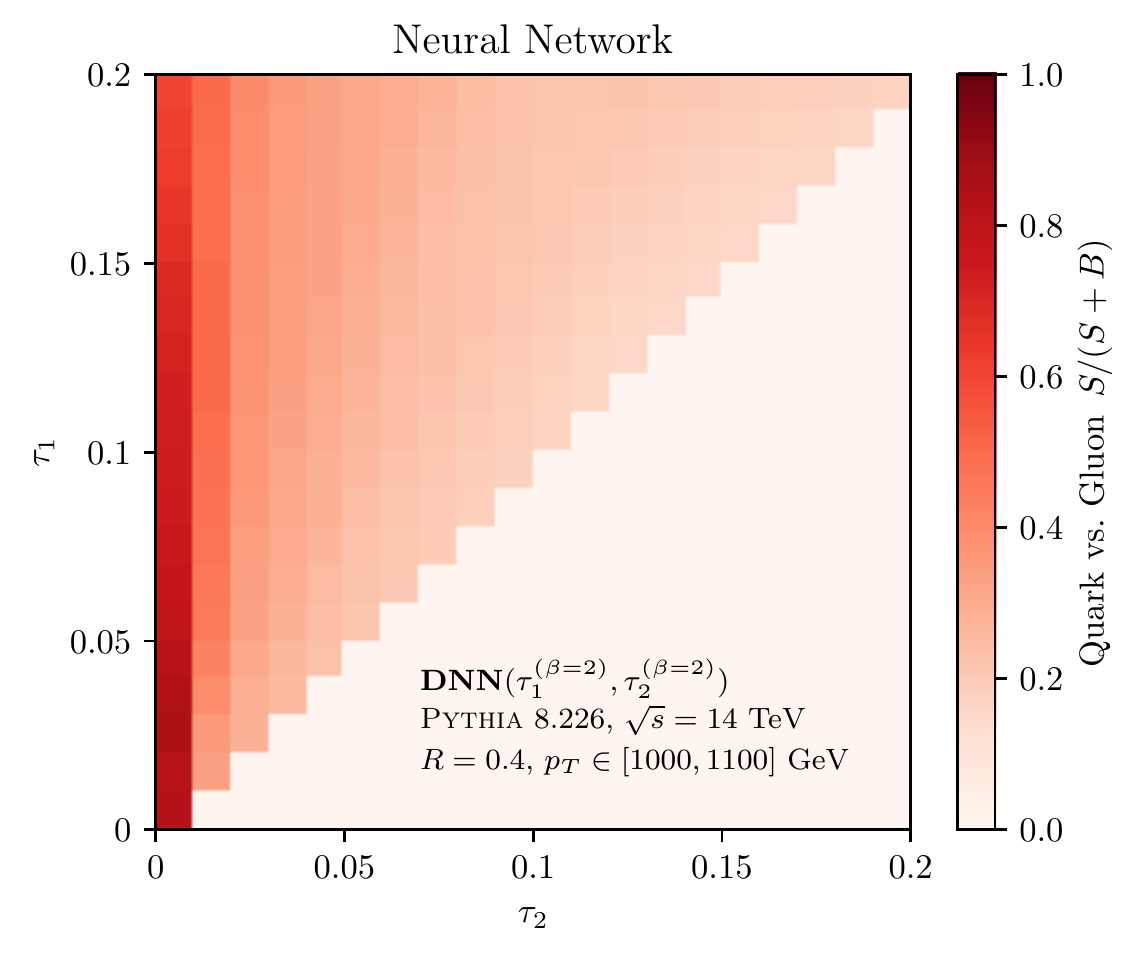}
	\label{fig:t1t2nn}
	}
\caption{\label{fig:tau1tau2ps} $S/(S+B)$ for quarks ($S$) vs.~gluons ($B$) in the $(\tau_1,\tau_2)$ phase space with $\beta=2$, determined by (a)  the prediction of \Eqs{eq:pqt12}{eq:pgt12} using $\alpha_s=0.118$, (b) Monte Carlo histogram counts, and (c) the output of a neural network trained to classify quarks and gluons.
The prediction successfully captures the qualitative features of the neural network decision boundaries and correctly predicts its saturation around $C_F^2/(C_F^2+C_A^2)\simeq 0.165$.
}
\end{figure}

This analysis can also be carried out for neural networks combining $N$-subjettiness observables that probe different numbers of emissions.
The output of a neural network that combines 1-subjettiness and 2-subjettiness is shown in \Fig{fig:tau1tau2ps}, compared with the corresponding theory prediction for $S/(S+B)$ and a Monte Carlo histogram estimate.
The values are shown only in the physical phase space of $\tau_2 \le \tau_1$.
Again, the neural network is well-described by $S/(S+B)$ and interpolates better than the binned histogram estimate.
The theory prediction provides a good description of the neural network decision boundaries and its saturation around
\begin{equation}
\frac{1}{1+\kappa_g(\tau_1,\tau_2)} = \frac{C_F^2}{C_F^2 + C_A^2}\simeq 0.165\,.
\end{equation}
%
%
More broadly, while neural networks themselves are black-box function approximators, the understanding of the optimum as $S/(S+B)$ yields a way to theoretically probe the decisions made by the model and to provide robust limits on its performance.

\begin{figure}[t]
\centering
\includegraphics[scale=0.9]{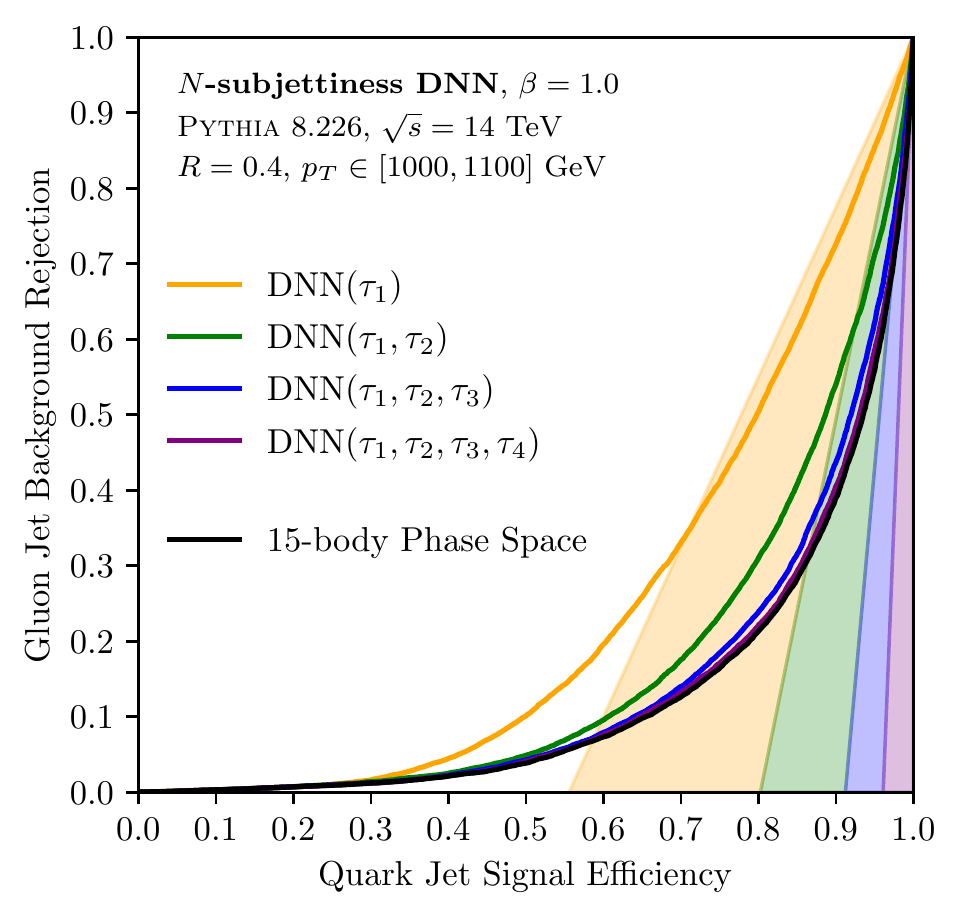}
\caption{\label{eq:DNNbeta1}
The ROC curves for neural networks combining $N$-subjettiness observables with $\beta=1$, together with the predictions for their asymptotic behavior in the high quark efficiency region based on $
\kappa_g = (C_F/C_A)^N$.
The classification performance increases as $N$-subjettiness observables are added, saturating at the performance of using $15$-body phase space.
The asymptotic classification performance is qualitatively well described by the analytical estimates via reducibility factors for $N$-emission sensitive observables.
}
\end{figure}

The ROC curves obtained by combining $N$-subjettiness values with a neural network are shown in \Fig{eq:DNNbeta1} through $N=4$ for $\beta=1$.
For comparison, we also show the results using 15-body phase space, namely all $N$-subjettiness values with $\beta\in\{0.5,1.0,2.0\}$ and $N$ up to 15.
This has been established to achieve competitive quark vs.~gluon classification performance with other machine learning methods, and so provides us with a proxy for absolute convergence of the ROC curve.
We see that combining a non-trivial number of $N$-subjettiness observables with a neural network indeed achieves comparable performance to general machine learning techniques.
Further, we see that the asymptotic classification performance at high quark efficiencies is indeed well-described by the bound based on our calculation of $\kappa_g = (C_F/C_A)^N$ for this feature space.

\begin{figure}[t]
\centering
\subfloat[]{
    \includegraphics[scale=0.675]{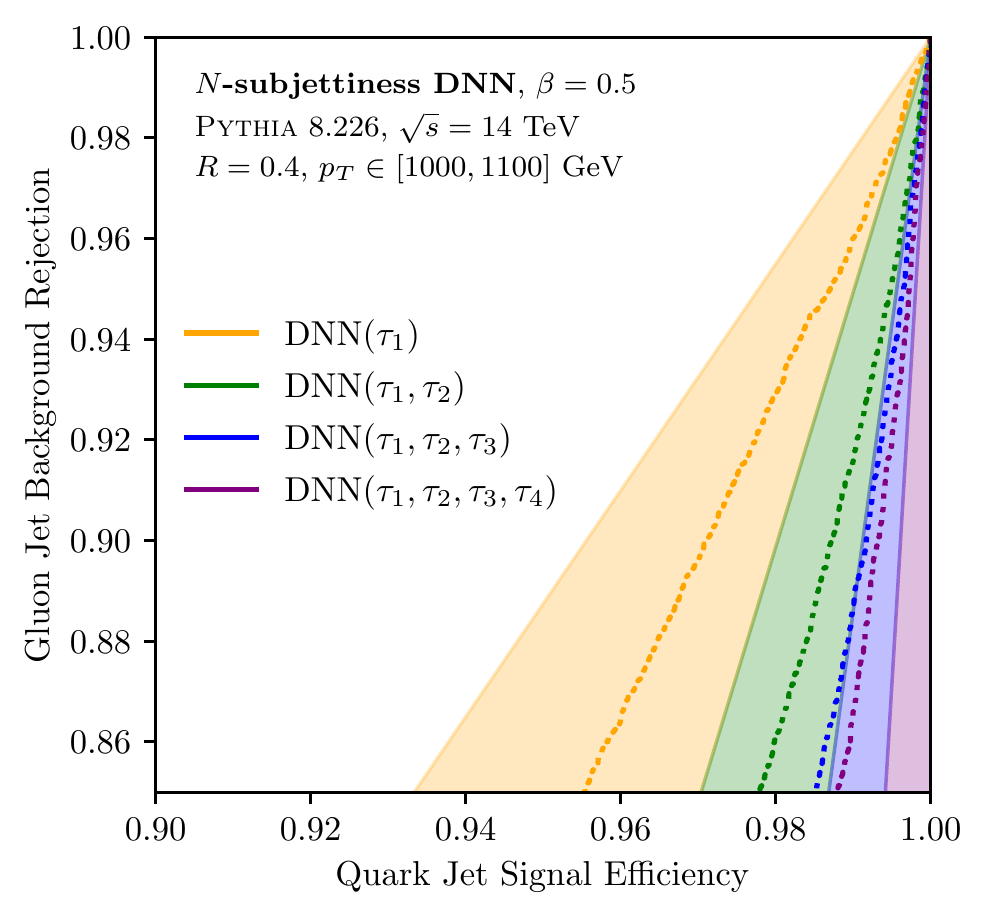}
	\label{fig:dnnb05}
	}
 \subfloat[]{
    \includegraphics[scale=0.675]{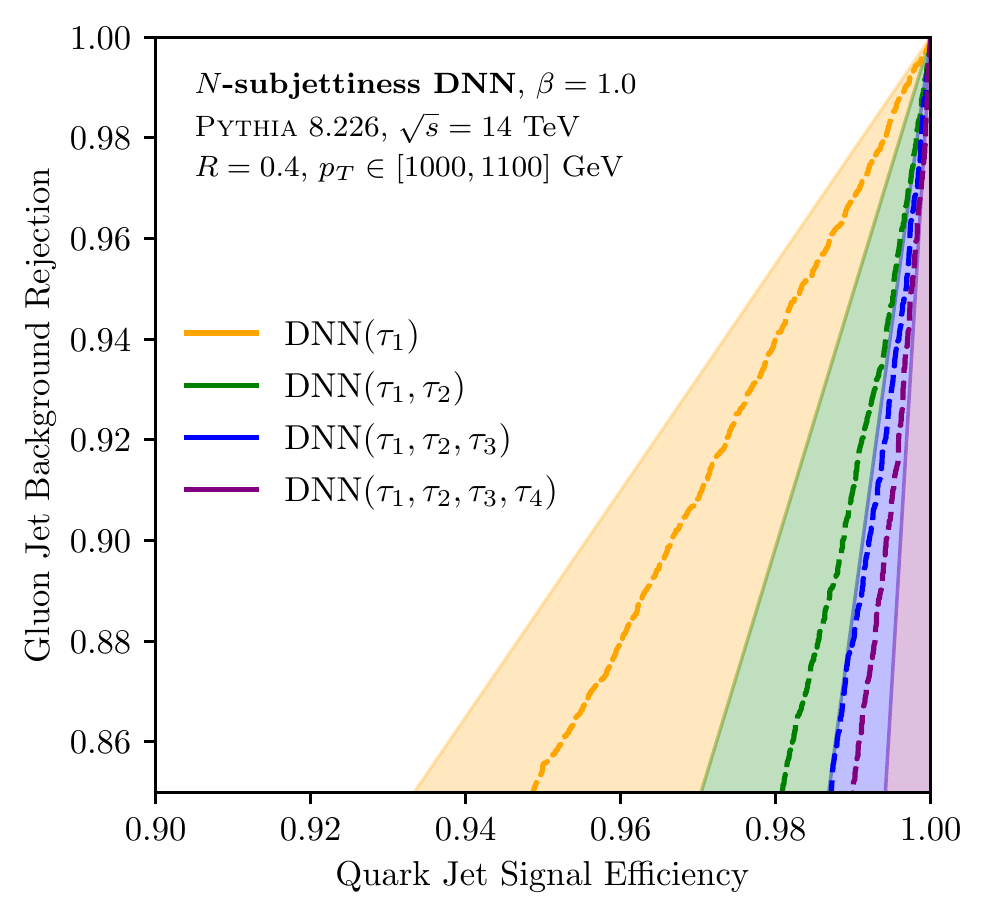}
	\label{fig:dnn10}
	}\\
 \subfloat[]{
    \includegraphics[scale=0.925]{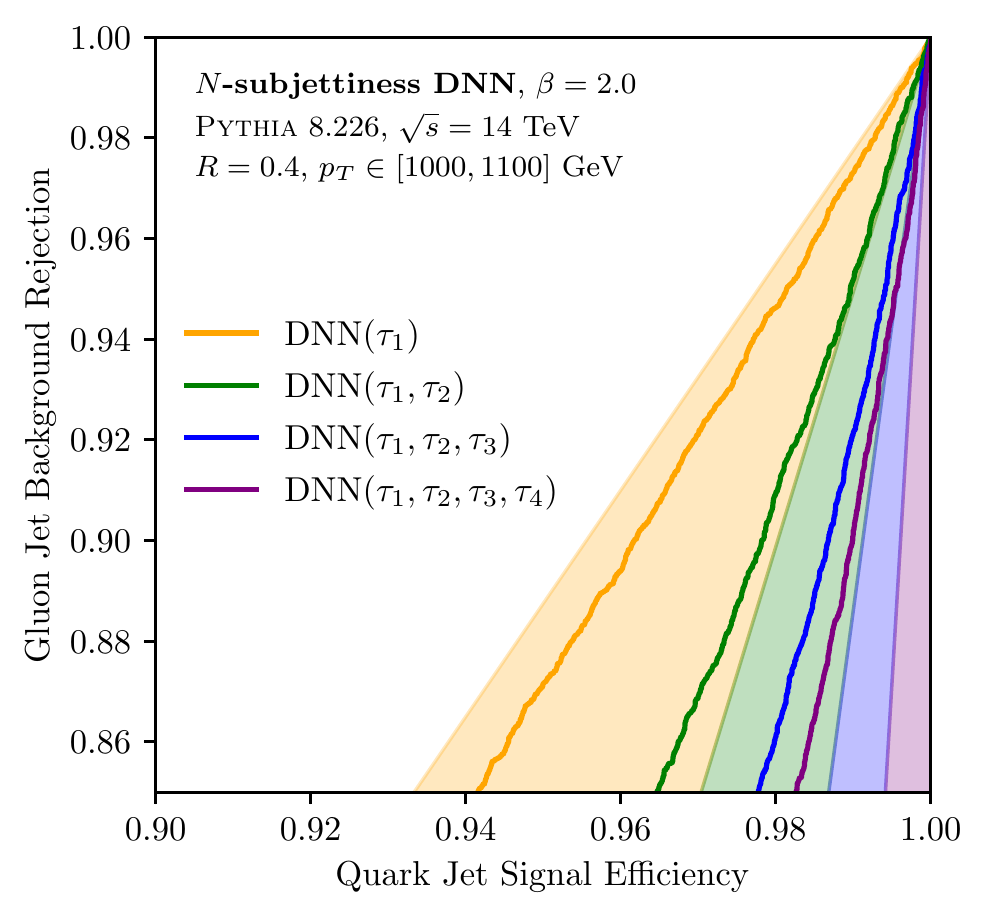}
	\label{fig:dnn20}
	}
\caption{\label{fig:dnnend} 
ROC curves for neural networks trained on collections of multiple $N$-subjettiness observables for (a) $\beta=0.5$, (b) $\beta=1.0$, and (c) $\beta=2.0$, focused on the high quark efficiency region.
The theoretical ROC bounds from $\kappa_g = (C_F/C_A)^N$ for those collections of observables are shown as the shaded regions.
We see that the bounds are indeed near saturated for $\beta=2.0$ with the best perturbative control.
Higher order effects increase the performance of lower $\beta$ values, though the hierarchy of performance remains.
}
\end{figure}

To probe our understanding of the parametric ROC curve performance, \Fig{fig:dnnend} shows the high quark efficiency region of $N$-subjettiness observables combined with neural networks for $\beta\in\{0.5,1.0,2.0\}$ and $N$ through 4.
For $\beta=2.0$ where we have the highest perturbative control, we see that indeed the parametric performance of the neural network is relatively well governed by these limits.
For smaller values of $\beta$, higher order effects become more important and classification performance is increased, but the relative hierarchy remains consistent.
Hence our understanding based solely on the quark- and gluon-enriched regions of phase space using our power counting rules has begun to provide a good qualitative and semi-quantitative understanding of neural network performance in high dimensions.

\section{Conclusions}\label{sec:conc}

The identification of the initiating particle of a jet and the discrimination of jets of different origins are central problems in the analysis of events at the LHC.
Due both to the importance of the problem and the abundance of data from the LHC, machine learning with DNNs, for example, has seen extensive use.
However, in most studies, the inputs to the DNN are low-level information such as individual particle four-momenta and so the dimensionality of the input can be tens or hundreds of numbers.
This enormous dimension is difficult to quantify and requires reliance on the DNN to tease out the important features.
Further, studies thus far have used simulation to train the models, which is not reality, and this risks learning the idiosyncrasies of the simulation, and not real physics.
Recent ideas for training directly on the data~\cite{Dery:2017fap,Cohen:2017exh,Metodiev:2017vrx,Komiske:2018oaa} are closely related to the notions of power counting and parametric discrimination power developed here~\cite{Metodiev:2018ftz,Komiske:2018vkc}.
More generally, in order to trust the output of the model and identify the relevant physics that drives the discrimination power, first-principles theoretical calculations that parallel the machine as best as possible are necessary.

Here, we performed these calculations to double-logarithmic accuracy within the context of quark vs.~gluon discrimination.
We explicitly considered the measurement of the IRC-safe and additive $N$-subjettiness observables to resolve emissions, which enable straightforward resummation and a sufficient number of them measured on a jet is in one-to-one correspondence with $M$-body phase space.
For a binary discrimination problem, a machine outputs an estimate of the likelihood ratio as a function of the training data and the classification performance can be quantified via the AUC.
With our predicted resummed probability distributions, the likelihood is just the ratio of signal and background distributions and the AUC is calculated through an ordered integral over the distributions.
Further, limits of the likelihood quantify the achievable sample purity through reducibility factors.
This has a close relationship to power counting and enables the identification of powerful discrimination observables without the necessity of a detailed calculation.
This established power counting method and our explicit calculations demonstrate that sensitivity to a large number of emissions in the jet produces a good quark/gluon discriminant and that, surprisingly, the likelihood is itself an IRC safe observable.
These predictions are exhibited in Monte Carlo parton shower simulations, providing an understanding of what a machine trained on simulation is learning.

This is a first step in a theoretical effort to deconstruct machine learning for particle physics.
This new field is becoming increasingly sophisticated and performance metrics are more well-established, providing concrete goals for theoretical studies.
Binary classification, like the case studied here, is an old problem within the field of jet substructure.
However, signal and background are not necessarily so well-defined, and so more general problems include multi-label classification in which a given sample is divided into more than two categories. 
In searching for new physics signals, the problem of anomaly detection or anti-tagging is relevant, in which deviations from a fiducial distribution (that predicted by the Standard Model), are of interest.
These problems are just now being studied from the machine learning angle \cite{Conway:2016caq,Aguilar-Saavedra:2017rzt,Collins:2018epr,Hajer:2018kqm,Heimel:2018mkt,Farina:2018fyg,Collins:2019jip,Roy:2019jae}, and theoretical efforts are necessary to identify the individual observables, techniques, and signatures that are most sensitive to the goals.

Establishing uncertainties and demonstrating robustness from machine learning is challenging due to the high-dimensionality of the inputs.
However, even in a simplified, but theoretically well-defined, approximation, if individual observables can be identified that perform comparable to the output of a DNN they are preferred.
The definition of such an observable would not rely on the details of Monte Carlo parton shower modeling and the physics of its performance would be well-understood. 
Such efforts work toward the goal of opening up the black box and shining a new light on the physics of jets.

\acknowledgments

The authors would like to thank Jesse Thaler for helpful conversations and detailed criticisms of the manuscript, as well as Patrick Komiske and Ben Nachman for comments on the manuscript.
The work of EMM is supported by the Office of Nuclear Physics of the U.S. Department of Energy (DOE) under grant DE-SC-0011090 and the DOE Office of High Energy Physics under grant DE-SC-0012567.
EMM benefited from the hospitality of the Harvard Center for the Fundamental Laws of Nature.

\appendix

\section{Up vs.~Down Quark Classification}

Our theoretical tools and results for quark vs.~gluon classification can be directly translated to a number of other interesting collider physics problems.
\Sec{sec:Zboson} discussed the calculations and implications for hadronically-decaying boosted $Z$ boson discrimination.
As another explicit demonstration, we consider up vs.~down quark classification using the photon radiation pattern within the jet.

Probing the electric charge of a jet, i.e.~discriminating jets initiated by up-type quarks from those initiated by down-type quarks, has been an ambitious and interesting goal of great theoretical~\cite{Field:1977fa,Krohn:2012fg,Waalewijn:2012sv} and experimental~\cite{Albanese:1984nv,Decamp:1991se,Nachman:2014qma,Aad:2015cua,Sirunyan:2017tyr} interest for many decades.
Recent work has also used machine learning to attack the problem~\cite{Fraser:2018ieu}.
Most strategies make use of manifestly infrared- and collinear-unsafe information, such as the energy-weighted charges of the constituents of the jet, making theoretical understanding more challenging.
Here, we will study this problem restricted to perturbatively accessible information: the radiation pattern of emitted photons, which has previously been used to disentangle up-type quark from down-type quark contributions to the $Z$ width~\cite{Mattig:1990wp,Abbiendi:2003ke}.
While we focus on up and down quarks, the lessons apply more broadly to all light up-type and down-type quarks and anti-quarks.

The principal difference between up and down quarks is their electric charge, $Q_u=+2/3$ and $Q_d=-1/3$.
The singular piece of the probability for a quark to radiate a photon at angle $\theta$ and energy fraction $z$ is:
\begin{equation}
dP_{q\to q\gamma} = \frac{\alpha_e Q_q^2}{2\pi} \frac{d\theta}{\theta} \frac{dz}{z},
\end{equation}
where $Q_q$ is the electric charge of the quark and $\alpha_e$ is the electromagnetic coupling constant.

Already, we can see that this problem mirrors the case of classifying quark vs.~gluon jets using their gluon radiation patterns, for which the relevant differences are the color factors.
Due to their parallel soft and collinear singularity structures, we can lift our quark vs.~gluon results to the up vs.~down quark case by the replacement $C_F\to Q_d^2$ and $C_A\to Q_u^2$.
Since $Q_u^2/Q_d^2=4$ whereas $C_A/C_F=9/4=2.25$, each perturbative photon emission will be significantly more valuable for distinguishing up and down quarks than a gluon emission in the analogous quark vs.~gluon classification case.

Observables $\tau_1$ which probe a single photon emission in the jet will, analogously to Casimir scaling, have cumulative distributions which scale as:
\begin{equation}
\Sigma_u(\tau_1) = \left(\Sigma_d(\tau_1)\right)^{Q_u^2/Q_d^2}.
\end{equation}
%
%
The up and down reducibility factors for such observables can then be computed to be:
\begin{equation}
\kappa_d = 0,\quad\quad \kappa_u =\frac{Q_d^2}{Q_u^2} = \frac14.
\end{equation}
For observables $\tau_1,\cdots,\tau_n$ probing up to $n$ photon emissions, the up vs.~down reducibility factors for the multi-differential phase space are:
\begin{equation}
\kappa_d = 0,\quad\quad \kappa_u =\left(\frac{Q_d^2}{Q_u^2}\right)^n = \frac{1}{2^{2n}}.
\end{equation}

%
%

Hence up and down quarks are only mutually irreducible in their photon radiation pattern in the limit of probing many emissions.
For instance, a selection of jets with an energetic photon will necessarily be contaminated by down quarks by a relative amount $Q_d^2/Q_u^2$.
In practice, one can probe the electromagnetic aspect of quark jet physics using isolated photon subjets, as studied in detail in \Ref{Hall:2018jub}.
There are several experimental complications that we do not consider here, such as backgrounds from $\pi^0\to\gamma\gamma$, that would limit the sensitivity to perturbative photon emissions and hence further degrade classification performance.
Even so, using our results we are able to obtain a theoretical understanding of and determine strict limits on the up vs.~down quark discrimination performance based on the photon radiation pattern.
Theoretical investigation of these ideas is important to extend operational jet (and event) flavor definitions~\cite{Metodiev:2018ftz,Komiske:2018vkc} beyond solely ``quark'' and ``gluon'' categories.

\bibliography{qg_bib}

\providecommand{\href}[2]{#2}\begingroup\raggedright\begin{thebibliography}{10}

\bibitem{Nilles:1980ys}
H.~P. Nilles and K.~H. Streng, {\it {Quark - Gluon Separation in Three Jet
  Events}},  {\em Phys. Rev.} {\bf D23} (1981) 1944.

\bibitem{Jones:1988ay}
L.~M. Jones, {\it {Tests for Determining the Parton Ancestor of a Hadron Jet}},
   {\em Phys. Rev.} {\bf D39} (1989) 2550.

\bibitem{Fodor:1989ir}
Z.~Fodor, {\it {How to See the Differences Between Quark and Gluon Jets}},
  {\em Phys. Rev.} {\bf D41} (1990) 1726.

\bibitem{Jones:1990rz}
L.~Jones, {\it {Towards a systematic jet classification}},  {\em Phys. Rev.}
  {\bf D42} (1990) 811--814.

\bibitem{Pumplin:1991kc}
J.~Pumplin, {\it {How to tell quark jets from gluon jets}},  {\em Phys. Rev.}
  {\bf D44} (1991) 2025--2032.

\bibitem{Gallicchio:2011xc}
J.~Gallicchio and M.~D. Schwartz, {\it {Pure Samples of Quark and Gluon Jets at
  the LHC}},  {\em JHEP} {\bf 10} (2011) 103,
  [\href{http://arxiv.org/abs/1104.1175}{{\tt arXiv:1104.1175}}].

\bibitem{Gallicchio:2011xq}
J.~Gallicchio and M.~D. Schwartz, {\it {Quark and Gluon Tagging at the LHC}},
  {\em Phys. Rev. Lett.} {\bf 107} (2011) 172001,
  [\href{http://arxiv.org/abs/1106.3076}{{\tt arXiv:1106.3076}}].

\bibitem{Gallicchio:2012ez}
J.~Gallicchio and M.~D. Schwartz, {\it {Quark and Gluon Jet Substructure}},
  {\em JHEP} {\bf 04} (2013) 090, [\href{http://arxiv.org/abs/1211.7038}{{\tt
  arXiv:1211.7038}}].

\bibitem{FerreiradeLima:2016gcz}
D.~Ferreira~de Lima, P.~Petrov, D.~Soper, and M.~Spannowsky, {\it {Quark-Gluon
  tagging with Shower Deconstruction: Unearthing dark matter and Higgs
  couplings}},  {\em Phys. Rev.} {\bf D95} (2017), no.~3 034001,
  [\href{http://arxiv.org/abs/1607.06031}{{\tt arXiv:1607.06031}}].

\bibitem{Frye:2017yrw}
C.~Frye, A.~J. Larkoski, J.~Thaler, and K.~Zhou, {\it {Casimir Meets Poisson:
  Improved Quark/Gluon Discrimination with Counting Observables}},  {\em JHEP}
  {\bf 09} (2017) 083, [\href{http://arxiv.org/abs/1704.06266}{{\tt
  arXiv:1704.06266}}].

\bibitem{Davighi:2017hok}
J.~Davighi and P.~Harris, {\it {Fractal based observables to probe jet
  substructure of quarks and gluons}},  {\em Eur. Phys. J.} {\bf C78} (2018),
  no.~4 334, [\href{http://arxiv.org/abs/1703.00914}{{\tt arXiv:1703.00914}}].

\bibitem{Komiske:2018cqr}
P.~T. Komiske, E.~M. Metodiev, and J.~Thaler, {\it {Energy Flow Networks: Deep
  Sets for Particle Jets}},  {\em JHEP} {\bf 01} (2019) 121,
  [\href{http://arxiv.org/abs/1810.05165}{{\tt arXiv:1810.05165}}].

\bibitem{Banfi:2006hf}
A.~Banfi, G.~P. Salam, and G.~Zanderighi, {\it {Infrared safe definition of jet
  flavor}},  {\em Eur. Phys. J.} {\bf C47} (2006) 113--124,
  [\href{http://arxiv.org/abs/hep-ph/0601139}{{\tt hep-ph/0601139}}].

\bibitem{Frye:2016aiz}
C.~Frye, A.~J. Larkoski, M.~D. Schwartz, and K.~Yan, {\it {Factorization for
  groomed jet substructure beyond the next-to-leading logarithm}},  {\em JHEP}
  {\bf 07} (2016) 064, [\href{http://arxiv.org/abs/1603.09338}{{\tt
  arXiv:1603.09338}}].

\bibitem{Gras:2017jty}
P.~Gras, S.~H{\"o}che, D.~Kar, A.~Larkoski, L.~L{\"o}nnblad, S.~Pl{\"a}tzer,
  A.~Si{\'o}dmok, P.~Skands, G.~Soyez, and J.~Thaler, {\it {Systematics of
  quark/gluon tagging}},  {\em JHEP} {\bf 07} (2017) 091,
  [\href{http://arxiv.org/abs/1704.03878}{{\tt arXiv:1704.03878}}].

\bibitem{Metodiev:2018ftz}
E.~M. Metodiev and J.~Thaler, {\it {Jet Topics: Disentangling Quarks and Gluons
  at Colliders}},  {\em Phys. Rev. Lett.} {\bf 120} (2018), no.~24 241602,
  [\href{http://arxiv.org/abs/1802.00008}{{\tt arXiv:1802.00008}}].

\bibitem{Komiske:2018vkc}
P.~T. Komiske, E.~M. Metodiev, and J.~Thaler, {\it {An operational definition
  of quark and gluon jets}},  {\em JHEP} {\bf 11} (2018) 059,
  [\href{http://arxiv.org/abs/1809.01140}{{\tt arXiv:1809.01140}}].

\bibitem{Larkoski:2014pca}
A.~J. Larkoski, J.~Thaler, and W.~J. Waalewijn, {\it {Gaining (Mutual)
  Information about Quark/Gluon Discrimination}},  {\em JHEP} {\bf 11} (2014)
  129, [\href{http://arxiv.org/abs/1408.3122}{{\tt arXiv:1408.3122}}].

\bibitem{Bhattacherjee:2015psa}
B.~Bhattacherjee, S.~Mukhopadhyay, M.~M. Nojiri, Y.~Sakaki, and B.~R. Webber,
  {\it {Associated jet and subjet rates in light-quark and gluon jet
  discrimination}},  {\em JHEP} {\bf 04} (2015) 131,
  [\href{http://arxiv.org/abs/1501.04794}{{\tt arXiv:1501.04794}}].

\bibitem{Mo:2017gzp}
J.~Mo, F.~J. Tackmann, and W.~J. Waalewijn, {\it {A case study of quark-gluon
  discrimination at NNLL' in comparison to parton showers}},  {\em Eur. Phys.
  J.} {\bf C77} (2017), no.~11 770,
  [\href{http://arxiv.org/abs/1708.00867}{{\tt arXiv:1708.00867}}].

\bibitem{Sakaki:2018opq}
Y.~Sakaki, {\it {Quark jet rates and quark/gluon discrimination in multi-jet
  final states}},  \href{http://arxiv.org/abs/1807.01421}{{\tt
  arXiv:1807.01421}}.

\bibitem{Lonnblad:1990qp}
L.~Lonnblad, C.~Peterson, and T.~Rognvaldsson, {\it {Using neural networks to
  identify jets}},  {\em Nucl. Phys.} {\bf B349} (1991) 675--702.

\bibitem{Komiske:2016rsd}
P.~T. Komiske, E.~M. Metodiev, and M.~D. Schwartz, {\it {Deep learning in
  color: towards automated quark/gluon jet discrimination}},  {\em JHEP} {\bf
  01} (2017) 110, [\href{http://arxiv.org/abs/1612.01551}{{\tt
  arXiv:1612.01551}}].

\bibitem{Cheng:2017rdo}
T.~Cheng, {\it {Recursive Neural Networks in Quark/Gluon Tagging}},  {\em
  Comput. Softw. Big Sci.} {\bf 2} (2018), no.~1 3,
  [\href{http://arxiv.org/abs/1711.02633}{{\tt arXiv:1711.02633}}].

\bibitem{Luo:2017ncs}
H.~Luo, M.-x. Luo, K.~Wang, T.~Xu, and G.~Zhu, {\it {Quark jet versus gluon
  jet: deep neural networks with high-level features}},
  \href{http://arxiv.org/abs/1712.03634}{{\tt arXiv:1712.03634}}.

\bibitem{Kasieczka:2018lwf}
G.~Kasieczka, N.~Kiefer, T.~Plehn, and J.~M. Thompson, {\it {Quark-Gluon
  Tagging: Machine Learning meets Reality}},
  \href{http://arxiv.org/abs/1812.09223}{{\tt arXiv:1812.09223}}.

\bibitem{Berger:2003iw}
C.~F. Berger, T.~Kucs, and G.~F. Sterman, {\it {Event shape / energy flow
  correlations}},  {\em Phys. Rev.} {\bf D68} (2003) 014012,
  [\href{http://arxiv.org/abs/hep-ph/0303051}{{\tt hep-ph/0303051}}].

\bibitem{Almeida:2008yp}
L.~G. Almeida, S.~J. Lee, G.~Perez, G.~F. Sterman, I.~Sung, and J.~Virzi, {\it
  {Substructure of high-$p_T$ Jets at the LHC}},  {\em Phys. Rev.} {\bf D79}
  (2009) 074017, [\href{http://arxiv.org/abs/0807.0234}{{\tt
  arXiv:0807.0234}}].

\bibitem{Ellis:2010rwa}
S.~D. Ellis, C.~K. Vermilion, J.~R. Walsh, A.~Hornig, and C.~Lee, {\it {Jet
  Shapes and Jet Algorithms in SCET}},  {\em JHEP} {\bf 11} (2010) 101,
  [\href{http://arxiv.org/abs/1001.0014}{{\tt arXiv:1001.0014}}].

\bibitem{Larkoski:2014wba}
A.~J. Larkoski, S.~Marzani, G.~Soyez, and J.~Thaler, {\it {Soft Drop}},  {\em
  JHEP} {\bf 05} (2014) 146, [\href{http://arxiv.org/abs/1402.2657}{{\tt
  arXiv:1402.2657}}].

\bibitem{Larkoski:2014gra}
A.~J. Larkoski, I.~Moult, and D.~Neill, {\it {Power Counting to Better Jet
  Observables}},  {\em JHEP} {\bf 12} (2014) 009,
  [\href{http://arxiv.org/abs/1409.6298}{{\tt arXiv:1409.6298}}].

\bibitem{Larkoski:2014zma}
A.~J. Larkoski, I.~Moult, and D.~Neill, {\it {Building a Better Boosted Top
  Tagger}},  {\em Phys. Rev.} {\bf D91} (2015), no.~3 034035,
  [\href{http://arxiv.org/abs/1411.0665}{{\tt arXiv:1411.0665}}].

\bibitem{Moult:2016cvt}
I.~Moult, L.~Necib, and J.~Thaler, {\it {New Angles on Energy Correlation
  Functions}},  {\em JHEP} {\bf 12} (2016) 153,
  [\href{http://arxiv.org/abs/1609.07483}{{\tt arXiv:1609.07483}}].

\bibitem{Neyman289}
J.~Neyman and E.~S. Pearson, {\it Ix. on the problem of the most efficient
  tests of statistical hypotheses},  {\em Philosophical Transactions of the
  Royal Society of London A: Mathematical, Physical and Engineering Sciences}
  {\bf 231} (1933), no.~694-706 289--337.

\bibitem{Stewart:2010tn}
I.~W. Stewart, F.~J. Tackmann, and W.~J. Waalewijn, {\it {N-Jettiness: An
  Inclusive Event Shape to Veto Jets}},  {\em Phys. Rev. Lett.} {\bf 105}
  (2010) 092002, [\href{http://arxiv.org/abs/1004.2489}{{\tt
  arXiv:1004.2489}}].

\bibitem{Thaler:2010tr}
J.~Thaler and K.~Van~Tilburg, {\it {Identifying Boosted Objects with
  N-subjettiness}},  {\em JHEP} {\bf 03} (2011) 015,
  [\href{http://arxiv.org/abs/1011.2268}{{\tt arXiv:1011.2268}}].

\bibitem{Thaler:2011gf}
J.~Thaler and K.~Van~Tilburg, {\it {Maximizing Boosted Top Identification by
  Minimizing N-subjettiness}},  {\em JHEP} {\bf 02} (2012) 093,
  [\href{http://arxiv.org/abs/1108.2701}{{\tt arXiv:1108.2701}}].

\bibitem{Tkachov:1994as}
F.~V. Tkachov, {\it {Measuring the number of hadronic jets}},  {\em Phys. Rev.
  Lett.} {\bf 73} (1994) 2405--2408,
  [\href{http://arxiv.org/abs/hep-ph/9901332}{{\tt hep-ph/9901332}}].

\bibitem{Tkachov:1995kk}
F.~V. Tkachov, {\it {Measuring multi - jet structure of hadronic energy flow or
  What is a jet?}},  {\em Int. J. Mod. Phys.} {\bf A12} (1997) 5411--5529,
  [\href{http://arxiv.org/abs/hep-ph/9601308}{{\tt hep-ph/9601308}}].

\bibitem{Larkoski:2013eya}
A.~J. Larkoski, G.~P. Salam, and J.~Thaler, {\it {Energy Correlation Functions
  for Jet Substructure}},  {\em JHEP} {\bf 06} (2013) 108,
  [\href{http://arxiv.org/abs/1305.0007}{{\tt arXiv:1305.0007}}].

\bibitem{Komiske:2017aww}
P.~T. Komiske, E.~M. Metodiev, and J.~Thaler, {\it {Energy flow polynomials: A
  complete linear basis for jet substructure}},  {\em JHEP} {\bf 04} (2018)
  013, [\href{http://arxiv.org/abs/1712.07124}{{\tt arXiv:1712.07124}}].

\bibitem{Andersson:1988gp}
B.~Andersson, G.~Gustafson, L.~Lonnblad, and U.~Pettersson, {\it {Coherence
  Effects in Deep Inelastic Scattering}},  {\em Z. Phys.} {\bf C43} (1989) 625.

\bibitem{Datta:2017rhs}
K.~Datta and A.~Larkoski, {\it {How Much Information is in a Jet?}},  {\em
  JHEP} {\bf 06} (2017) 073, [\href{http://arxiv.org/abs/1704.08249}{{\tt
  arXiv:1704.08249}}].

\bibitem{Aad:2015rpa}
{\bf ATLAS} Collaboration, G.~Aad et~al., {\it {Identification of boosted,
  hadronically decaying W bosons and comparisons with ATLAS data taken at
  $\sqrt{s} = 8$ TeV}},  {\em Eur. Phys. J.} {\bf C76} (2016), no.~3 154,
  [\href{http://arxiv.org/abs/1510.05821}{{\tt arXiv:1510.05821}}].

\bibitem{Aaboud:2016qgg}
{\bf ATLAS} Collaboration, M.~Aaboud et~al., {\it {Search for dark matter
  produced in association with a hadronically decaying vector boson in $pp$
  collisions at $\sqrt{s} =$ 13 TeV with the ATLAS detector}},  {\em Phys.
  Lett.} {\bf B763} (2016) 251--268,
  [\href{http://arxiv.org/abs/1608.02372}{{\tt arXiv:1608.02372}}].

\bibitem{Aaboud:2018psm}
{\bf ATLAS} Collaboration, M.~Aaboud et~al., {\it {Performance of top-quark and
  $W$-boson tagging with ATLAS in Run 2 of the LHC}},  {\em Eur. Phys. J.} {\bf
  C79} (2019), no.~5 375, [\href{http://arxiv.org/abs/1808.07858}{{\tt
  arXiv:1808.07858}}].

\bibitem{Aaboud:2019aii}
{\bf ATLAS} Collaboration, M.~Aaboud et~al., {\it {Measurement of
  jet-substructure observables in top quark, $W$ boson and light jet production
  in proton-proton collisions at $\sqrt{s}=13$ TeV with the ATLAS detector}},
  {\em Submitted to: JHEP} (2019) [\href{http://arxiv.org/abs/1903.02942}{{\tt
  arXiv:1903.02942}}].

\bibitem{Dasgupta:2015lxh}
M.~Dasgupta, L.~Schunk, and G.~Soyez, {\it {Jet shapes for boosted jet
  two-prong decays from first-principles}},  {\em JHEP} {\bf 04} (2016) 166,
  [\href{http://arxiv.org/abs/1512.00516}{{\tt arXiv:1512.00516}}].

\bibitem{Ellis:1991qj}
R.~K. Ellis, W.~J. Stirling, and B.~R. Webber, {\it {QCD and collider
  physics}},  {\em Camb. Monogr. Part. Phys. Nucl. Phys. Cosmol.} {\bf 8}
  (1996) 1--435.

\bibitem{Soyez:2012hv}
G.~Soyez, G.~P. Salam, J.~Kim, S.~Dutta, and M.~Cacciari, {\it {Pileup
  subtraction for jet shapes}},  {\em Phys. Rev. Lett.} {\bf 110} (2013),
  no.~16 162001, [\href{http://arxiv.org/abs/1211.2811}{{\tt
  arXiv:1211.2811}}].

\bibitem{Larkoski:2013paa}
A.~J. Larkoski and J.~Thaler, {\it {Unsafe but Calculable: Ratios of
  Angularities in Perturbative QCD}},  {\em JHEP} {\bf 09} (2013) 137,
  [\href{http://arxiv.org/abs/1307.1699}{{\tt arXiv:1307.1699}}].

\bibitem{Larkoski:2014tva}
A.~J. Larkoski, I.~Moult, and D.~Neill, {\it {Toward Multi-Differential Cross
  Sections: Measuring Two Angularities on a Single Jet}},  {\em JHEP} {\bf 09}
  (2014) 046, [\href{http://arxiv.org/abs/1401.4458}{{\tt arXiv:1401.4458}}].

\bibitem{Procura:2014cba}
M.~Procura, W.~J. Waalewijn, and L.~Zeune, {\it {Resummation of
  Double-Differential Cross Sections and Fully-Unintegrated Parton Distribution
  Functions}},  {\em JHEP} {\bf 02} (2015) 117,
  [\href{http://arxiv.org/abs/1410.6483}{{\tt arXiv:1410.6483}}].

\bibitem{Procura:2018zpn}
M.~Procura, W.~J. Waalewijn, and L.~Zeune, {\it {Joint resummation of two
  angularities at next-to-next-to-leading logarithmic order}},  {\em JHEP} {\bf
  10} (2018) 098, [\href{http://arxiv.org/abs/1806.10622}{{\tt
  arXiv:1806.10622}}].

\bibitem{Larkoski:2014uqa}
A.~J. Larkoski, D.~Neill, and J.~Thaler, {\it {Jet Shapes with the Broadening
  Axis}},  {\em JHEP} {\bf 04} (2014) 017,
  [\href{http://arxiv.org/abs/1401.2158}{{\tt arXiv:1401.2158}}].

\bibitem{Bertolini:2013iqa}
D.~Bertolini, T.~Chan, and J.~Thaler, {\it {Jet Observables Without Jet
  Algorithms}},  {\em JHEP} {\bf 04} (2014) 013,
  [\href{http://arxiv.org/abs/1310.7584}{{\tt arXiv:1310.7584}}].

\bibitem{salamunp}
G.~Salam, ``Unpublished.''

\bibitem{Aad:2014gea}
{\bf ATLAS} Collaboration, G.~Aad et~al., {\it {Light-quark and gluon jet
  discrimination in $pp$ collisions at $\sqrt{s}=7\mathrm {\ TeV}$ with the
  ATLAS detector}},  {\em Eur. Phys. J.} {\bf C74} (2014), no.~8 3023,
  [\href{http://arxiv.org/abs/1405.6583}{{\tt arXiv:1405.6583}}].

\bibitem{Salam:2016yht}
G.~P. Salam, L.~Schunk, and G.~Soyez, {\it {Dichroic subjettiness ratios to
  distinguish colour flows in boosted boson tagging}},  {\em JHEP} {\bf 03}
  (2017) 022, [\href{http://arxiv.org/abs/1612.03917}{{\tt arXiv:1612.03917}}].

\bibitem{Napoletano:2018ohv}
D.~Napoletano and G.~Soyez, {\it {Computing $N$-subjettiness for boosted
  jets}},  {\em JHEP} {\bf 12} (2018) 031,
  [\href{http://arxiv.org/abs/1809.04602}{{\tt arXiv:1809.04602}}].

\bibitem{Hahn:2004fe}
T.~Hahn, {\it {CUBA: A Library for multidimensional numerical integration}},
  {\em Comput. Phys. Commun.} {\bf 168} (2005) 78--95,
  [\href{http://arxiv.org/abs/hep-ph/0404043}{{\tt hep-ph/0404043}}].

\bibitem{Sjostrand:2006za}
T.~Sjostrand, S.~Mrenna, and P.~Z. Skands, {\it {PYTHIA 6.4 Physics and
  Manual}},  {\em JHEP} {\bf 05} (2006) 026,
  [\href{http://arxiv.org/abs/hep-ph/0603175}{{\tt hep-ph/0603175}}].

\bibitem{Sjostrand:2014zea}
T.~Sjostrand, S.~Ask, J.~R. Christiansen, R.~Corke, N.~Desai, P.~Ilten,
  S.~Mrenna, S.~Prestel, C.~O. Rasmussen, and P.~Z. Skands, {\it {An
  Introduction to PYTHIA 8.2}},  {\em Comput. Phys. Commun.} {\bf 191} (2015)
  159--177, [\href{http://arxiv.org/abs/1410.3012}{{\tt arXiv:1410.3012}}].

\bibitem{Cacciari:2008gp}
M.~Cacciari, G.~P. Salam, and G.~Soyez, {\it {The anti-$k_t$ jet clustering
  algorithm}},  {\em JHEP} {\bf 04} (2008) 063,
  [\href{http://arxiv.org/abs/0802.1189}{{\tt arXiv:0802.1189}}].

\bibitem{Cacciari:2011ma}
M.~Cacciari, G.~P. Salam, and G.~Soyez, {\it {FastJet User Manual}},  {\em Eur.
  Phys. J.} {\bf C72} (2012) 1896, [\href{http://arxiv.org/abs/1111.6097}{{\tt
  arXiv:1111.6097}}].

\bibitem{fjcontrib}
``Fastjet contrib.'' \url{https://fastjet.hepforge.org/contrib/}.

\bibitem{Datta:2017lxt}
K.~Datta and A.~J. Larkoski, {\it {Novel Jet Observables from Machine
  Learning}},  {\em JHEP} {\bf 03} (2018) 086,
  [\href{http://arxiv.org/abs/1710.01305}{{\tt arXiv:1710.01305}}].

\bibitem{Datta:2019ndh}
K.~Datta, A.~Larkoski, and B.~Nachman, {\it {Automating the Construction of Jet
  Observables with Machine Learning}},
  \href{http://arxiv.org/abs/1902.07180}{{\tt arXiv:1902.07180}}.

\bibitem{keras}
F.~Chollet, ``Keras.'' \url{https://github.com/fchollet/keras}, 2015.

\bibitem{tensorflow}
M.~Abadi, P.~Barham, J.~Chen, Z.~Chen, A.~Davis, J.~Dean, M.~Devin,
  S.~Ghemawat, G.~Irving, M.~Isard, et~al., {\it Tensorflow: A system for
  large-scale machine learning.},  in {\em OSDI}, vol.~16, pp.~265--283, 2016.

\bibitem{relu}
V.~Nair and G.~E. Hinton, {\it Rectified linear units improve restricted
  boltzmann machines},  in {\em Proceedings of the 27th international
  conference on machine learning (ICML-10)}, pp.~807--814, 2010.

\bibitem{heuniform}
K.~He, X.~Zhang, S.~Ren, and J.~Sun, {\it Delving deep into rectifiers:
  Surpassing human-level performance on imagenet classification},  in {\em
  Proceedings of the IEEE international conference on computer vision},
  pp.~1026--1034, 2015.

\bibitem{adam}
D.~Kingma and J.~Ba, {\it Adam: A method for stochastic optimization},
  \href{http://arxiv.org/abs/1412.6980}{{\tt arXiv:1412.6980}}.

\bibitem{Dery:2017fap}
L.~M. Dery, B.~Nachman, F.~Rubbo, and A.~Schwartzman, {\it {Weakly Supervised
  Classification in High Energy Physics}},  {\em JHEP} {\bf 05} (2017) 145,
  [\href{http://arxiv.org/abs/1702.00414}{{\tt arXiv:1702.00414}}].

\bibitem{Cohen:2017exh}
T.~Cohen, M.~Freytsis, and B.~Ostdiek, {\it {(Machine) Learning to Do More with
  Less}},  {\em JHEP} {\bf 02} (2018) 034,
  [\href{http://arxiv.org/abs/1706.09451}{{\tt arXiv:1706.09451}}].

\bibitem{Metodiev:2017vrx}
E.~M. Metodiev, B.~Nachman, and J.~Thaler, {\it {Classification without labels:
  Learning from mixed samples in high energy physics}},  {\em JHEP} {\bf 10}
  (2017) 174, [\href{http://arxiv.org/abs/1708.02949}{{\tt arXiv:1708.02949}}].

\bibitem{Komiske:2018oaa}
P.~T. Komiske, E.~M. Metodiev, B.~Nachman, and M.~D. Schwartz, {\it {Learning
  to classify from impure samples with high-dimensional data}},  {\em Phys.
  Rev.} {\bf D98} (2018), no.~1 011502,
  [\href{http://arxiv.org/abs/1801.10158}{{\tt arXiv:1801.10158}}].

\bibitem{Conway:2016caq}
J.~S. Conway, R.~Bhaskar, R.~D. Erbacher, and J.~Pilot, {\it {Identification of
  High-Momentum Top Quarks, Higgs Bosons, and W and Z Bosons Using Boosted
  Event Shapes}},  {\em Phys. Rev.} {\bf D94} (2016), no.~9 094027,
  [\href{http://arxiv.org/abs/1606.06859}{{\tt arXiv:1606.06859}}].

\bibitem{Aguilar-Saavedra:2017rzt}
J.~A. Aguilar-Saavedra, J.~H. Collins, and R.~K. Mishra, {\it {A generic
  anti-QCD jet tagger}},  {\em JHEP} {\bf 11} (2017) 163,
  [\href{http://arxiv.org/abs/1709.01087}{{\tt arXiv:1709.01087}}].

\bibitem{Collins:2018epr}
J.~H. Collins, K.~Howe, and B.~Nachman, {\it {Anomaly Detection for Resonant
  New Physics with Machine Learning}},  {\em Phys. Rev. Lett.} {\bf 121}
  (2018), no.~24 241803, [\href{http://arxiv.org/abs/1805.02664}{{\tt
  arXiv:1805.02664}}].

\bibitem{Hajer:2018kqm}
J.~Hajer, Y.-Y. Li, T.~Liu, and H.~Wang, {\it {Novelty Detection Meets Collider
  Physics}},  \href{http://arxiv.org/abs/1807.10261}{{\tt arXiv:1807.10261}}.

\bibitem{Heimel:2018mkt}
T.~Heimel, G.~Kasieczka, T.~Plehn, and J.~M. Thompson, {\it {QCD or What?}},
  {\em SciPost Phys.} {\bf 6} (2019), no.~3 030,
  [\href{http://arxiv.org/abs/1808.08979}{{\tt arXiv:1808.08979}}].

\bibitem{Farina:2018fyg}
M.~Farina, Y.~Nakai, and D.~Shih, {\it {Searching for New Physics with Deep
  Autoencoders}},  \href{http://arxiv.org/abs/1808.08992}{{\tt
  arXiv:1808.08992}}.

\bibitem{Collins:2019jip}
J.~H. Collins, K.~Howe, and B.~Nachman, {\it {Extending the search for new
  resonances with machine learning}},  {\em Phys. Rev.} {\bf D99} (2019), no.~1
  014038, [\href{http://arxiv.org/abs/1902.02634}{{\tt arXiv:1902.02634}}].

\bibitem{Roy:2019jae}
T.~S. Roy and A.~H. Vijay, {\it {A robust anomaly finder based on
  autoencoder}},  \href{http://arxiv.org/abs/1903.02032}{{\tt
  arXiv:1903.02032}}.

\bibitem{Field:1977fa}
R.~D. Field and R.~P. Feynman, {\it {A Parametrization of the Properties of
  Quark Jets}},  {\em Nucl. Phys.} {\bf B136} (1978) 1. [,763(1977)].

\bibitem{Krohn:2012fg}
D.~Krohn, M.~D. Schwartz, T.~Lin, and W.~J. Waalewijn, {\it {Jet Charge at the
  LHC}},  {\em Phys. Rev. Lett.} {\bf 110} (2013), no.~21 212001,
  [\href{http://arxiv.org/abs/1209.2421}{{\tt arXiv:1209.2421}}].

\bibitem{Waalewijn:2012sv}
W.~J. Waalewijn, {\it {Calculating the Charge of a Jet}},  {\em Phys. Rev.}
  {\bf D86} (2012) 094030, [\href{http://arxiv.org/abs/1209.3019}{{\tt
  arXiv:1209.3019}}].

\bibitem{Albanese:1984nv}
{\bf European Muon} Collaboration, J.~P. Albanese et~al., {\it {Quark Charge
  Retention in Final State Hadrons From Deep Inelastic Muon Scattering}},  {\em
  Phys. Lett.} {\bf 144B} (1984) 302--308.

\bibitem{Decamp:1991se}
{\bf ALEPH} Collaboration, D.~Decamp et~al., {\it {Measurement of charge
  asymmetry in hadronic Z decays}},  {\em Phys. Lett.} {\bf B259} (1991)
  377--388.

\bibitem{Nachman:2014qma}
{\bf ATLAS} Collaboration, B.~Nachman, {\it {Jet Charge with the ATLAS Detector
  using $\sqrt{s}=8$ TeV $pp$ Collision Data}},  in {\em {Proceedings, 2nd
  Conference on Large Hadron Collider Physics Conference (LHCP 2014): New York,
  USA, June 2-7, 2014}}, 2014.
\newblock \href{http://arxiv.org/abs/1409.0318}{{\tt arXiv:1409.0318}}.

\bibitem{Aad:2015cua}
{\bf ATLAS} Collaboration, G.~Aad et~al., {\it {Measurement of jet charge in
  dijet events from $\sqrt{s}$=8 TeV pp collisions with the ATLAS detector}},
  {\em Phys. Rev.} {\bf D93} (2016), no.~5 052003,
  [\href{http://arxiv.org/abs/1509.05190}{{\tt arXiv:1509.05190}}].

\bibitem{Sirunyan:2017tyr}
{\bf CMS} Collaboration, A.~M. Sirunyan et~al., {\it {Measurements of jet
  charge with dijet events in pp collisions at $\sqrt{s}=8$ TeV}},  {\em JHEP}
  {\bf 10} (2017) 131, [\href{http://arxiv.org/abs/1706.05868}{{\tt
  arXiv:1706.05868}}].

\bibitem{Fraser:2018ieu}
K.~Fraser and M.~D. Schwartz, {\it {Jet Charge and Machine Learning}},  {\em
  JHEP} {\bf 10} (2018) 093, [\href{http://arxiv.org/abs/1803.08066}{{\tt
  arXiv:1803.08066}}].

\bibitem{Mattig:1990wp}
P.~Mattig and W.~Zeuner, {\it {Final state photon bremsstrahlung in e+ e-
  ---$>$ Z0 ---$>$ hadrons as a tool for a precise measurement of the weak
  quark couplings}},  {\em Z. Phys.} {\bf C52} (1991) 31--42.

\bibitem{Abbiendi:2003ke}
{\bf OPAL} Collaboration, G.~Abbiendi et~al., {\it {Measurement of the partial
  widths of the Z into up and down type quarks}},  {\em Phys. Lett.} {\bf B586}
  (2004) 167--182, [\href{http://arxiv.org/abs/hep-ex/0312043}{{\tt
  hep-ex/0312043}}].

\bibitem{Hall:2018jub}
Z.~Hall and J.~Thaler, {\it {Photon isolation and jet substructure}},  {\em
  JHEP} {\bf 09} (2018) 164, [\href{http://arxiv.org/abs/1805.11622}{{\tt
  arXiv:1805.11622}}].

\end{thebibliography}\endgroup

\end{document}